\documentclass[conference]{IEEEtran}
\IEEEoverridecommandlockouts
\usepackage{cite}
\usepackage{amsmath,amssymb,amsfonts}
\usepackage{algorithmic}
\usepackage{graphicx}
\usepackage{textcomp}
\usepackage{xcolor}

\usepackage{booktabs}
\usepackage{epsfig}
\usepackage{wrapfig}
\usepackage{subfig}
\usepackage{multirow,tabularx}

\def\BibTeX{{\rm B\kern-.05em{\sc i\kern-.025em b}\kern-.08em
    T\kern-.1667em\lower.7ex\hbox{E}\kern-.125emX}}
\begin{document}


\title{PAS: A Position-Aware Similarity Measurement for Sequential Recommendation
}


\author{
	\IEEEauthorblockN{Zijie Zeng$^{1,4}$, Jing Lin$^{2,3}$, Weike Pan$^{2,3}$, Zhong Ming$^{2,3}$ and Zhongqi Lu$^1$}
	\IEEEauthorblockA{$^1$ Shenzhen HuaYun ZhongSheng Technology Co., Ltd, Shenzhen, China}
	\IEEEauthorblockA{$^2$ College of Computer Science and Software Engineering, Shenzhen University, Shenzhen, China}
	\IEEEauthorblockA{$^3$ Guangdong Laboratory of Artificial Intelligence and Digital Economy (SZ), Shenzhen, China}
	\IEEEauthorblockA{$^4$ Centre for Learning Analytics at Monash (CoLAM), Monash University, Australia}
	\IEEEauthorblockA{zijie.zeng@monash.edu, linjing2018@email.szu.edu.cn,  \{panweike, mingz\}@szu.edu.cn, zhongqi@outlook.com}
}

\maketitle

\begin{abstract}
	
The common item-based collaborative filtering framework becomes a typical recommendation method when equipped with a certain item-to-item similarity measurement. On one hand, we realize that a well-designed similarity measurement is the key to providing satisfactory recommendation services. On the other hand, similarity measurements designed for sequential recommendation are rarely studied by the recommender systems community. Hence in this paper, we focus on devising a novel similarity measurement called position-aware similarity (PAS) for sequential recommendation. The proposed PAS is, to our knowledge, the first count-based similarity measurement that concurrently captures the sequential patterns from the historical user behavior data and from the item position information within the input sequences. We conduct extensive empirical studies on four public datasets, in which our proposed PAS-based method exhibits competitive performance even compared to the state-of-the-art sequential recommendation methods, including a very recent similarity-based method and two GNN-based methods.
\end{abstract}


\section{Introduction}\label{sec:intro}


In this information-overload age, recommender systems have been playing a more and more important role in our daily lives. With recommender systems, people can have easier access to things or information they need than in the past. Behind all these conveniences are the cores of recommender systems, i.e., recommendation algorithms. Historical user-item interactions are crucial source data that can reveal both the users' preferences and items' properties. Sequential recommendation, a research problem that has drawn many attentions from the community of recommender systems in recent years~\cite{zhang2018next,zhang2019feature,DLSR-HUI-2020,fan2021continuous,zhang2021causerec,peng2021ham}, is the task of leveraging such historical user-item interactions to predict which item(s) are the most likely to be chosen/purchased by a user in the near future~\cite{fan2021continuous,zhang2021causerec,peng2021ham}. Sequential recommendation is also referred to as next-item recommendation (or next-item prediction) to emphasize when there is only one item needed to be predicted for the next time step~\cite{fan2021lighter,Zijie2019Next,wang2021time}.



Following the classification in ~\cite{Zijie2019Next}, we discuss two classes of collaborative filtering methods that have been adopted for sequential recommendation, namely model-based  methods and neighborhood-based methods. On one hand, factorization-based methods~\cite{WWW2010-Rendle-FPMC,ICDM2016-RuiningHE-Fossile} are among the most representative model-based methods. For instance, factorizing personalized Markov chain (FPMC)~\cite{WWW2010-Rendle-FPMC} is a sequential recommendation method that applies first-order Markov chains to the process of factorizing a user-item interaction matrix. Similarly, Fossil~\cite{ICDM2016-RuiningHE-Fossile} is proposed to fuse factored item-to-item similarity model (FISM)~\cite{KDD2013-FISM} with N-order Markov chains in order to tackle the sparsity problem existing in real-world applications. On the other hand, studies with regard to deep learning-based sequential recommendation have become increasingly popular~\cite{DLSR-HUI-2020}. For example, GRU4Rec~\cite{ICLR2016-RNN-GRU2Rec}, a variant of RNN with a pairwise ranking loss, is among the earliest works that apply deep learning techniques for sequential recommendation.    SR-GNN~\cite{wu2019session} models session sequences as session graphs, from which graph neural networks are applied to capture complex transitions of items. Target-aware sequential recommendation methods argue that the user representation can be influenced by not only historical items but also the target items. TAGNN~\cite{yu2020tagnn} is a state-of-the-art GNN-based sequential recommendation model which is equipped with a target-attentive module in order to learn different interest representation for different target items.  TGSRec~\cite{fan2021continuous} introduces a Transformer-based layer to capture temporal collaborative signals, which are unified with sequential patterns to boost the recommendation performance.

Although model-based methods have the advantages of being able to capture the underlying patterns~\cite{weiss2011unsupervised,lu2020algorithms} within a data and can provide impressive performance even under sparse conditions~\cite{yang2018enhancing,pirasteh2015exploiting}, the item-based collaborative framework~\cite{TOIS2004-ItemBasedCF} with a certain similarity measurement is still a practical choice in industry due to its interpretability and ease of implementation~\cite{xue2019deep,he2018nais}. It is also worth noticing that such a framework is so flexible that one can apply different similarity measurements to it to obtain different recommendation methods. Similarity measurements vary from count-based similarity measurements, e.g., Jaccard index, cosine similarity and bidirectional item similarity (BIS)~\cite{Zijie2019Next} to factorization-based similarity, e.g., FISM~\cite{KDD2013-FISM}. Some recent works go further to model high-order item-to-item similarity based on the attention mechanism, e.g., NAIS~\cite{he2018nais}. We notice that some item-to-item similarity measurements have been proposed to take into consideration of the temporal information within a data. The similarity based on the influential neighbors~\cite{JoS2013-GuangfuSUN-sequential} is calculated with a timespan threshold to filter out the item pairs $(x,y)$ where $x$ and $y$ concurrently appear in one item list of a specific user but the gap between their timestamps are not close enough. CF methods that leverage absolute timestamps are referred to as time-aware collaborative filtering methods~\cite{Zijie2019Next}, while methods utilizing only the position information of item sequences are regarded as sequence-aware methods~\cite{CSUR2018-Sequence,Zijie2019Next}. Similar to the similarity based on the influential neighbors~\cite{JoS2013-GuangfuSUN-sequential}, the bidirectional item similarity~\cite{Zijie2019Next} adopts a valid distance with respect to the gap between their item positions in ordered item sequences as a threshold to filter out the irrelevant item pairs and further introduce a reverse factor to tolerate the noisy data and maintain robustness. Although the item-based collaborative filtering model with bidirectional item similarity has achieved the state-of-the-art performance for sequential recommendation, we realize that such a method ignores the valuable item position information within the current input sequence. Hence in this paper we focus on devising a novel similarity measurement to cope with the aforementioned issues. Specifically, we summarize the main contributions of this paper as follows:
\begin{itemize}
	
	\item We devise a position-aware similarity (PAS), which concurrently captures the sequential patterns from the historical user behavior data and from the item position information within the current input sequences.	
	
	\item We propose a novel collaborative filtering model based on the proposed PAS to address the sequential recommendation problem.
	
	\item We conduct extensive empirical studies on four public datasets, in which our proposed collaborative filtering model shows competitive performance compared to the baseline methods. We also conduct experiments to explore (\romannumeral1) the influence of the tradeoff parameter $\lambda$ on the proposed PAS-based collaborative filtering method, and (\romannumeral2) the influence of different scaling functions on the proposed PAS-based collaborative filtering method.

\end{itemize} 

%


\section{Preliminaries}\label{sec:background}
In this section, we discuss related works that can serve as the preliminary knowledge for better understanding the proposed method. 

\subsection{Bidirectional Item Similarity}\label{sec:background-bis}
Bidirectional item similarity~\cite{Zijie2019Next}
is a novel similarity measurement that can effectively capture sequential patterns from sequences of user-item interaction data even under noisy conditions. The bidirectional item similarity from item $i'$ to item $i$ can be written as follows:

\begin{eqnarray}\label{eq:BIS}
s^{(\ell,\rho)}_{i' \rightarrow i} =
\frac{ \sum\limits_{ v \in \mathcal{U} }
	\delta(i,i' \in \mathcal{I}_v)  
	\delta_{p_{vii'}}^{(\ell,\rho)}
}
{ |\mathcal{U}_{i'} \cup \mathcal{U}_i| },
\end{eqnarray}
where $\delta_{ p_{vii'} }^{ (\ell,\rho) } = \delta( - \rho \ell \leq p_v(i) - p_v(i') \leq \ell )$ is a binary indicator function with $\ell$ as the valid distance between the positions of item $i$ and item $i'$, i.e., $p_v(i)$ and $p_v(i')$. $\mathcal{U}_{i}$ represents the set of users who have interacted with item $i$ and $\mathcal{I}_v$ denotes the items that have been interacted by user $v$. By restricting the gap between their item positions to the interval $[- \rho \ell , \ell]$, i.e., $- \rho \ell\leq p_v(i) - p_v(i')\leq\ell$,
BIS defines a valid co-occurrence as a pair of items whose item positions  w.r.t. the specific item sequence $\mathcal{L}_u$ are close enough. Besides, by introducing the reverse factor $\rho \in (0,1)$, BIS can capture sequential patterns even from a noisy data~\cite{Zijie2019Next}. 

\subsection{ Item-based Collaborative Filtering with  Bidirectional Item Similarity  }
With the bidirectional item similarity, the prediction rule w.r.t. the preference of user $u$ towards item $i$ can naturally be written as follows:

\begin{eqnarray}\label{eq:0.5}
\hat{r}_{ui}
&=&\sum_{i' \in \mathcal{I}_u\cap \mathcal{N}_i} s^{(\ell,\rho)}_{i' \rightarrow i},
\end{eqnarray}
where $\mathcal{N}_i$ is a set of nearest neighbors of item $i$ w.r.t. the bidirectional item similarity and $\mathcal{I}_u$ denotes the items that have been interacted by user $u$. To represent the user $ u $'s historical preference, we adopt the latest
active session window~\cite{mobasher2002using,Zijie2019Next} to preserve only the latest previous $k$ items w.r.t. item $i$ in $\mathcal{L}_u$, i.e.,  $\mathcal{I}_{u, \text{latest}(i)}^k$,  reaching the following prediction rule:

\begin{eqnarray}\label{eq:1}
\hat{r}_{ui}
&=&\sum_{i' \in \mathcal{I}_{u,\text{latest}(i)}^{k}} s^{(\ell,\rho)}_{i' \rightarrow i},
\end{eqnarray}
where we manually set $s^{(\ell,\rho)}_{i' \rightarrow i}=0$ when item $i'$ does not belong to the nearest neighborhood of item $i$, i.e., $i'\notin  \mathcal{N}_i$.

\section{Our Solution}\label{sec:solution}

\subsection{Disadvantages of Bidirectional Item Similarity}\label{sec:disadvantage of BIS}

\begin{figure}[!htb]
	\begin{center}
		\begin{tabular}{c}
			\psfig{figure = 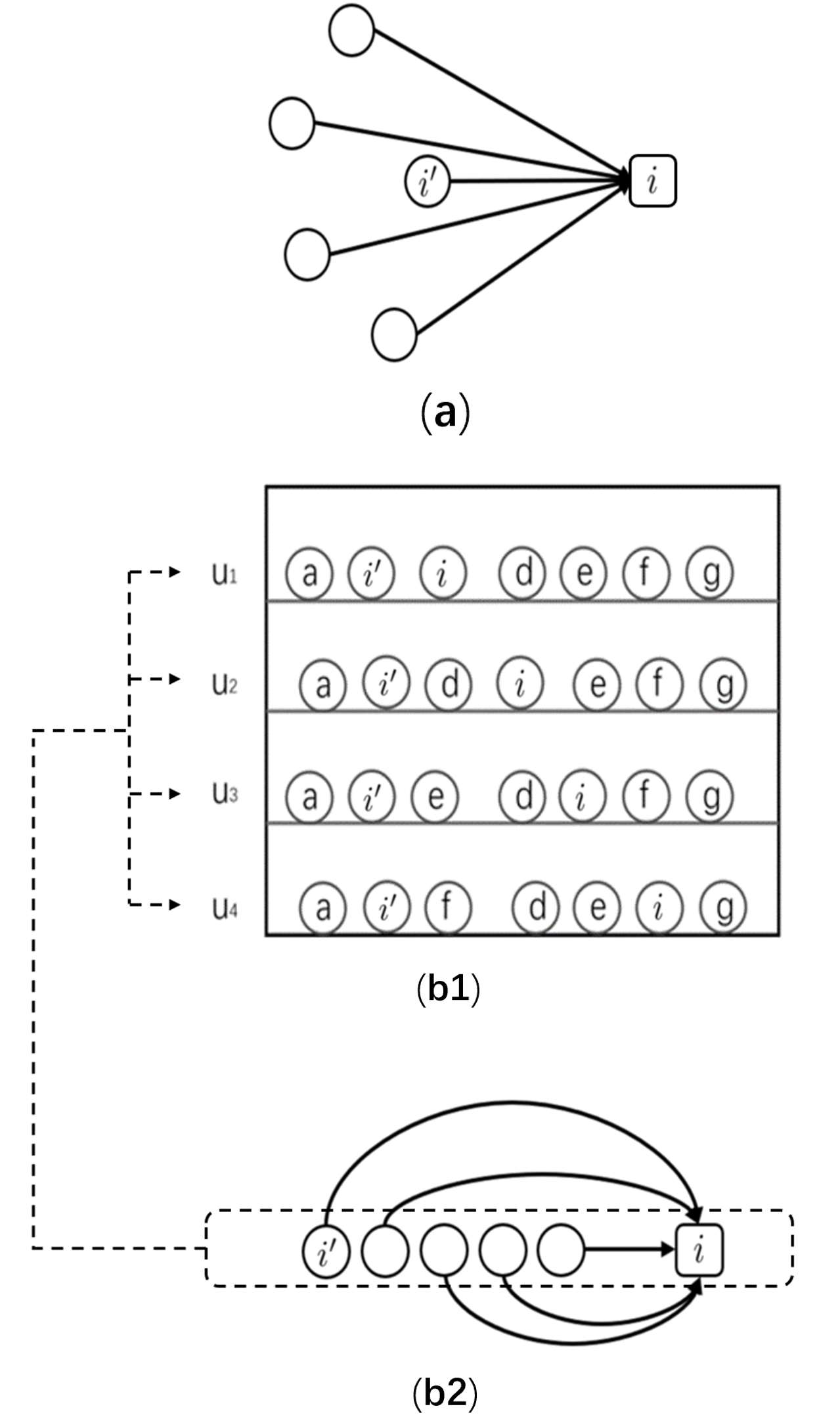, height = 3.4in, width = 2.4in}
			
		\end{tabular}
	\end{center}
	\caption{An illustration to show the disadvantages of using BIS in Equation (\ref{eq:1}) and an idea of improvement. From the angle of $\mathcal{I}_{u,\text{latest}(i)}^{k}$ being a set (see subfigure (\textbf{a})), the bidirectional item similarity  $s^{(\ell,\rho)}_{i' \rightarrow i}$ is calculated independent of the item position $p_u(i')$. However, from a sequence perspective (see subfigure (\textbf{b2})), we can have some extra information about the corresponding item positions, i.e., $p_u(i) - p_u(i') \geq 4$, for the current input sequence in this case. To utilize such information, we first scan over the historical sequences of the user data (subfigure (\textbf{b1})) and then calculate the specific item-to-item similarity basing on how frequent the case  $p_v(i) - p_v(i') \geq 4$ can be retrieved from the historical data, where $v\in\mathcal{U}$. Notice that $\mathcal{U}$ denotes the user set in the historical data.
	}
	\label{fig:1}
\end{figure}

Generally speaking, the whole process of calculating the preference of user $ u $ towards item $ i $ using Equation (\ref{eq:1})
can be divided into two steps:
\begin{itemize}
	\item Step 1: Calculate the bidirectional item similarity $s^{(\ell,\rho)}_{i' \rightarrow i}$ between two items $i'$ and $i$,  where $i' \in\mathcal{I}_{u,\text{latest}(i)}^{k}$.
	\item Step 2: Accumulate the calculated similarities to get $\hat{r}_{ui}$.
\end{itemize}
Note that the reason for Equation (\ref{eq:1}) being regarded as a sequential model mainly lies in Step 1, i.e., using the bidirectional item similarity as its similarity measurement. While admitting that BIS has shown its effectiveness in capturing sequential patterns, we also notice that the value of BIS, i.e., the value of $s^{(\ell,\rho)}_{i' \rightarrow i}$,  is fixed once the parameters $\ell$, $\rho$ and item pair $(i', i)$ are given. As a result, in Step 2 we can see the model does not consider the position information w.r.t. the items in the input sequence. We further present an illustration in Figure \ref{fig:1} to help illustrate the disadvantages of using BIS in Equation (\ref{eq:1}). Figure \ref{fig:1}(\textbf{a}) symbolically represents the prediction stage (Step 2), where we can see the user profile consists of a set of unsorted items from $\mathcal{I}_{u,\text{latest}(i)}^{k}$ (with $k=5$). On one hand, we are fully aware that $\mathcal{I}_{u,\text{latest}(i)}^{k}$ is, by definition, an ordered sequence of the latest $k$ items w.r.t. item $i$ in user $u$'s interacted item list. On the other hand, the way that $\mathcal{I}_{u,\text{latest}(i)}^{k}$ is used in Equation (\ref{eq:1}) is merely about its properties of being a set, ignoring its properties of being an ordered sequence. We take the case of item $i'$ in the user profile and the target item $i$, as illustrated in Figure \ref{fig:1}. From the angle of $\mathcal{I}_{u,\text{latest}(i)}^{k}$ being a set (see Figure \ref{fig:1}(\textbf{a})), the bidirectional item similarity  $s^{(\ell,\rho)}_{i' \rightarrow i}$ is calculated independent of the current input sequence, or independent of the item position $p_u(i')$, to be exact. However, from a sequence perspective (see Figure \ref{fig:1}(\textbf{b2})), we can have extra information about the corresponding item positions, i.e., $p_u(i) - p_u(i') \geq 4$ for the current input sequence in this case. We believe such item position information within the input sequence is informative. Different from the common practice of calculating the item-to-item similarity in a way which is independent of the current input sequence, we devise a new type of similarity measurement that concurrently captures the sequential patterns from the historical user behavior data as well as the item position information within the current input sequence.

\subsection{Position-Aware Similarity}\label{sec:PAS}
In this subsection, we formally define the position-aware similarity (PAS). Mathematically, our PAS from item $i'$ to item $i$ is defined as follows:
\begin{small}
	\begin{eqnarray}\label{eq:PAS}
	s^{(\ell,\rho,\lambda,L_u(i'))}_{i' \rightarrow i} =
	\frac{ 
		\sum\limits_{ v \in \mathcal{U} }
		\delta(i,i' \in \mathcal{I}_v) 			
		[
		(1-\lambda)  		
		\delta_{ p_{vii'} }^{ (\ell,\rho) }
		+
		\lambda
		\widetilde{\delta}_{ p_{vii'} }^{ (\ell, L_u(i')) }
		]
	}
	{ |\mathcal{U}_{i'} \cup \mathcal{U}_i| },
	\end{eqnarray}
\end{small}where $\widetilde{\delta}_{ p_{vii'} }^{ (\ell, L_u(i')) } = \delta( h(k - L_u(i')) < p_v(i) - p_v(i')\leq \ell ) 
$ is a position-aware binary indicator that enables our PAS to leverage the position information within the input sequence. $ L_u(i') \in \{1,2,...,k\} $ denotes the position of item $i'$ w.r.t. the current user $u$'s input sequence, i.e., $\mathcal{I}_{u,\text{latest}(i)}^{k}$. Function $h(\cdot)$ is introduced to scale the value of $k - L_u(i')$ so as to adapt to different datasets and  alleviate the potential sparsity problem, which will be discussed later in this subsection. For simplicity, let us start with $h(x)=x$. By devising the position-aware binary indicator, we hold the belief that the similarity from $i'$ to $i$ should be calculated only after when we are informed of the position of item $i'$ w.r.t. the current user $u$'s input sequence. Again, we refer to Figure \ref{fig:1}(\textbf{b}) to see the  reasonability of the position-aware binary indicator. Firstly, our goal is to calculate the similarity from $i'$ to $i$ and what we have already known is $p_u(i) - p_u(i') \geq 4$ for the current input sequence (Figure \ref{fig:1}(\textbf{b2})). Secondly, we scan over the historical sequences of the user data (Figure \ref{fig:1}(\textbf{b1})) and find that $p_v(i) - p_v(i') \geq 4$ never happens and the position-aware binary indicator is always zero. We argue that the current case $p_u(i) - p_u(i') \geq 4$ should be of low probability to exist if similar case cannot be retrieved from the historical data. Our design of the position-aware binary indicator naturally reflects our consideration.

We also incorporate $\delta_{ p_{vii'} }^{ (\ell,\rho) }$ (defined in Section \ref{sec:background-bis}) from BIS into our similarity framework to alleviate the sparsity problem caused by the aforementioned position-aware binary indicator with $h(x) = x$. To fully understand the reasonability of our proposed PAS, we go further to discuss two of its special cases:

\begin{itemize}
	\item When $\lambda = 0$, the general PAS can include BIS as a special case (see Equation (\ref{eq:BIS})).	
	
	\item When $\lambda = 1$, the general PAS loses its bidirectionality and reduces to unidirectional position-aware similarity, or PAS(uni) for short:

	\begin{equation}
	\begin{aligned}	
	s^{(\ell,\rho,1,L_u(i'))}_{i' \rightarrow i} 
	&=&	
	\frac{ \sum\limits_{ v \in \mathcal{U} }
		\delta(i,i' \in \mathcal{I}_v) 		 
		\widetilde{\delta}_{ p_{vii'} }^{ (\ell, L_u(i')) }				
	}
	{ |\mathcal{U}_{i'} \cup \mathcal{U}_i| }\label{eq:unipabis}.	
	\end{aligned}
	\end{equation}

\end{itemize}
From what has been discussed above, we can see clearly that our PAS aims to achieve a good balance between these two special cases. Intuitively, such a scheme is expected to be capable of leveraging both of their advantages. Specifically, by introducing the reverse factor $\rho$, BIS is able to capture the sequential patterns even under noisy conditions. At the same time, the disadvantage of BIS is that it does not consider the item position information w.r.t. the current input sequence, as has been discussed elaborately in Section \ref{sec:disadvantage of BIS}. As for PAS(uni), we can see $L_u(i')$ from the term $\delta( h(k - L_u(i')) < p_v(i) - p_v(i')\leq \ell )$ in Equation (\ref{eq:unipabis}) that the position information within the input sequence is considered. As a result of $\lambda=1$, it loses its bidirectionality and becomes unidirectional. Furthermore, we realize that PAS(uni) with $h(x) = x$ may suffer from a sparsity problem in practice. In general, for $ L_u(i')=r, r+1$, where $r\in [1, k-1] $, the proposed PAS(uni) is subject to the following rule:
\begin{small}	
	\begin{equation}
	\begin{aligned}
	s^{(\ell,\rho,1,r+1)}_{i' \rightarrow i} 
	&=	
	s^{(\ell,\rho,1,r)}_{i' \rightarrow i} 
	+	
	\frac{ \sum\limits_{ v \in \mathcal{U} }
		\delta(i,i' \in \mathcal{I}_v) 		 
		\delta( p_v(i) - p_v(i')=k-r)				
	}
	{ |\mathcal{U}_{i'} \cup \mathcal{U}_i| }\\
	&\geq s^{(\ell,\rho,1,r)}_{i' \rightarrow i}.		
	\end{aligned}
	\end{equation}	
\end{small}Therefore we have:
\begin{equation}\label{eq:sparsity}
s^{(\ell,\rho,1,1)}_{i' \rightarrow i}
\leq  s^{(\ell,\rho,1,2)}_{i' \rightarrow i} 
\leq \cdots \leq
s^{(\ell,\rho,1,k-1)}_{i' \rightarrow i}
\leq 
s^{(\ell,\rho,1,k)}_{i' \rightarrow i}.
\end{equation}
For items located at the head of the input sequence (usually with small item position), $k - L_u(i')$ is relatively large and cases with $\delta( k - L_u(i')<p_v(i) - p_v(i')\leq \ell ) = 1$ can be rare, causing the calculated value of PAS(uni) to be small and not reliable. We therefore introduce different forms of function $h(\cdot)$, a series of monotonic non-decreasing functions that satisfy the inequality $h(x)\leq x$ and it aims to make the true condition of the position-aware binary indicator easier to satisfy. We adopt the following three specific forms of function $h(\cdot)$ for our empirical studies:
\begin{equation}\label{eq:h_function}
h_a(x) = x, \qquad h_b(x) = \frac{1}{w}x, \qquad h_c(x) = w\lfloor \frac{1}{w}x \rfloor,
\end{equation}where $w>1$. Interestingly, such a sparsity problem does not exist in BIS because the binary indicator of BIS, i.e., $\delta_{ p_{vii'} }^{ (\ell,\rho) }$ (defined in Section \ref{sec:background-bis}), is not position-aware.

We summarize the contributions of our proposed PAS:
\begin{itemize}
	\item The design of our PAS follows a novel idea that
	the calculation of the specific item-to-item similarity should refer to both the historical user behavior sequences and the current input sequence.
	
	\item We add flexibility to PAS by introducing a monotonic non-decreasing   function $h(x)$ to scale the threshold of the true condition of the position-aware binary indicator.
	
	\item We incorporate the binary indicator $\delta_{ p_{vii'} }^{ (\ell,\rho) }$ (defined in Section \ref{sec:background-bis}) from BIS into our similarity framework to further alleviate the sparsity problem.
\end{itemize}


\subsection{Collaborative Filtering with Position-Aware Similarity}
With the proposed PAS, we reach a new collaborative filtering model with the following prediction rule:

\begin{eqnarray}\label{eq:5}
\hat{r}_{ui}
&=&\sum_{i' \in \mathcal{I}_{u,\text{latest}(i)}^{k}} s^{(\ell,\rho,\lambda,L_u(i'))}_{i' \rightarrow i},
\end{eqnarray}
where we manually set $s^{(\ell,\rho,\lambda,L_u(i'))}_{i' \rightarrow i}=0$ when $i'$ does not belong to the nearest neighborhood of item $i$, i.e.,  $i'\notin  \mathcal{N}_i$. What we need to point out is that although the value of our PAS is affected by the item position information within the current input sequence, we can actually fulfill the nearest neighborhood construction before a specific input sequence is provided. Considering that $ L_u(i') \in \{1,2,...,k\}$, we can calculate $s^{(\ell,\rho,\lambda,t)}_{i' \rightarrow i}$, $ t=1,2,\cdots,k $ for each item pair $(i',i)$, which will lead to higher time complexity of the training stage while the prediction stage remains unaffected.

Finally, while the benefit of introducing a time decay function into the prediction rule has been demonstrated in many works~\cite{CIKM2005-YiDING-Weight-CF,PiecewiseDecay2010,Zijie2019Next}, in this paper we focus on further exploiting the potential of the bidirectional item similarity and thus neither the proposed method nor the baseline methods will consider such a time decay scheme.

\section{Experiments}\label{sec:experiments and discussion}
\subsection{Datasets and Evaluation Metrics}

We conduct empirical studies on four public datasets, i.e., ML10M, Netflix, Beauty and Steam. ML10M (MovieLens 10M)~\cite{TIIS2015-MovieLensDataset} and Netflix\footnote{http://www.netflix.com/.} are two famous public datasets of movie rating records. Beauty is a categorized dataset from Amazon, while Steam is a review dataset from the eponymous game platform Steam. For rating records from ML10M, Netflix and Beauty, we follow~\cite{IEEE-IS2016-TJSL,TOIS2019-RoToR} and preserve only the records with a rating value equal to 5 so as to simulate positive feedback. For the review dataset Steam, we simply use review records as positive feedback. Note that in all of the four datasets, a timestamp is assigned to each record (rating or review), denoting exactly when the corresponding user-item interaction happened. For dataset processing, we first remove duplicate records so that only the earliest one is preserved if multiple records exist for a single (user, item) pair. Then we strictly follow~\cite{Zijie2019Next} to construct the training, test, validation data for our empirical studies. Note that an extra step is implemented for the processed Netflix and Steam datasets, i.e., randomly sampling a subset with no more than $ 20,000 $ users. We adopt NDCG@K ~\cite{NDCG2002cumulated}  and 1-call@K~\cite{shi2012climf} as the evaluation metrics.

\begin{table}[!htb]
	
	\caption{
		\MakeUppercase{Statistics of the processed datasets.}	
	}
	
	\label{tbl:dataset}
	\begin{center}
		
		
		\begin{tabular}{ c|rrrc }
			\toprule
			
			Dataset&\#Users &\#Items &\#Records & Average Length \\ \midrule
			ML10M & $ 40,600 $ &	$ 8,625 $ &	$ 1,411,225  $& $ 34.76 $\\
			Netflix & $ 20,000 $ & $ 14,395 $ & $ 968,498 $  & $ 67.15 $\\
			Beauty & $ 5,195 $ &	$ 41,727 $ &	$ 87,492 $ & $ 16.84 $\\
			Steam &$ 20,000  $&	$ 11,955 $ &	$ 454,180 $ & $ 22.71 $\\				
			\bottomrule
			
		\end{tabular}
	\end{center}
\end{table}

\subsection{A Sparsity Problem in Unidirectional Position-Aware Similarity}\label{sec:sparsity}
\begin{figure}[!htb]
	\begin{center}
		\begin{tabular}{ccc}
			\psfig{figure=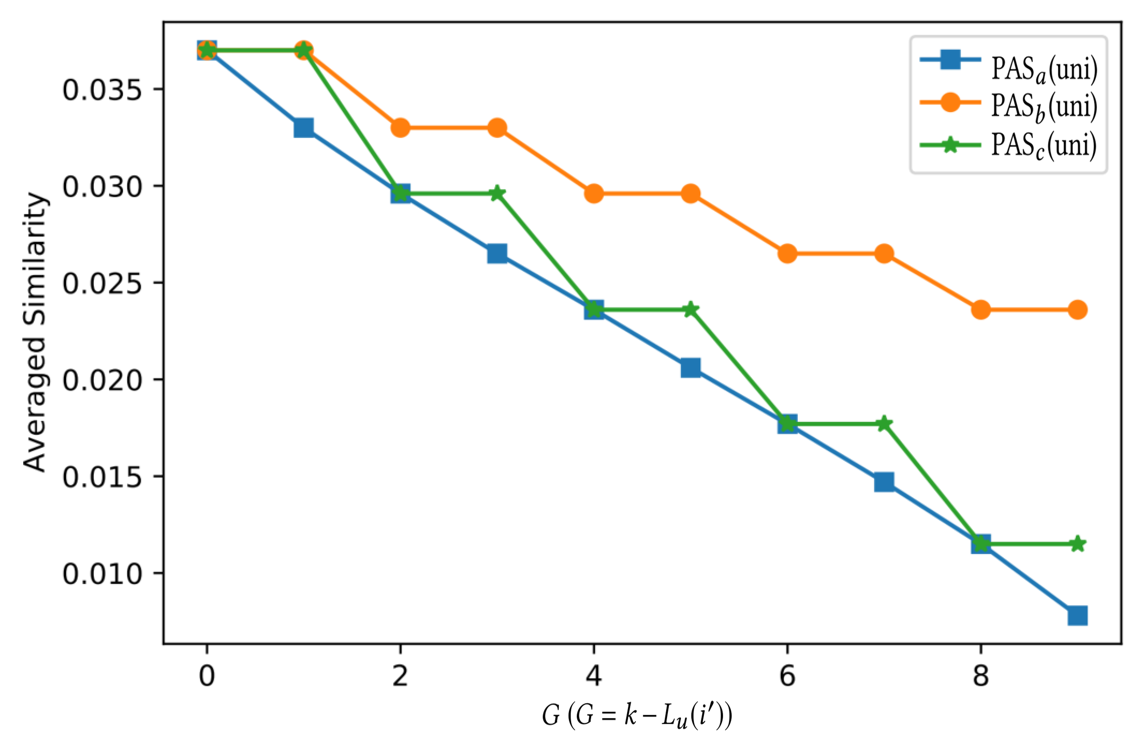,height=0.85in,width=1.2in}& &
			\psfig{figure=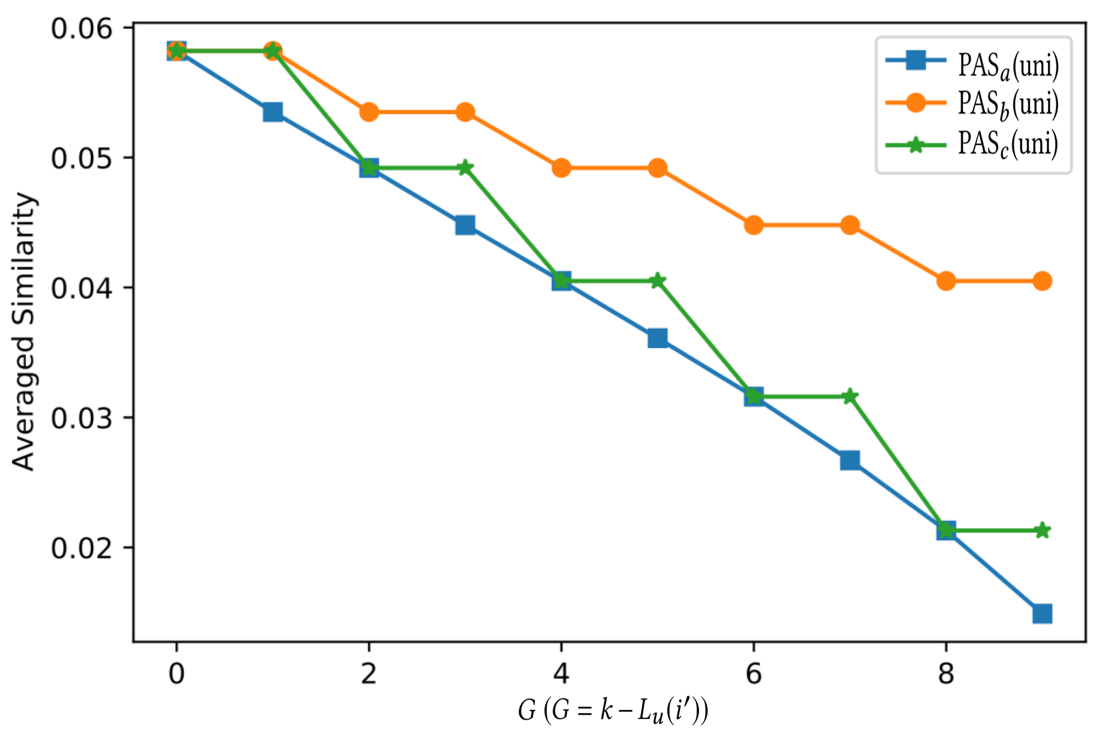,height=0.85in,width=1.2in}\\ML10M && Netflix\\\\
			\psfig{figure=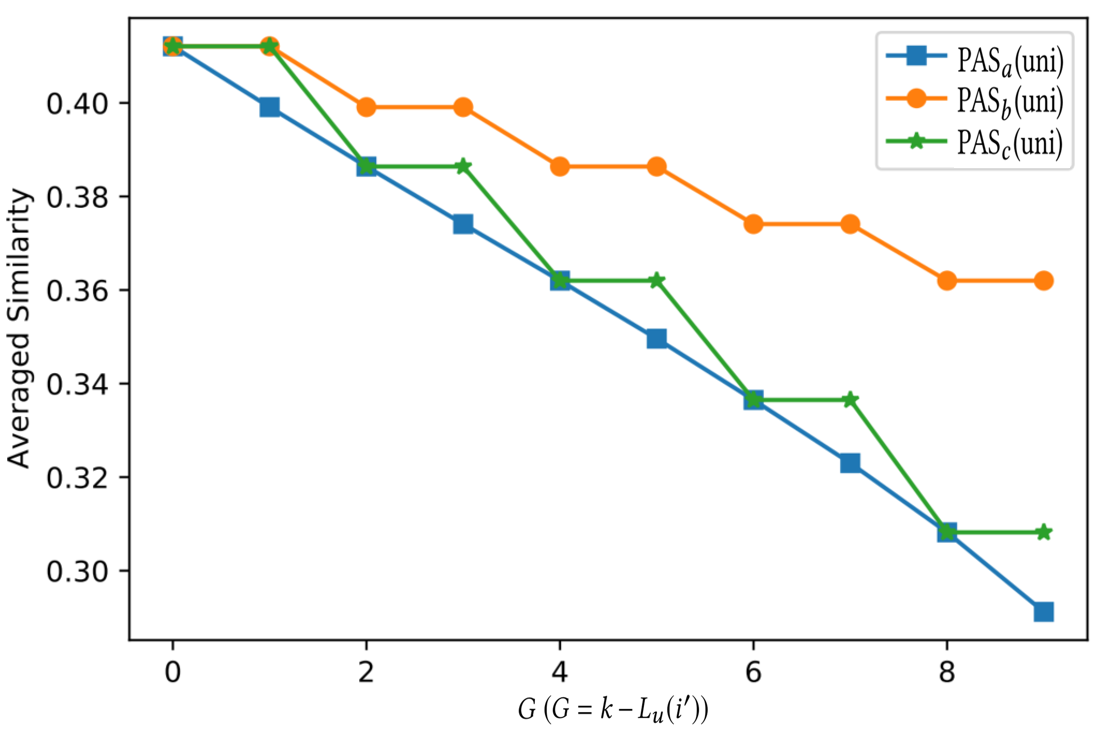,height=0.85in,width=1.2in}& &	\psfig{figure=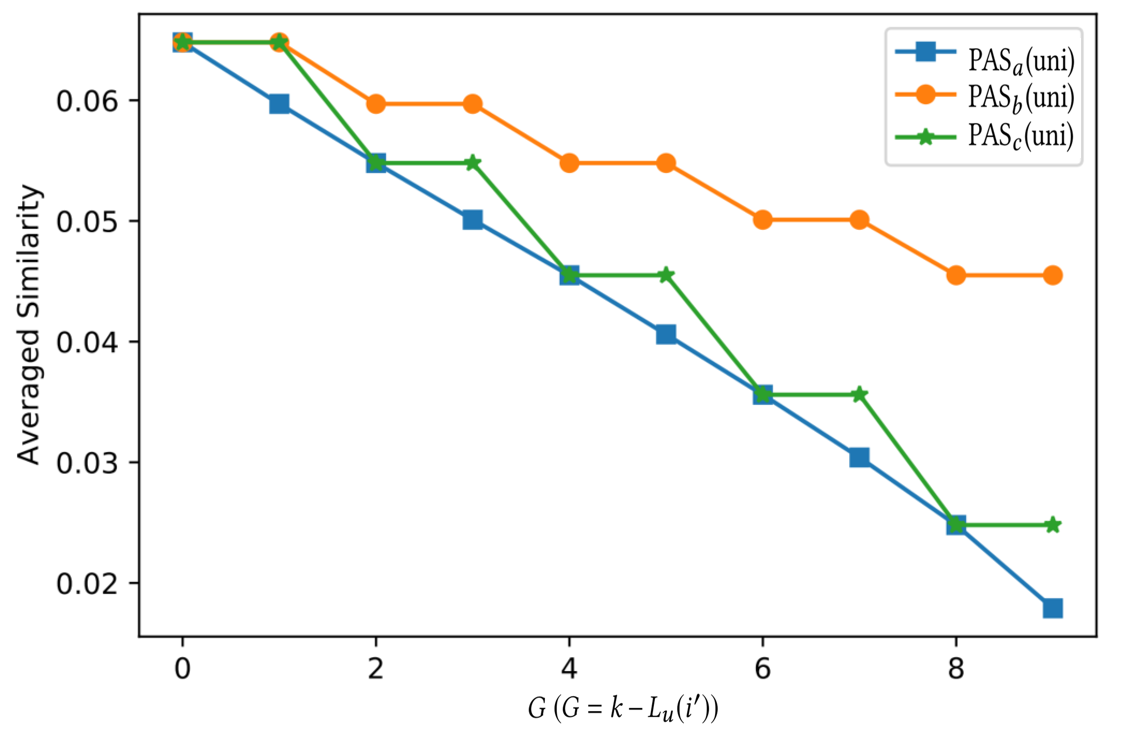,height=0.85in,width=1.2in}\\Beauty & & Steam	
			
		\end{tabular}
	\end{center}
	\caption{The average value of the unidirectional position-aware similarity (PAS(uni)) with different scaling functions $h(\cdot)\in \{h_a(\cdot),h_b(\cdot) , h_c(\cdot)\}$, i.e., PAS$ _a $(uni), PAS$ _b $(uni) and PAS$ _c $(uni). Generally, we can see that the average value decreases as $G$ increases ($G = k - L_u(i')$). Moreover, we notice that the speed of the decreasing varies with different scaling functions.
	}
	\label{fig:2}
\end{figure}

As has been indicated by Equation (\ref{eq:sparsity}), PAS(uni)  with scaling function $h(x)=x$ may suffer from a problem of sparsity and become unreliable, especially for items located at the head of the input sequence. In this section, we provide some empirical statistics so as to give an intuitive and direct understanding about the aforementioned sparsity problem. Specifically, for PAS(uni) with different scaling functions, we fix the length of the input sequence $k$  to $10$, and calculate the average value of PAS(uni) for $L_u(i') = 1,2,\cdots,10$:
$$
avg(
s^{(\ell, \rho, 1, L_u(i'))}_{i' \rightarrow i}), \quad
i \in \mathcal{I}, i'\in \mathcal{N}_i.
$$
Note that we only preserve the nearest $ 20 $ neighbors for each item $ i\in \mathcal{I} $, i.e., $|\mathcal{N}_i| \leq 20$  and fix $\ell$ as 10.  It is also noteworthy that we do not specify the value for  $\rho$ because it is meaningless to PAS(uni). Figure \ref{fig:2} shows clearly how the averaged similarity decays as $G$ increases (or $L_u(i')$ decreases), where  $k=10$ is the length of the input sequence and $G = k - L_u(i')$. As expected, the speed of decreasing varies with different scaling functions. Specifically, PAS$_a$(uni) decreases with the fastest pace and PAS$_b$(uni) decreases with the slowest pace. In fact, the scaling function serves as a way to control the decreasing speed of PAS(uni), which will eventually have influence on the recommendation performance of the proposed method (see Section \ref{sec:scaling}).

%

\subsection{Baselines and Parameter Configurations}\label{sec:settings}
In this section, we briefly introduce the proposed collaborative filtering method as well as several baseline methods with which we conduct our empirical studies:
\begin{itemize}
	\item CS: Item-based collaborative filtering with cosine similarity as the similarity measurement.
	\item BPR-MF: Bayesian personalized ranking~\cite{UAI2009-Rendle-BPR} is a typical factorization-based collaborative filtering method that	factorizes the user-item interaction matrix via a pair-wise loss.
	\item FPMC: Factorizing personalized Markov chain~\cite{WWW2010-Rendle-FPMC} is a sequence-aware recommendation method that applies first-order Markov chains to the process of factorizing a user-item interaction matrix. The basket size $ b $ is chosen from $ \{ 1, 2, 3 \} $ via the performance on the validation data.
	\item GRU4Rec: GRU4Rec~\cite{ICLR2016-RNN-GRU2Rec} is an RNN-based collaborative filtering model for sequential recommendation. We follow~\cite{ICLR2016-RNN-GRU2Rec} and fix the hidden size, batch size and dropout  as $ 100 $, $ 50 $, $ 0.2 $ respectively, and choose the learning rate $ lr $ in the Adam strategy from $ \{0.005, 0.001, 0.0005\} $.
	\item BIS: Collaborative filtering using bidirectional item similarity. The model can also be described in Equation (\ref{eq:5}) with the tradeoff parameter $\lambda=0$.

	\item SR-GNN: SR-GNN~\cite{wu2019session} is a recently proposed graph neural network model for sequential recommendation. We fix the hidden size as $20$ and select the length of a session from $\{5,10,20,40\}$. We adopt the default settings\footnote{https://github.com/CRIPAC-DIG/SR-GNN/.} for the remaining parameters.
	
	\item TAGNN: TAGNN~\cite{yu2020tagnn} is a target attentive version of SR-GNN. We fix the hidden size as $20$ and select the length of a session from $\{5,10,20,40\}$. We adopt the default settings\footnote{https://github.com/CRIPAC-DIG/TAGNN/.} for the remaining parameters' configuration.
	
	\item PAS($\lambda$): Collaborative filtering using PAS. The model is also described in Equation (\ref{eq:5}) with tradeoff parameter $\lambda\in\{0.5,1\}$. We refer to  PAS($\lambda)$ with $h_a(\cdot)$ as PAS$_a(\lambda)$.  PAS$_b(\lambda)$ and PAS$_c(\lambda)$ are defined similarly. We set $w=2$ for $w$ in $h(\cdot)$.

\end{itemize}

For PAS($\lambda = 0.5, 1$) and BIS, we follow~\cite{Zijie2019Next} and fix the reverse factor as $0.2$. We keep the length of the input sequence $k$ equivalent to the valid distance $\ell$, i.e., $k=\ell$. Following~\cite{Zijie2019Next} we select the best $\ell$ (or $k$) from $\{5,10,20,40\}$. The number of the nearest neighbors preserved for each item is set to $20$. 

For BPR-MF and FPMC, we fix the number of latent dimensions to $20$. We choose the tradeoff parameter on regularization terms $\alpha$ from $\{0.001,0.01,0.1\}$ and the iteration number $T$ is from $\{100, 500, 1000\}$. For GRU4Rec, we use a sliding window of size $ 8 $, moving step $ 1 $, maximum iteration number $ 50 $, and adopt an early stopping strategy by checking the performance in 10 continuous iterations. Finally, we search the best parameter configuration for each of the aforementioned recommendation methods, according to their performance on validation data under $\text{1-call}@5$ metric~\cite{shi2012climf}, which is consistent with \cite{Zijie2019Next}, and report their performance on the test set.

\subsection{Recommendation Performance}\label{sec:results}

\begin{table*}[!htb]
	
	\caption{
		\MakeUppercase{Recommendation performance of our proposed PAS($\lambda=0.5,1$) and other methods on four datasets, i.e., ML10M, B\MakeLowercase{eauty}, N\MakeLowercase{etflix} and S\MakeLowercase{team}. Note that the best results are marked in bold and the second-best results are underlined. Paired t-tests were employed on $\text{1-call}@5$ across four datasets to examine whether PAS($\lambda=0.5,1$) significantly outperforms the state-of-the-art count-based method, i.e., BIS. The test results are shown in the last column, using p-value to indicate the significance level.	P-values less than 0.05 are marked in bold.		
			}		
	}
	
	\label{tbl:main-result}
	\begin{center}
		
		\resizebox{0.80\textwidth}{!}{

			\begin{tabular}{c|cccccccc|c} 
				\toprule
				\multirow{2}{*}{Method} & \multicolumn{2}{c}{ML10M} & \multicolumn{2}{c}{Beauty} & \multicolumn{2}{c}{Netflix} & \multicolumn{2}{c}{Steam}  \\
				& $\text{NDCG}@5$  & $\text{1-call}@5$         & $\text{NDCG}@5$  & $\text{1-call}@5$         & $\text{NDCG}@5$  & $\text{1-call}@5$          & $\text{NDCG}@5$  & $\text{1-call}@5$ & p-value         \\ 
				\midrule

				CS                      & 0.0354 & 0.0558           & 0.0119 & 0.0167            & 0.0488 & 0.0723             & 0.0198 & 0.0304  &     -     \\
				BIS                     & 0.0585 & 0.0854           & 0.0183 & 0.0274            & 0.0692 & 0.1019             & 0.0203 & 0.0324   &    -     \\				
				PAS$_a$($\lambda=1$)              & 0.0586 & 0.0862           & 0.0180 & 0.0266            & 0.0708 & 0.1022             & 0.0212 & 0.0327     &   0.687    \\
				PAS$_b$($\lambda=1$)             & 0.0586 & 0.0868           & 0.0187 & 0.0277            & 0.0684 & 0.1006             & \underline{0.0233} & 0.0348       &  0.442   \\			
				PAS$_c$($\lambda=1$)             & 0.0594 & 0.0878           & 0.0187 & 0.0274            & 0.0719 & 0.1044             & 0.0214 & 0.0327      &  0.146    \\				
				PAS$_a$($\lambda=0.5$)              & \underline{0.0611} & \textbf{0.0907}           & \textbf{0.0218} & \underline{0.0300}            & 0.0746 & 0.1083             & 0.023 & 0.0349   &    \textbf{0.023}     \\				
				PAS$_b$($\lambda=0.5$)             & 0.0600 & 0.0890           & 0.0202 & \textbf{0.0304}            & 0.0701 & 0.1041             & 0.0227 & 0.0346         & \textbf{0.004}  \\
				PAS$_c$($\lambda=0.5$)             & \underline{0.0611} & \underline{0.0901}           & \underline{0.0209} & \underline{0.0300}            & 0.0751 & 0.1099             & 0.0224 & 0.0342     &   0.054    \\\midrule
				
				GRU4REC                 & 0.0399 & 0.0629           & 0.0119 & 0.0171            & 0.0505 & 0.0768             & 0.0097 & 0.0153      &   -   \\
				BPR-MF                  & 0.0317 & 0.0509           & 0.0085 & 0.0133            & 0.0318 & 0.0511             & 0.0169 & 0.0281      &   -   \\
				FPMC                    & 0.0529 & 0.0825           & 0.0115 & 0.0186            & 0.0433 & 0.0674             & 0.0230 & \underline{0.0368}      &   -   \\
				SR-GNN                  & 0.0475 & 0.0741           & 0.0053 & 0.0095            & \underline{0.0760} & \underline{0.1123}             & 0.0221 & 0.0367       &  -   \\
				TAGNN                   & 0.0411 & 0.0662           & 0.0044 & 0.0076            & \textbf{0.0795} & \textbf{0.1169 }            & \underline{0.0233} & \textbf{0.0380}      &   -   \\
				\bottomrule
			\end{tabular}

		}
		
	\end{center}
	\vspace{-4mm}
\end{table*}

We conduct comprehensive empirical studies on the aforementioned four datasets (see Table \ref{tbl:main-result} for the results), from which we have the following observations:

\begin{itemize}
	\item \textbf{The improvement of PAS over BIS}: It is observed that PAS($\lambda=1$) exhibits higher performance than BIS on 3 out of 4 datasets, regardless of which scaling function $h(\cdot)$ is equipped. More importantly, PAS($\lambda=0.5$) performs better than PAS($\lambda=1$) in 11 out of 12
	cases of performance comparison. We conclude that (\romannumeral1) introducing the position information can be beneficial, and (\romannumeral2) the idea of combining BIS and PAS(uni) into one similarity framework is effective. Considering that BIS is, as has been indicated in Section \ref{sec:PAS}, a special case of PAS, we believe the proposed PAS is a general yet effective similarity measurement. We also go further to explore the influence of $\lambda$ in Section \ref{sec:lambda}.	
	
	\item \textbf{Comparison with GNN-based methods}: GNN-based models are popular methods that have achieved the state-of-the-art performance in sequential recommendation and session-based recommendation. According to the empirical results, the proposed PAS($\lambda=0.5$) beats both SR-GNN~\cite{wu2019session} and TAGNN~\cite{yu2020tagnn} on 2 out of 4 datasets (ML10M and Beauty). On one hand, we admit the advantages of GNN-based methods in that they leverage graph neural networks to capture complex transitions of items from session graphs. On the other hand, it should be pointed out that the main contribution of this paper is to propose a novel position-aware similarity measurement. We believe that the proposed PAS($\lambda$) is on the whole a competitive and practical model, considering its robustness, interpretability and ease of implementation.

	\item \textbf{Comparison with other baseline methods}: On one hand, it is easy to understand that CS and BPR-MF achieve unsatisfactory performance  compared to the proposed method because they are non-sequential collaborative filtering recommendation methods. On the other hand, the proposed PAS($\lambda=0.5$) stays competitive even compared to sequential CF baseline methods such as GRU4REC and FPMC. 
\end{itemize}


\subsection{Influence of the Scaling Function $ h(\cdot) $}\label{sec:scaling}
\begin{figure}[!htb]
	\begin{center}
		\vspace{-2mm}
		\begin{tabular}{ccc}
			\psfig{figure=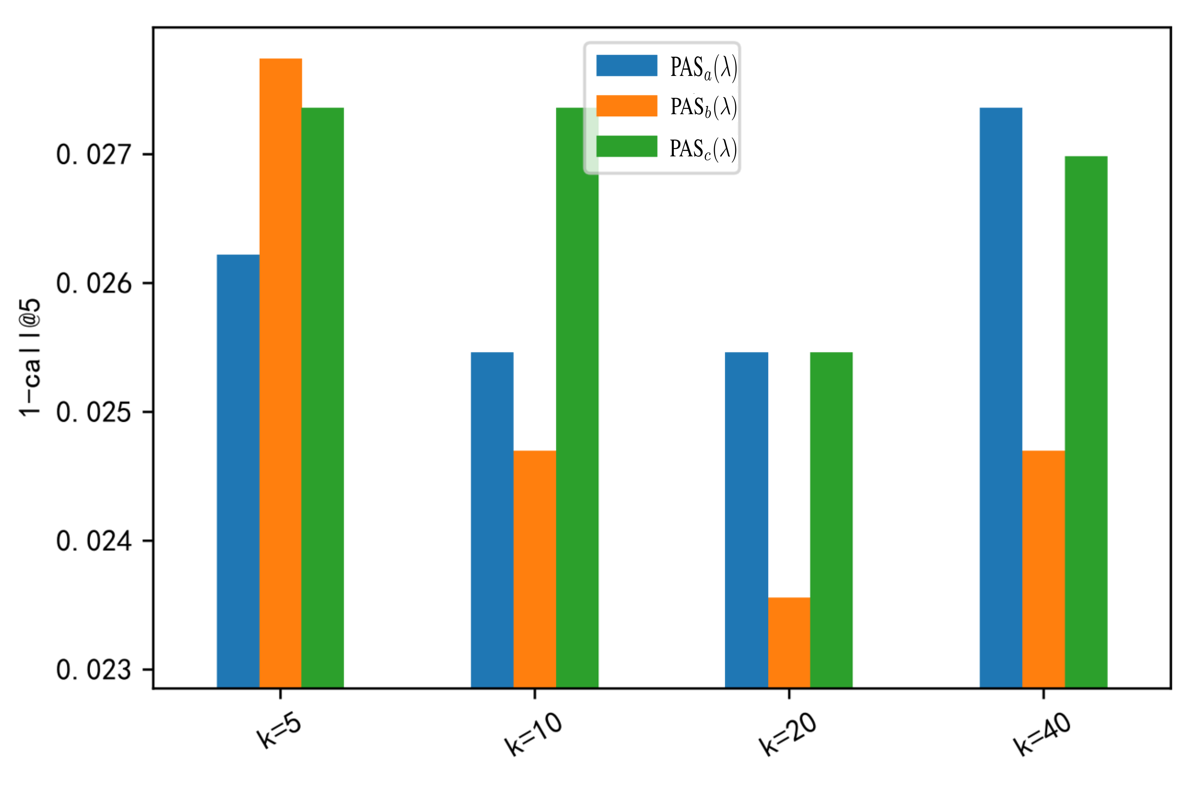,height=0.8in,width=1.20in}&&\psfig{figure=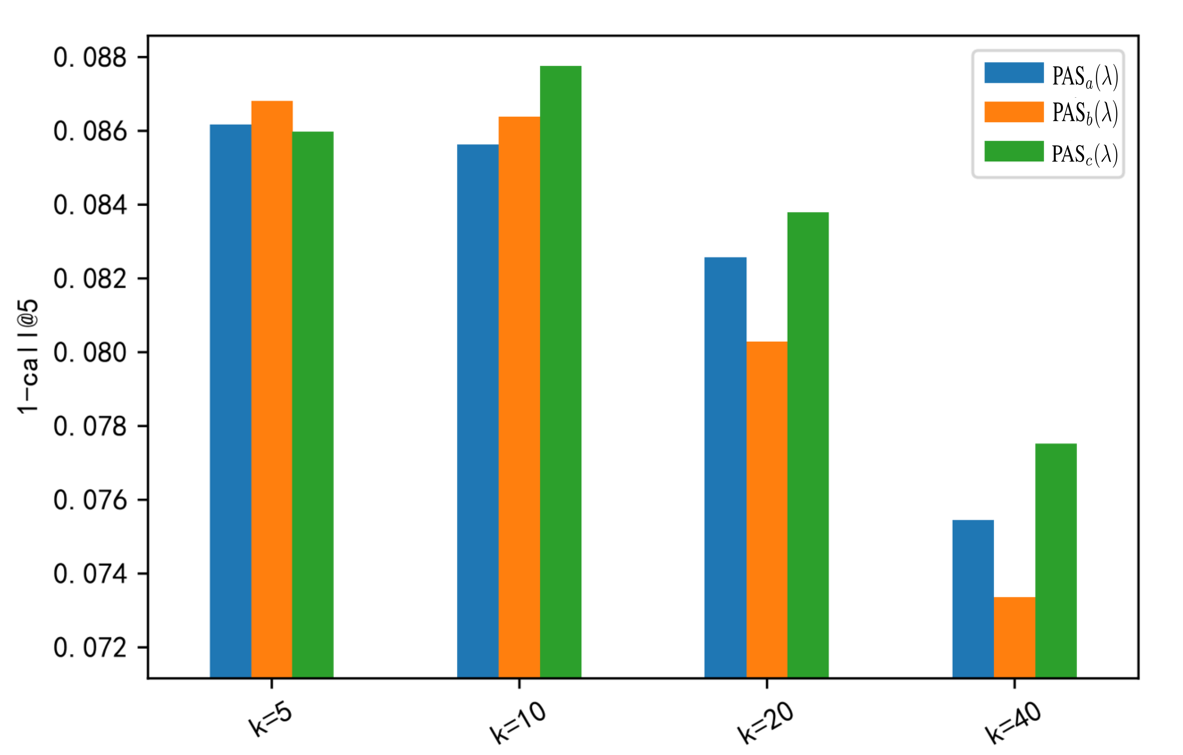,height=0.8in,width=1.20in}\\
			Beauty&&ML10M\\\\			
			\psfig{figure=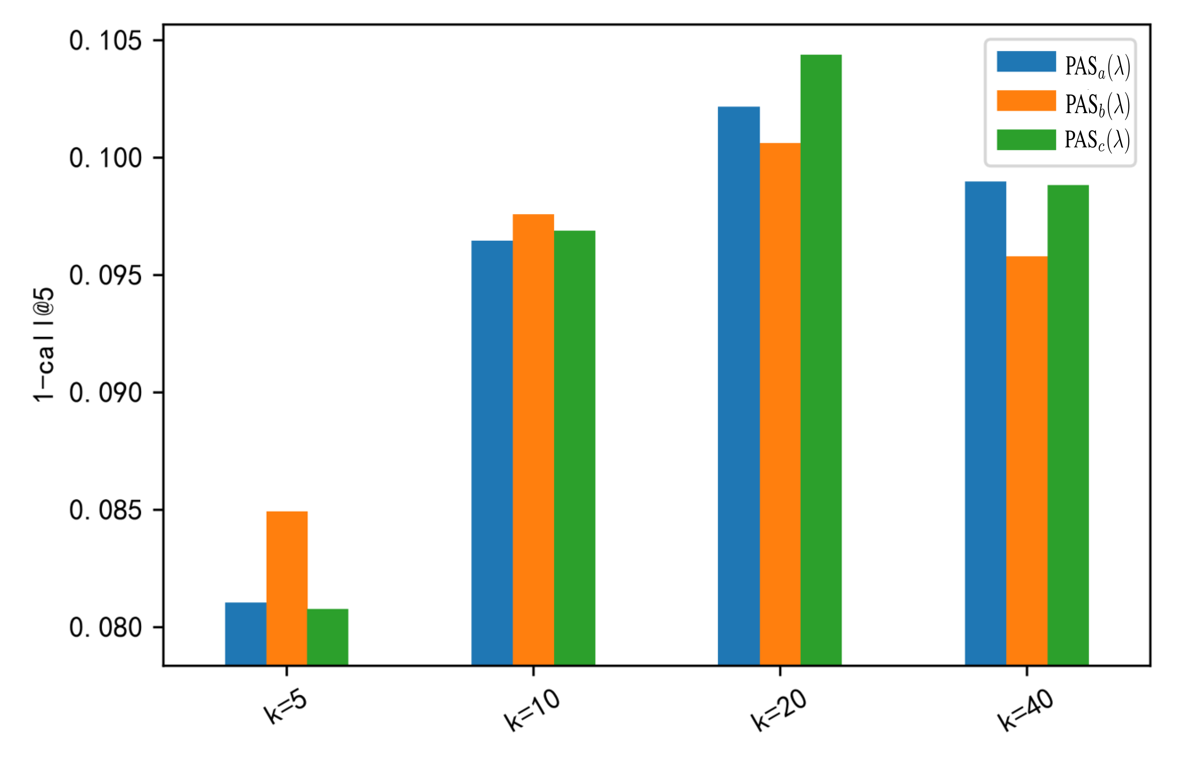,height=0.8in,width=1.20in}
			&&\psfig{figure=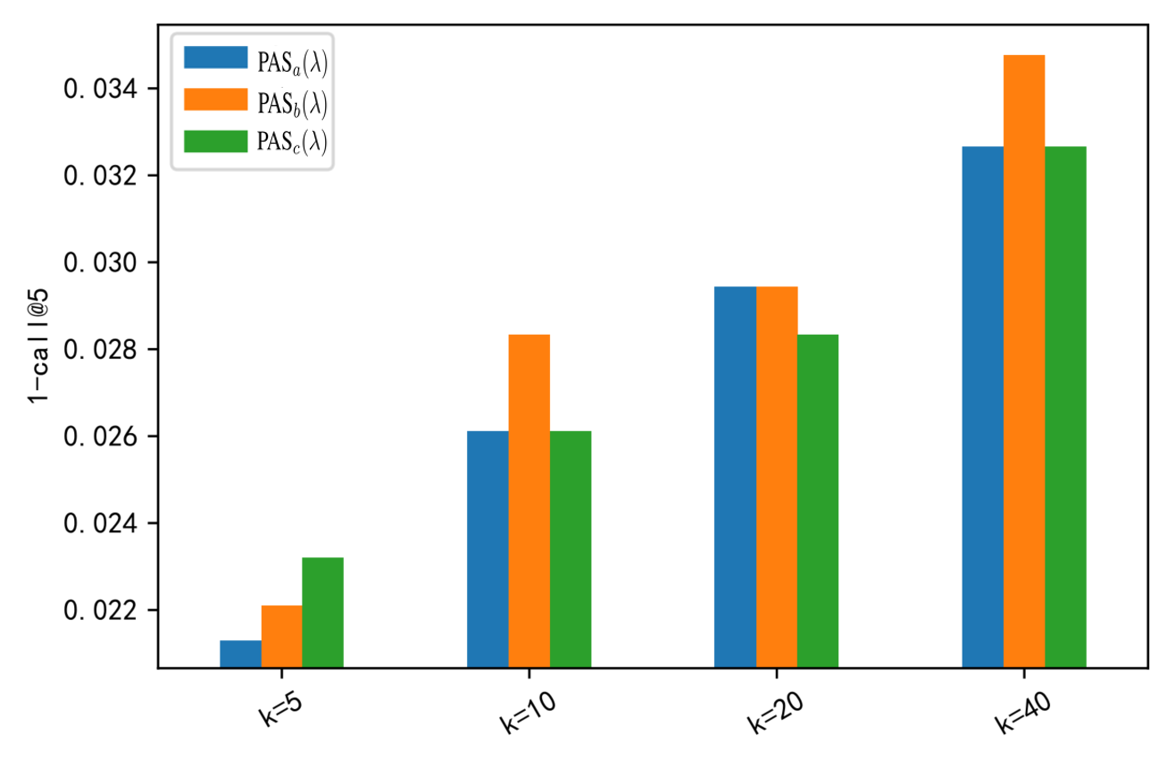,height=0.8in,width=1.20in}\\
			Netflix && Steam						
			
		\end{tabular}
	\end{center}
	\caption{
		Exploration of the influence of different position-aware functions $h(\cdot)\in \{h_a(\cdot), h_b(\cdot), h_c(\cdot)\} $ on the proposed PAS-based collaborative filtering method. 
	}\label{fig:experiment4-2}
\end{figure}

In this section, we explore how different forms of scaling function $h(\cdot)$ (as described in Equation (\ref{eq:h_function})) will affect the recommendation performance of our proposed method, i.e., PAS($\lambda$). Specifically, we follow the experimental settings described in Section \ref{sec:settings}, except that we fix $\lambda$ as $1.0$ and report the performance for all $k\in\{5,10,20,40\}$ on test data. The main results are presented in Figure \ref{fig:experiment4-2}. We refer to PAS($\lambda$) with $h(x) = x$ as PAS($\lambda$) of \textit{Non-Scaling} scheme.  PAS$(\lambda)$ with scaling function other than $h(x) = x$ is referred to as PAS$(\lambda)$ of \textit{Scaling} scheme. For example, PAS$_a(\lambda)$ adopts the \textit{Non-Scaling} scheme. Both PAS$_b(\lambda)$ and PAS$_c(\lambda)$ adopt the \textit{Scaling} scheme. We group the empirical results according to the length of the input sequence $k$. Each group consists of the empirical results of PAS$(\lambda)$ with three different $h(\cdot)$ functions. 

We summarize our observations in Table \ref{tbl:summary}. Specifically, we notice that the average $\text{1-call}@5$ of the \textit{Non-Scaling} PAS$_a(\lambda)$  across all $ 16 $ groups is $ 0.0577  $. The average $\text{1-call}@5$ of the \textit{Scaling} PAS$_b(\lambda)$ across all $ 16 $ groups is $ 0.0576 $. The average $\text{1-call}@5$ of \textit{Scaling} PAS$_c(\lambda)$ across all $ 16 $ groups is $ 0.0583 $. We also observe that the \textit{Non-Scaling} PAS$_a(\lambda)$ achieves the best performance in merely $12.5\% (2 \text{ out of } 16) $ groups of empirical results. The \textit{Scaling} PAS$_b(\lambda)$ achieves the best performance in $37.5\% (6 \text{ out of } 16) $ groups of empirical results. The \textit{Scaling} PAS$_c(\lambda)$ achieves the best performance in $37.5\% (6 \text{ out of } 16) $ groups of empirical results. Note that groups with multiple best performances are excluded. To summarize, we do not observe a scaling function that can outperform others in every case. However, we notice that the proposed method with a \textit{Scaling} scheme (PAS$_b(\lambda)$ and PAS$_c(\lambda)$) achieves the best in $75\% (12 \text{ out of } 16) $ groups of empirical results, indicating the advantages of a \textit{Scaling} scheme to some extent and it is worth consideration in practical use.

\begin{table}[!htb]
	
	
	\caption{
		\MakeUppercase{Statistics observed from Figure \ref{fig:experiment4-2}. The average $\text{\MakeLowercase{1-call}}@5$ of the \textit{Non-Scaling} PAS$_\MakeLowercase{a}(\lambda)$  across all $ 16 $ groups is $ 0.0577  $. The \textit{Non-Scaling} PAS$_\MakeLowercase{a}(\lambda)$ achieves the best performance in merely $2$ out of $ 16 $ groups of empirical results. Note that groups with multiple best performances are excluded. The observations of PAS$_\MakeLowercase{b}(\lambda)$ and PAS$_\MakeLowercase{c}(\lambda)$ can be interpreted similarly.}	
	}
	
	\label{tbl:summary}
	\begin{center}
		
		\begin{tabular}{ c|ccc  }
			\toprule
			& PAS$_a(\lambda)$ &PAS$_b(\lambda)$&PAS$_c(\lambda)$\\\midrule
			
			$\text{1-call}@5$& 0.0577 & 0.0576 &0.0583\\
			Winning$\%$& $12.5\%(2/16)$  & $37.5\%(6/16)$ & $37.5\%(6/16)$ \\
			\bottomrule
			
		\end{tabular}
		
	\end{center}
	\vspace{-3mm}
\end{table}


%
%

\subsection{Influence of the Tradeoff Parameter}\label{sec:lambda}
In this subsection, we explore how the tradeoff parameter $\lambda$ affects the performance of our proposed method. Specifically, we follow the experimental settings described in Section \ref{sec:settings}, except that we report the performance for all parameter combinations $(k,\lambda, h(\cdot))$ on test sets, where $k\in\{5,10,20,40\}$ is the length of the input sequence. $\lambda\in\{0,0.2,0.4,\cdots,1.0\}$ denotes range of the tradeoff parameter. Scaling function $ h(\cdot)$ is defined in Equation (\ref{eq:h_function}). 

\begin{figure*}[!htb]
	\begin{center}
		\resizebox{0.97\textwidth}{!}{
		\begin{tabular}{cccccccccccccc}
			\psfig{figure=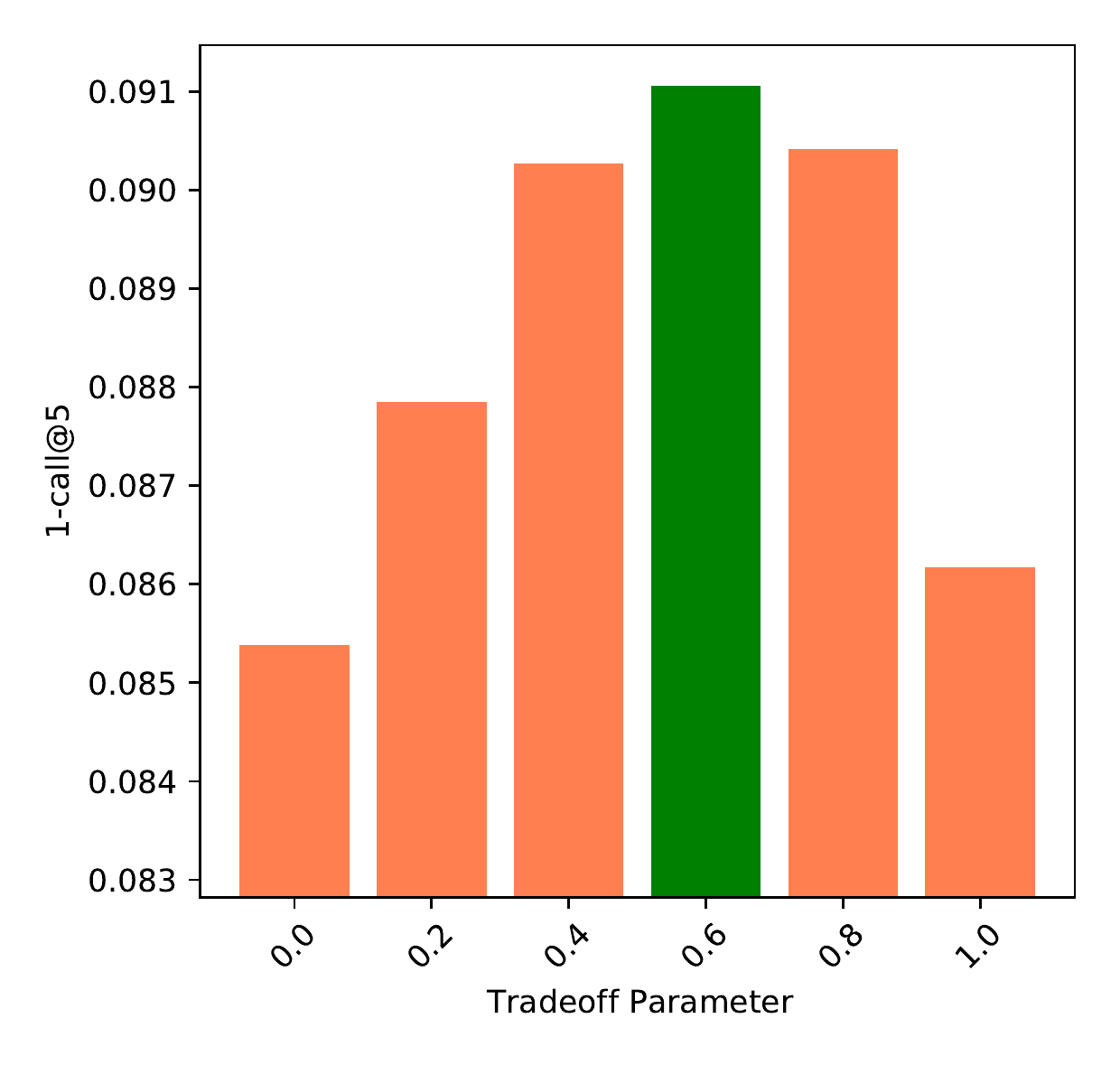,height=0.60in,width=0.41in }&
			\psfig{figure=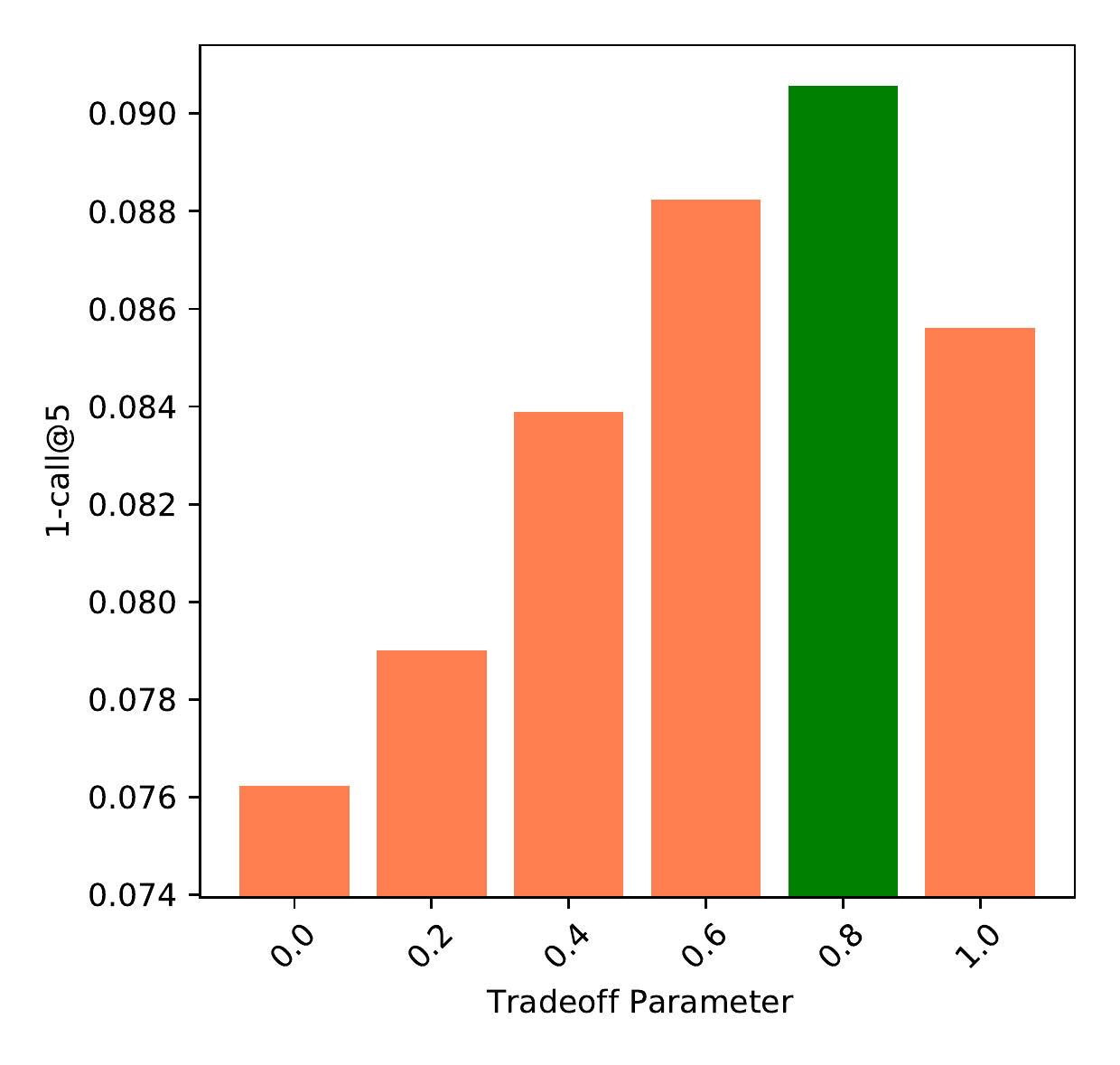,height=0.60in,width=0.41in }&
			\psfig{figure=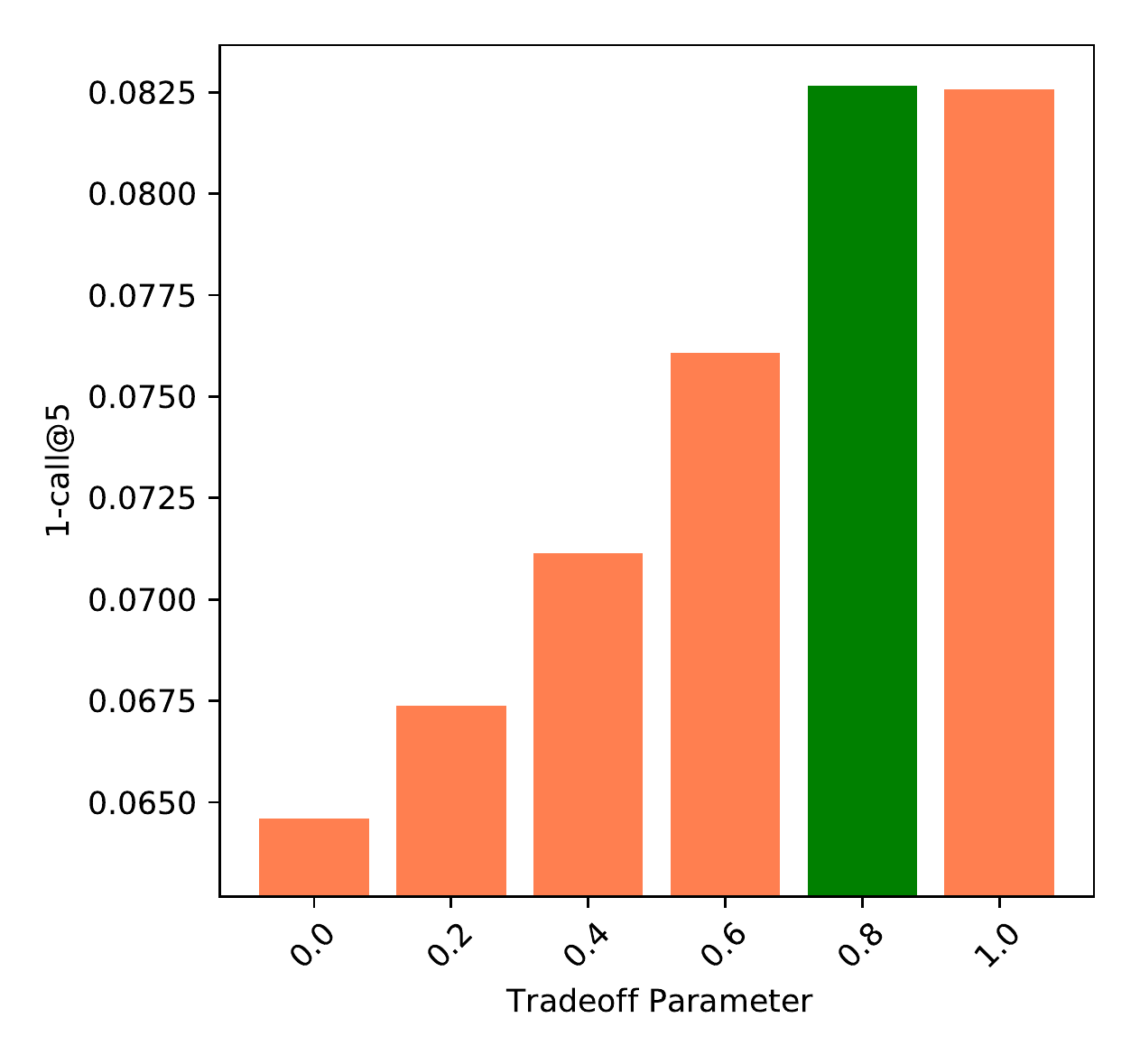,height=0.60in,width=0.41in }&
			\psfig{figure=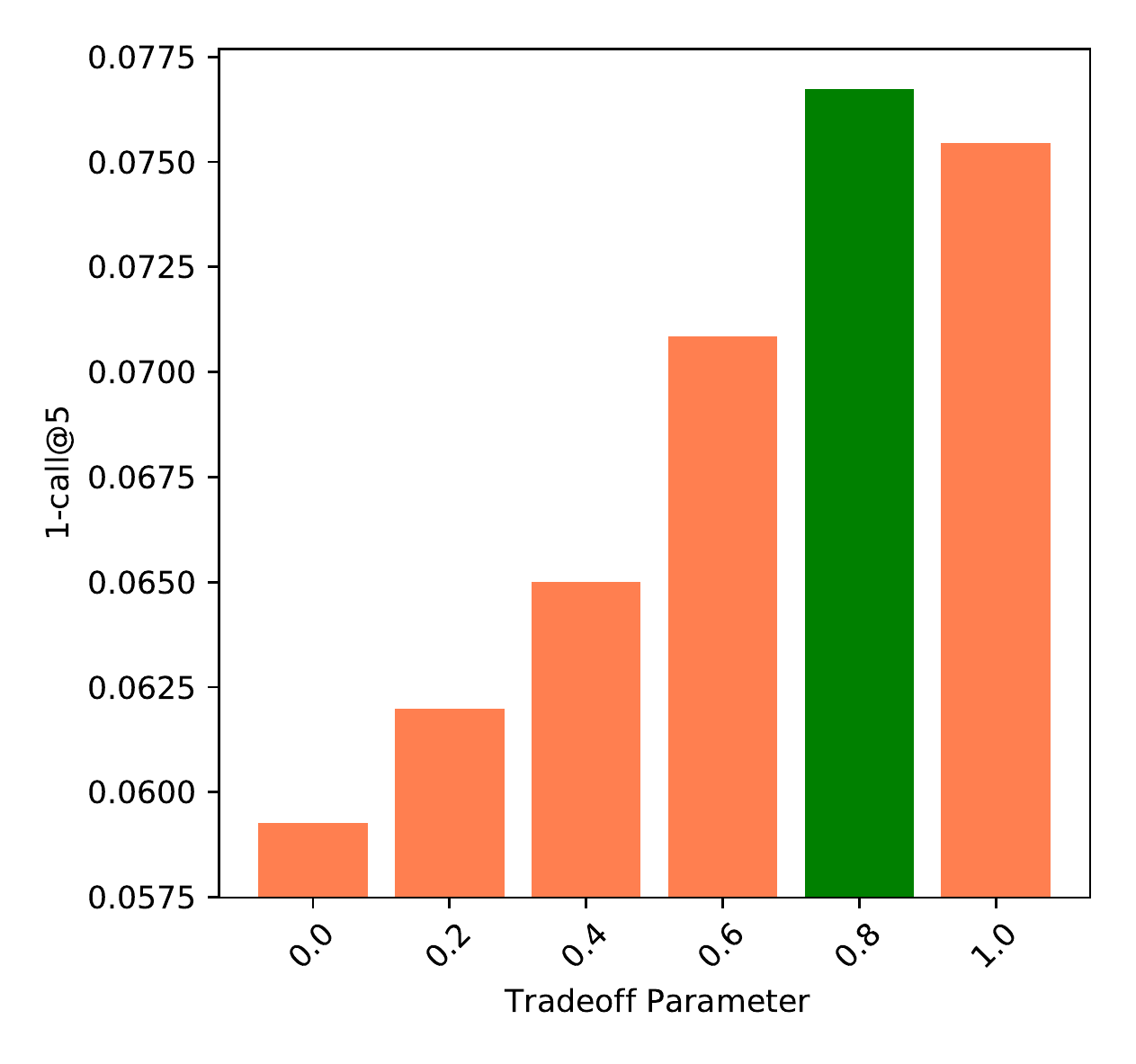,height=0.60in,width=0.41in }&
			
			&
			
			\psfig{figure=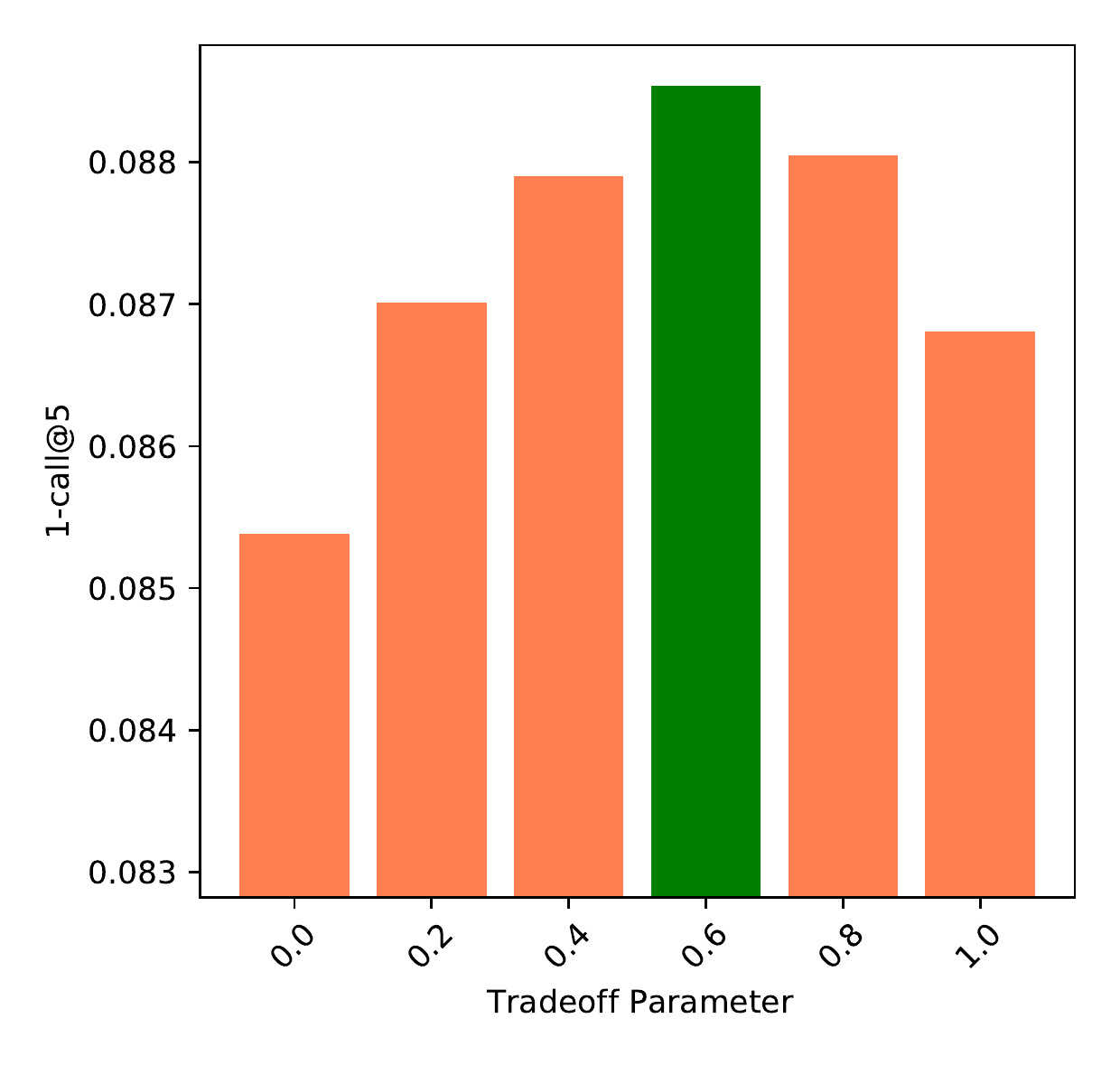,height=0.60in,width=0.41in }&
			\psfig{figure=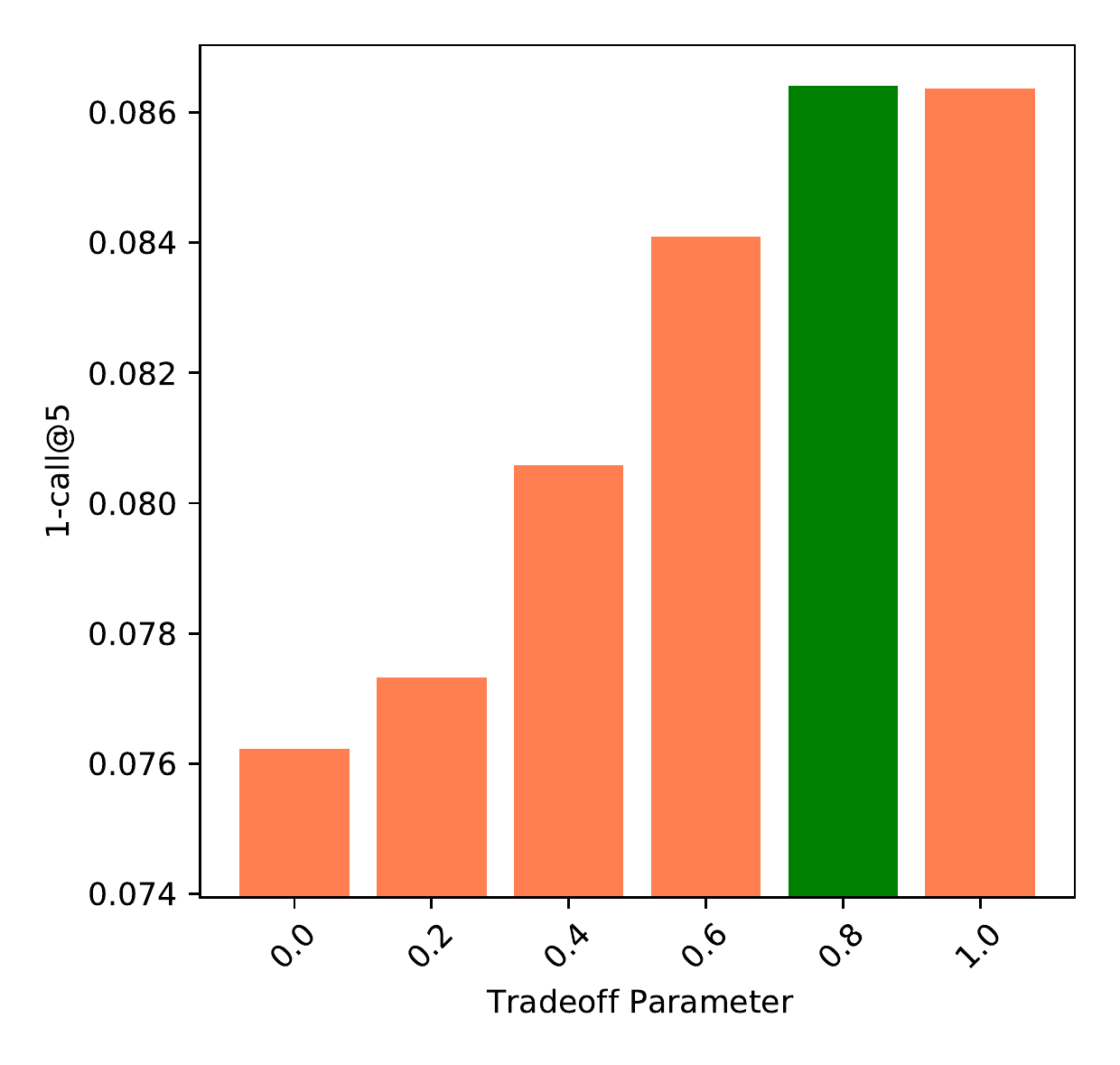,height=0.60in,width=0.41in }&
			\psfig{figure=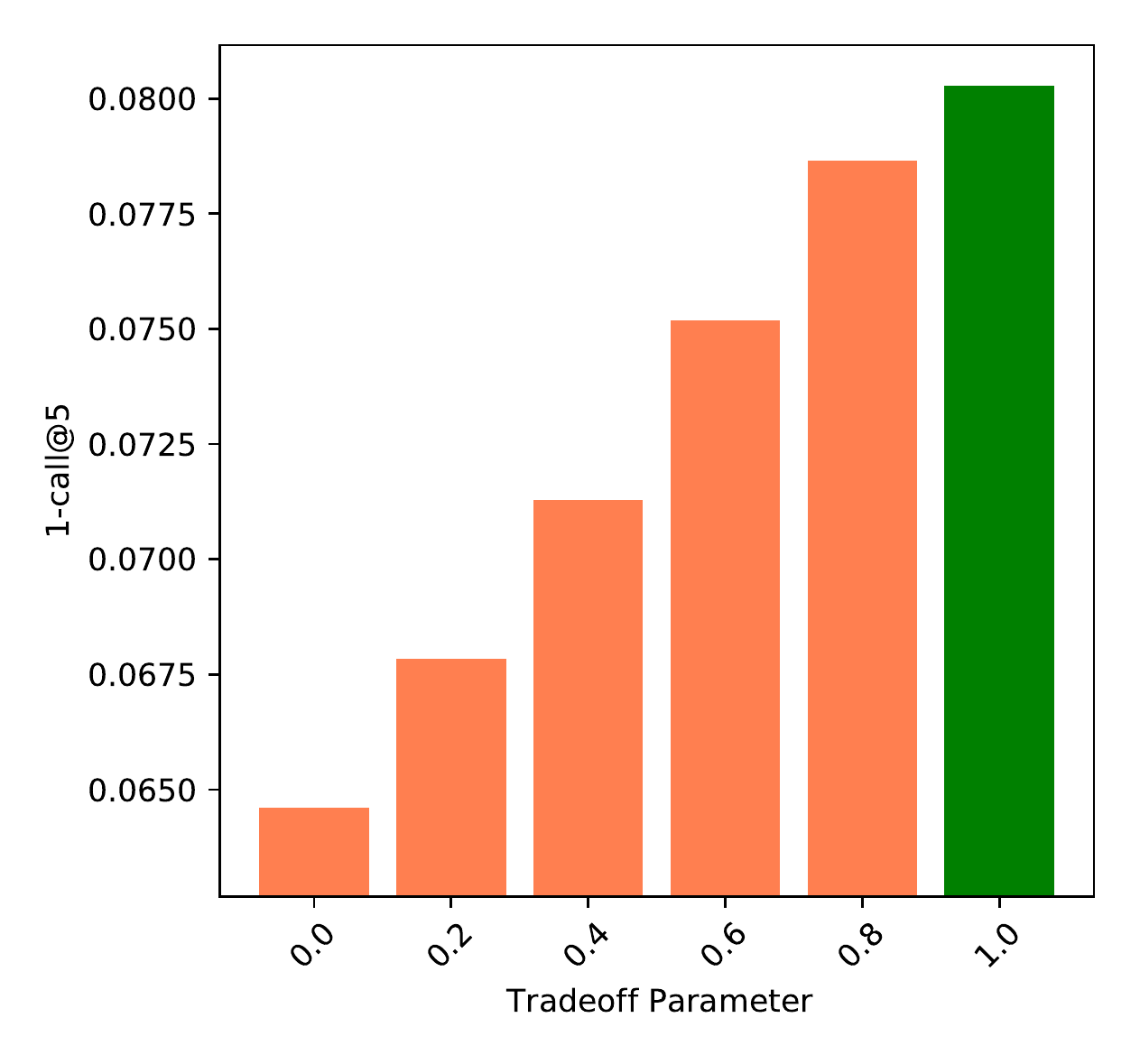,height=0.60in,width=0.41in }&
			\psfig{figure=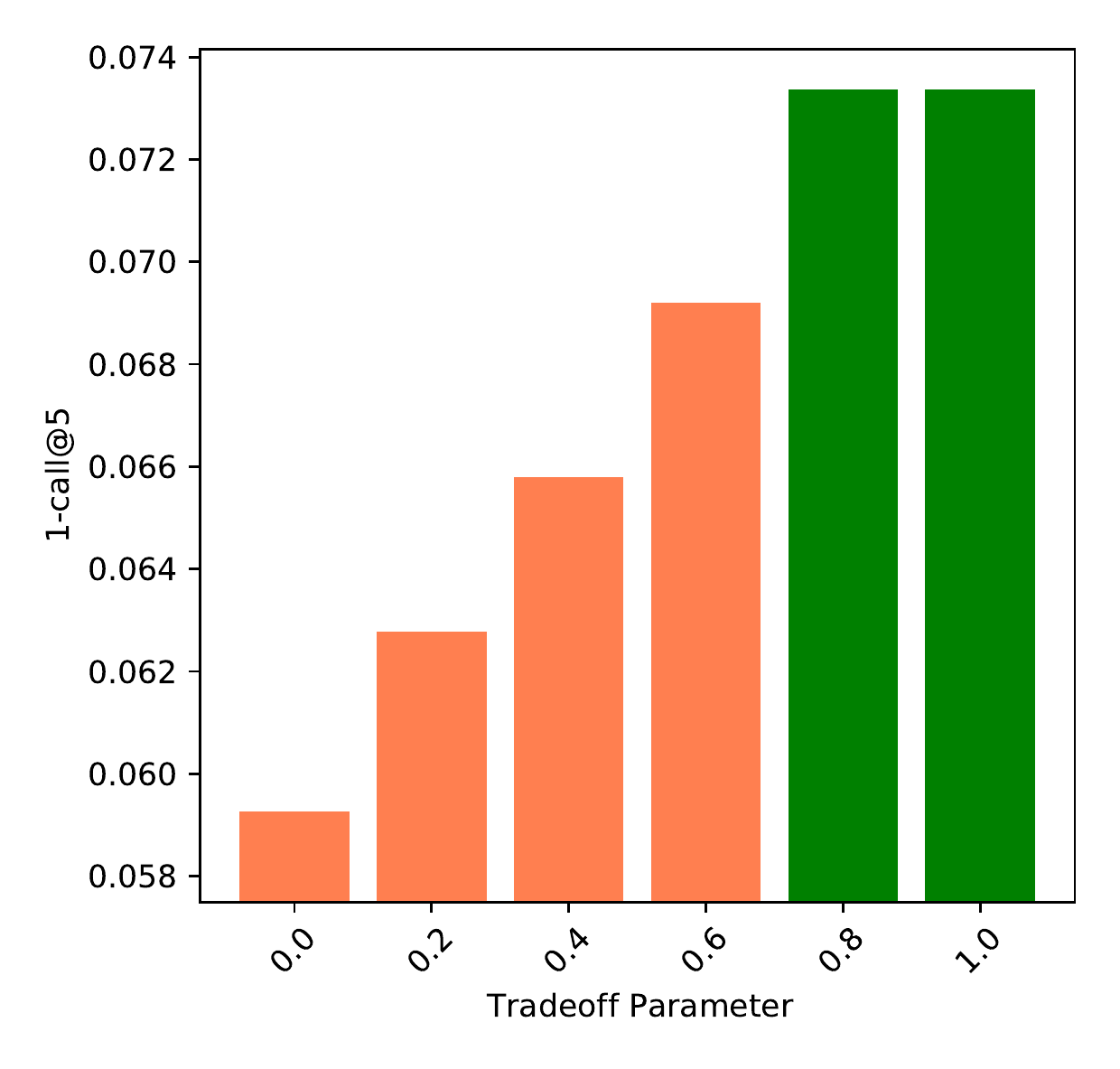,height=0.60in,width=0.41in }&
			
			&
			
			\psfig{figure=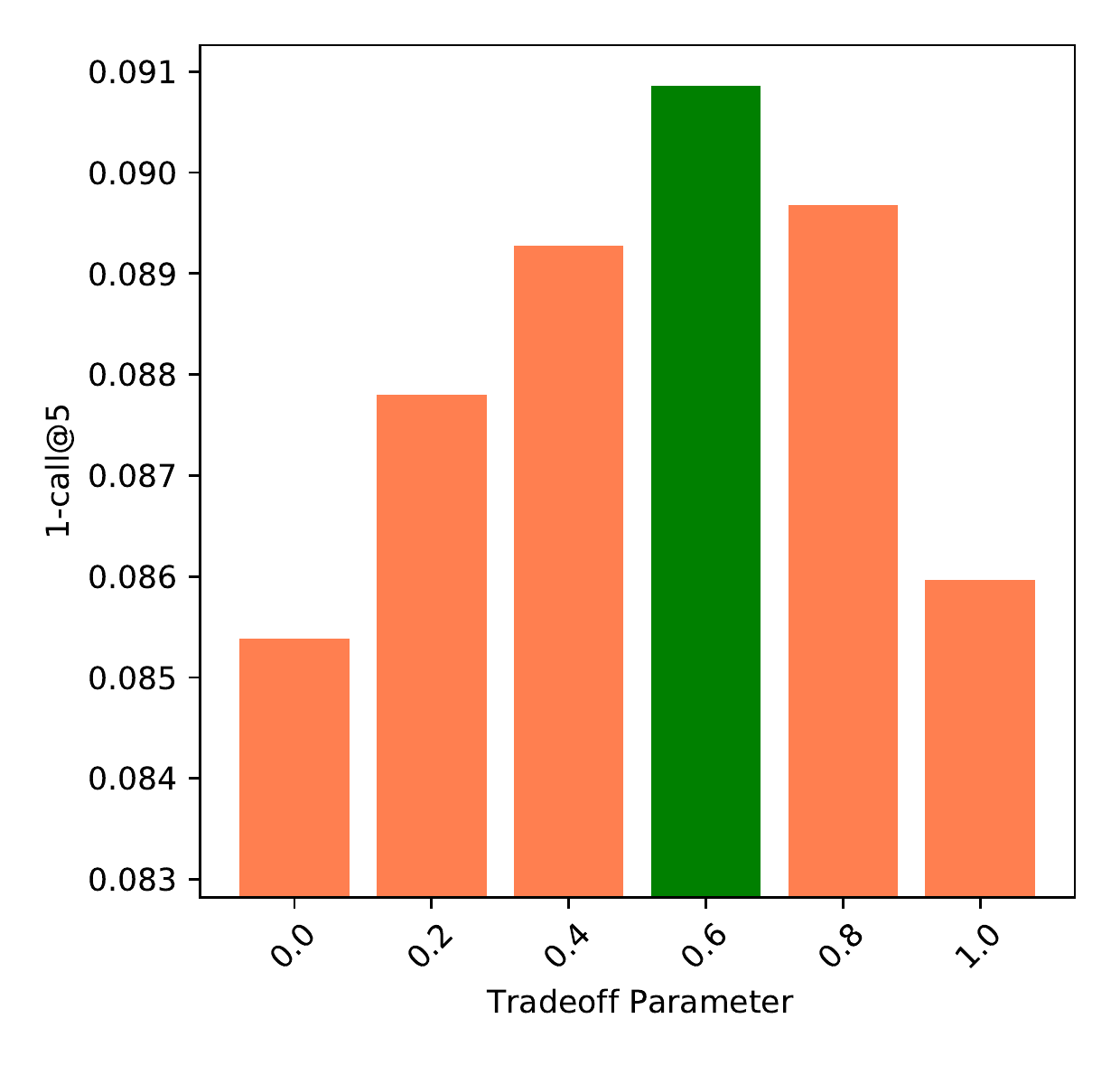,height=0.60in,width=0.41in }&
			\psfig{figure=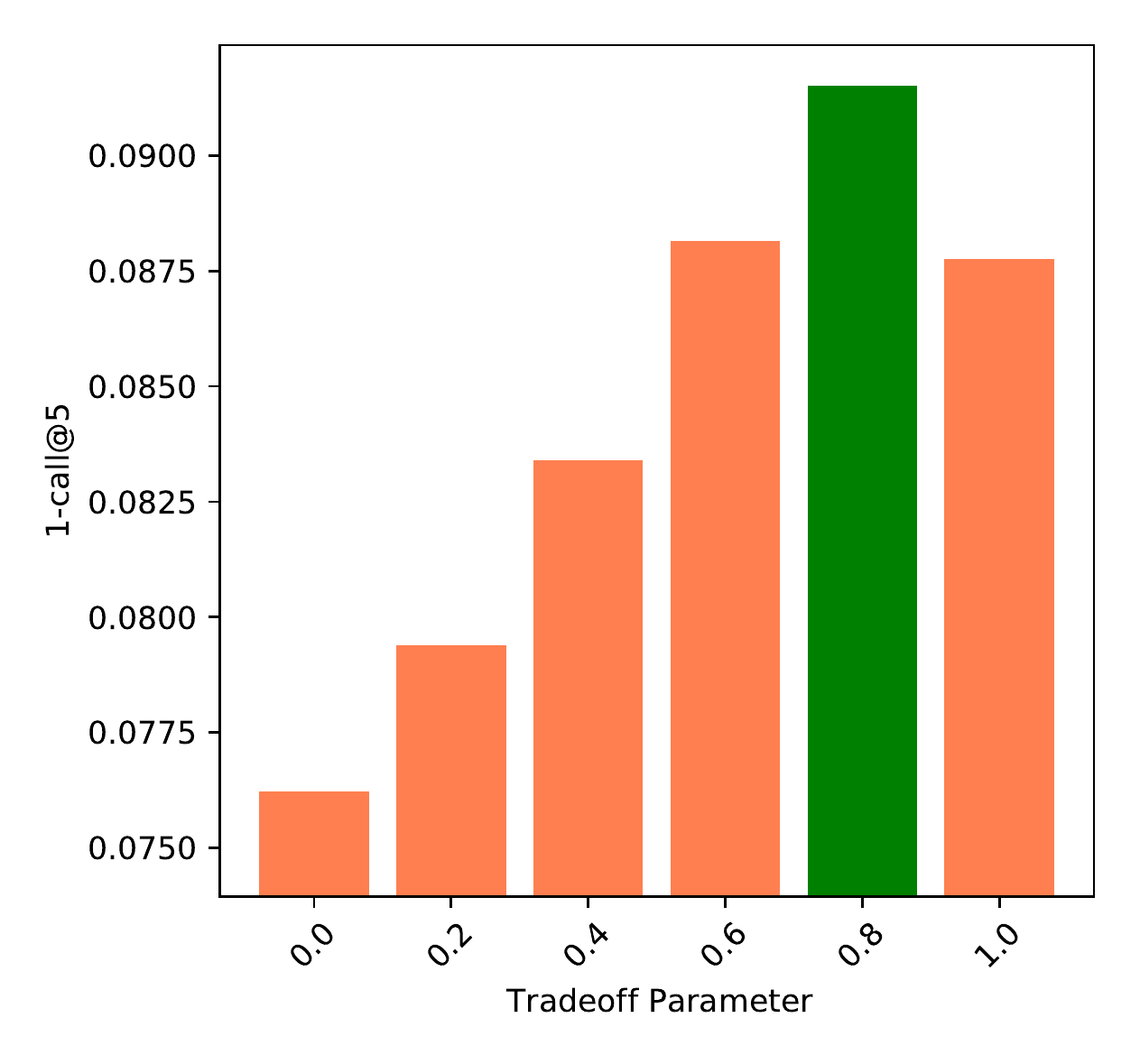,height=0.60in,width=0.41in }&
			\psfig{figure=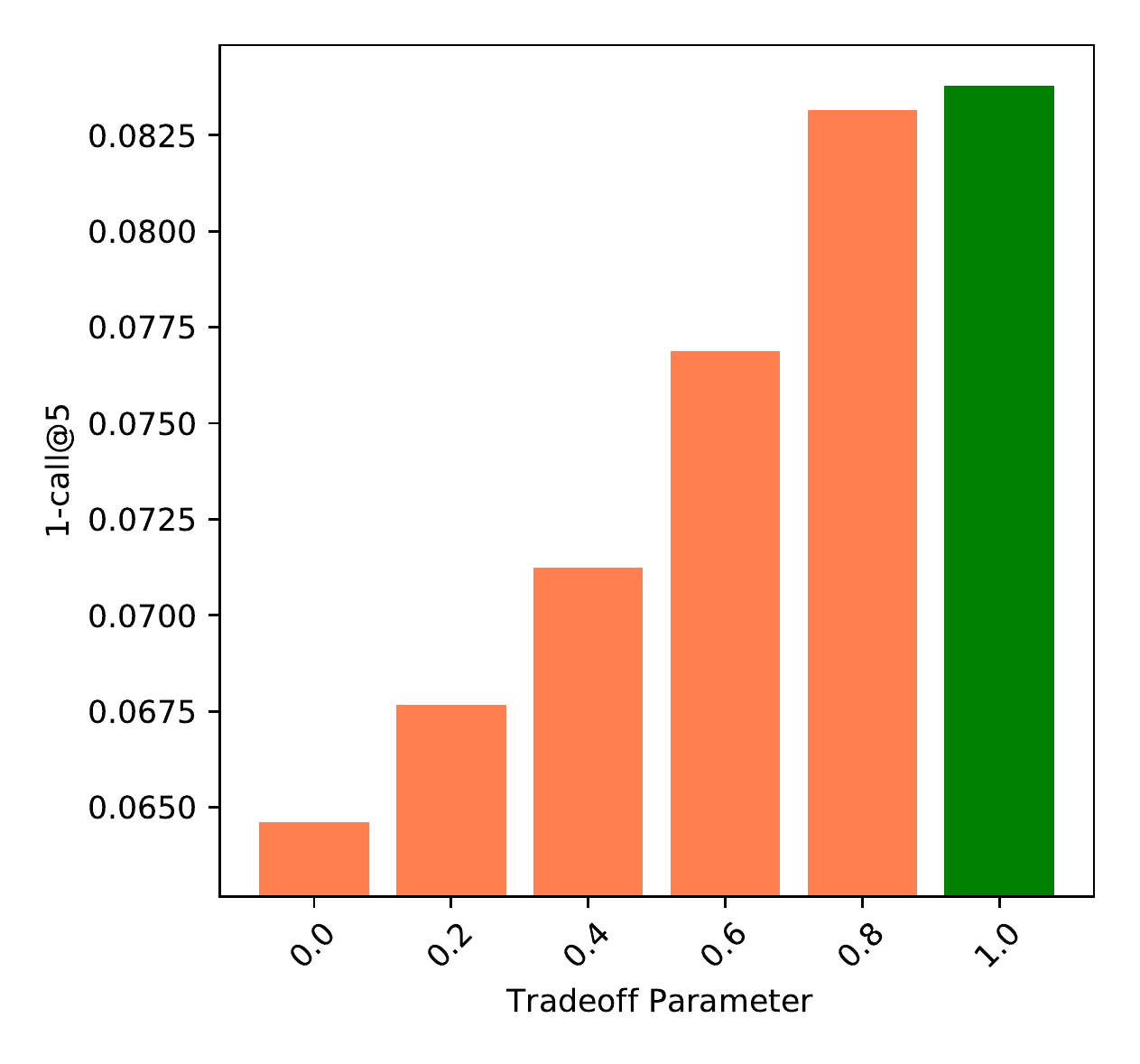,height=0.60in,width=0.41in }&
			\psfig{figure=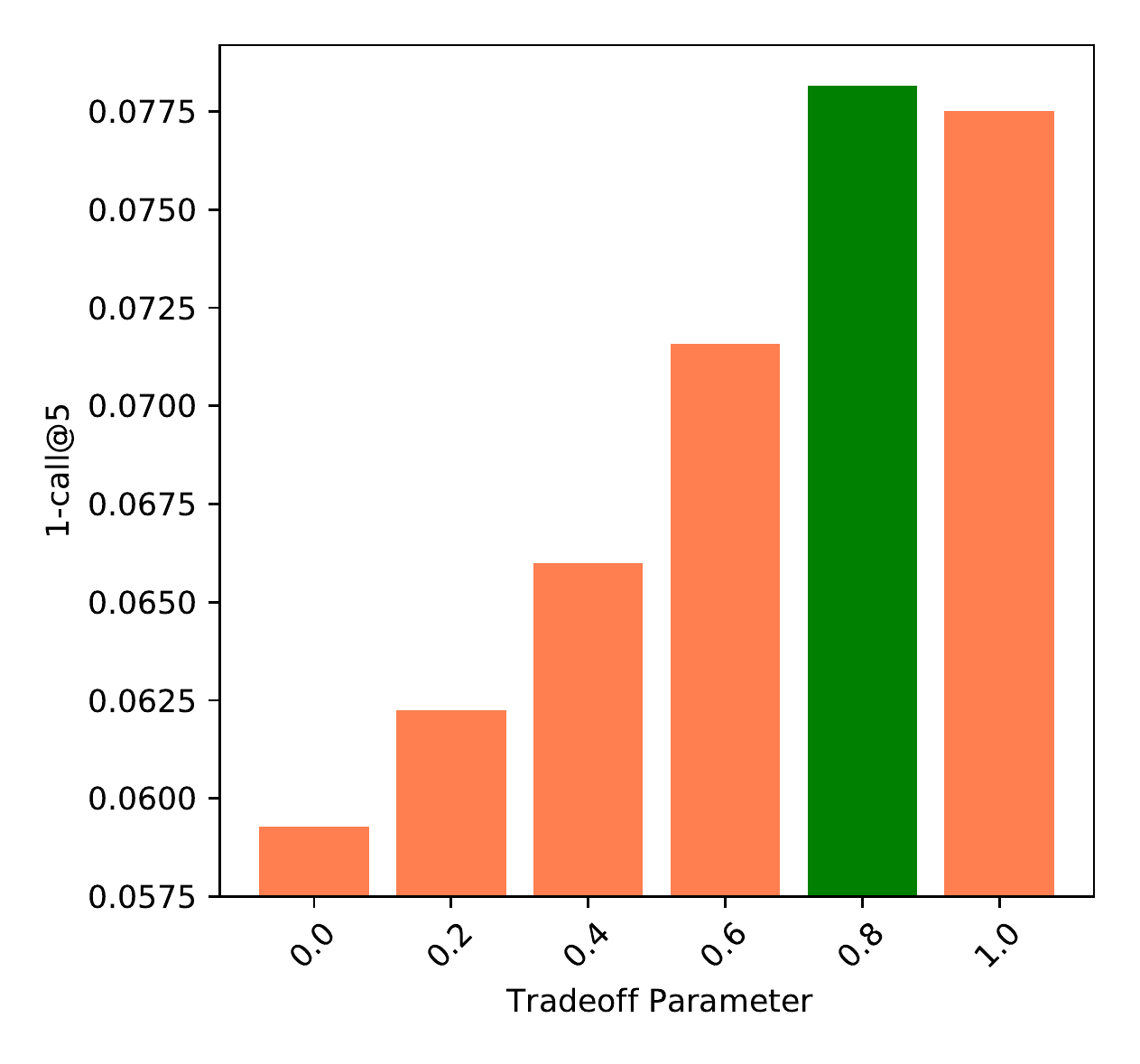,height=0.60in,width=0.41in }
			
			\\
			
			\psfig{figure=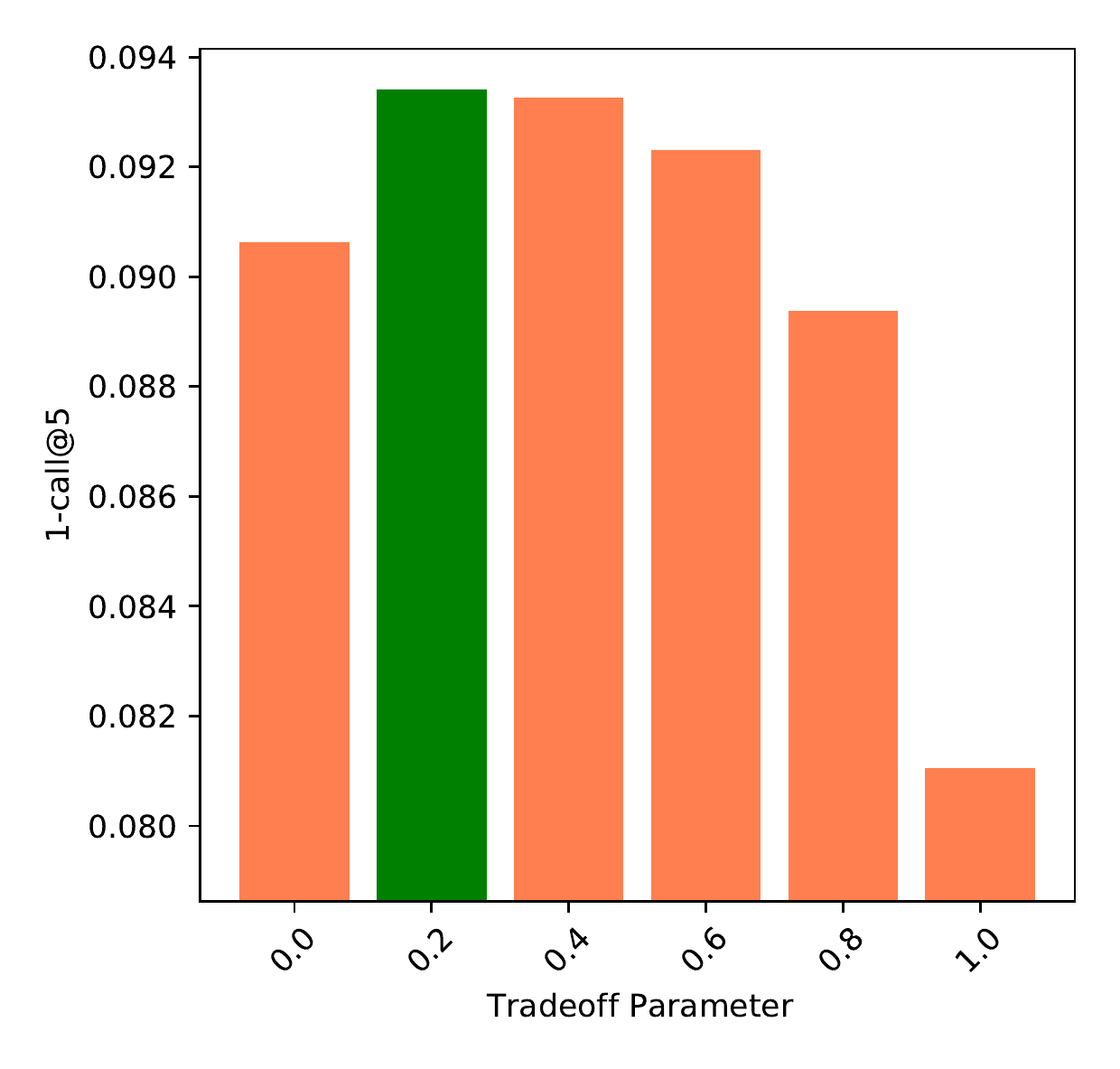,height=0.60in,width=0.41in }&
			\psfig{figure=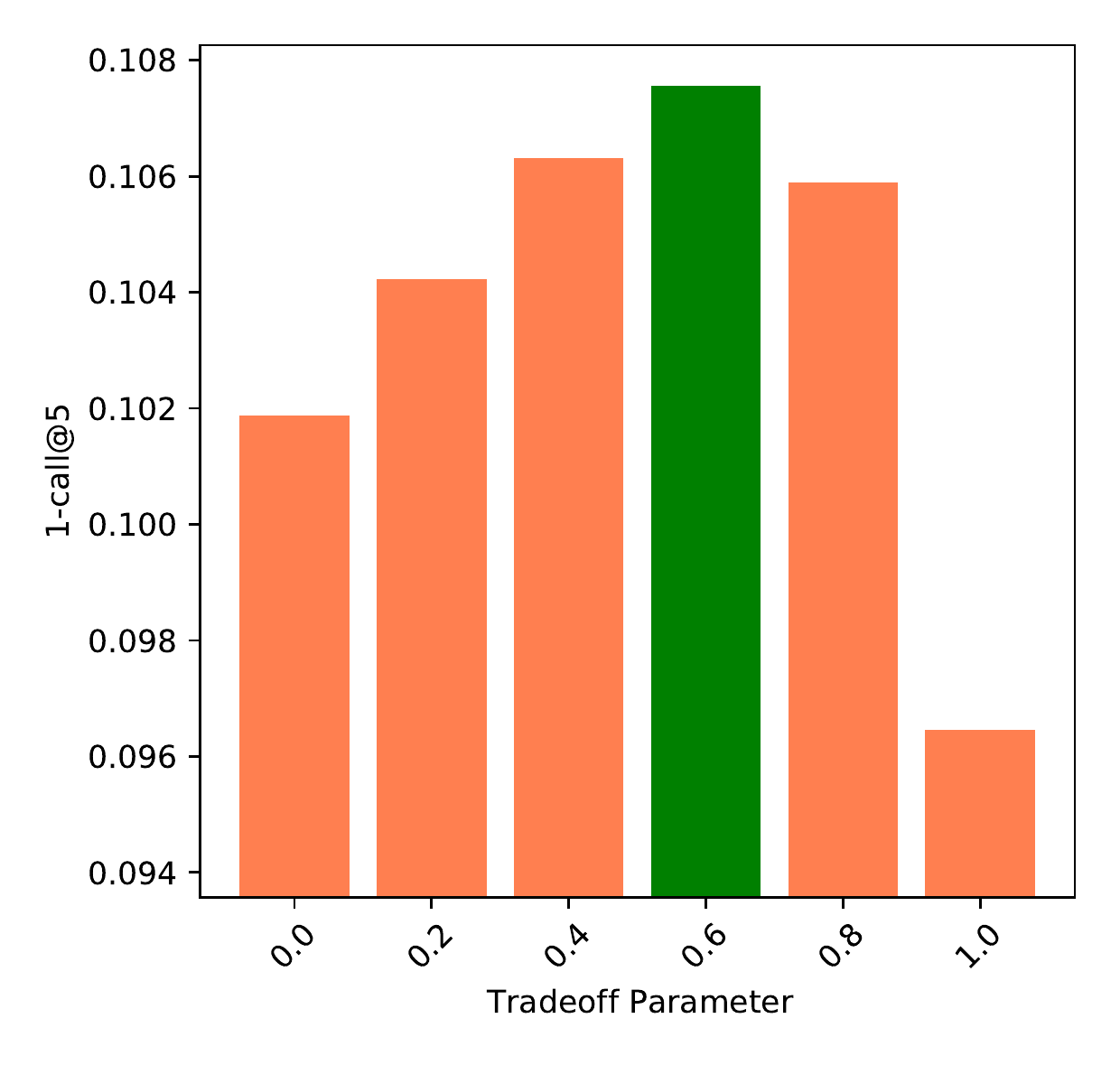,height=0.60in,width=0.41in }&
			\psfig{figure=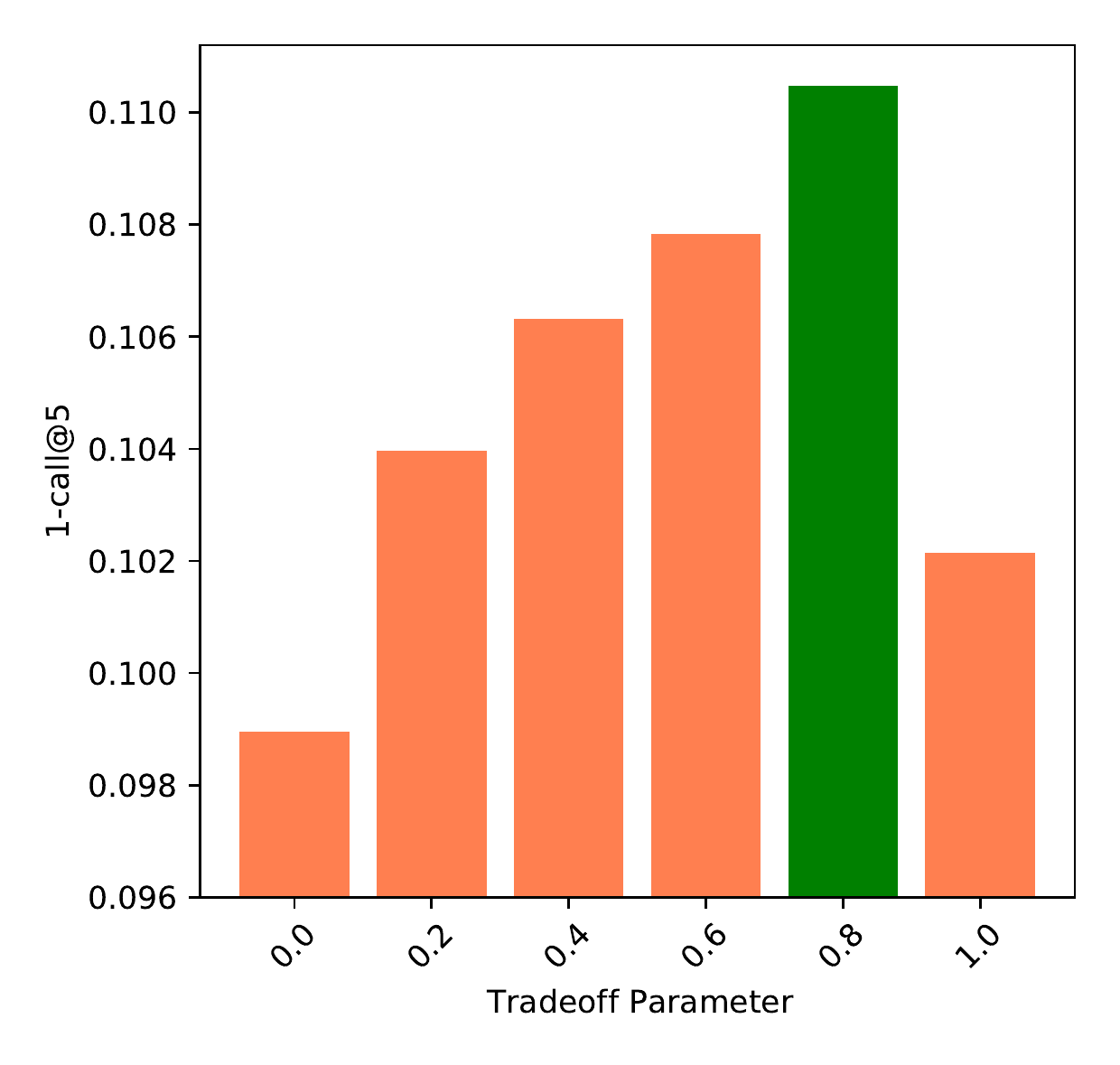,height=0.60in,width=0.41in }&
			\psfig{figure=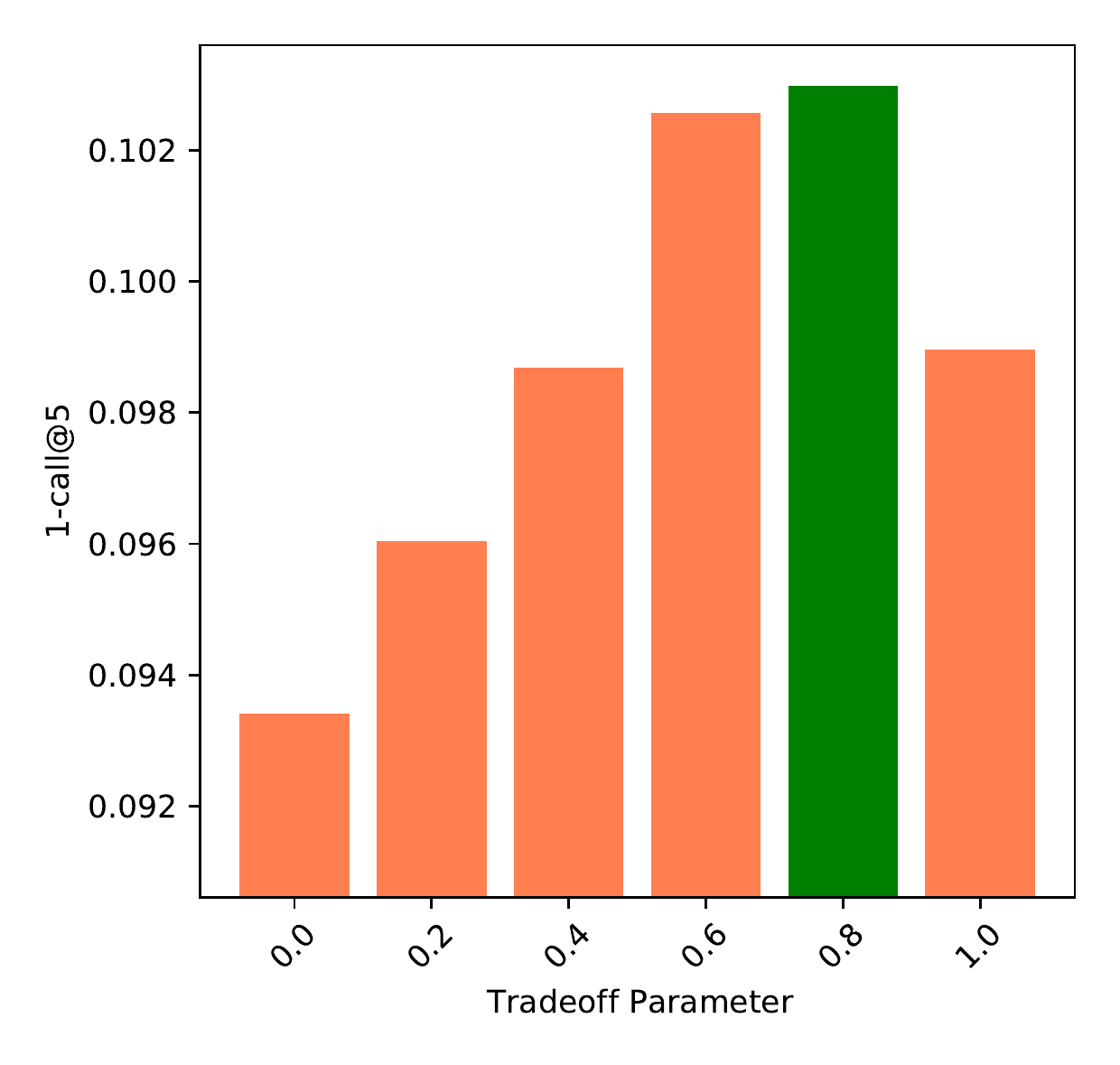,height=0.60in,width=0.41in }&
			
			&
			
			\psfig{figure=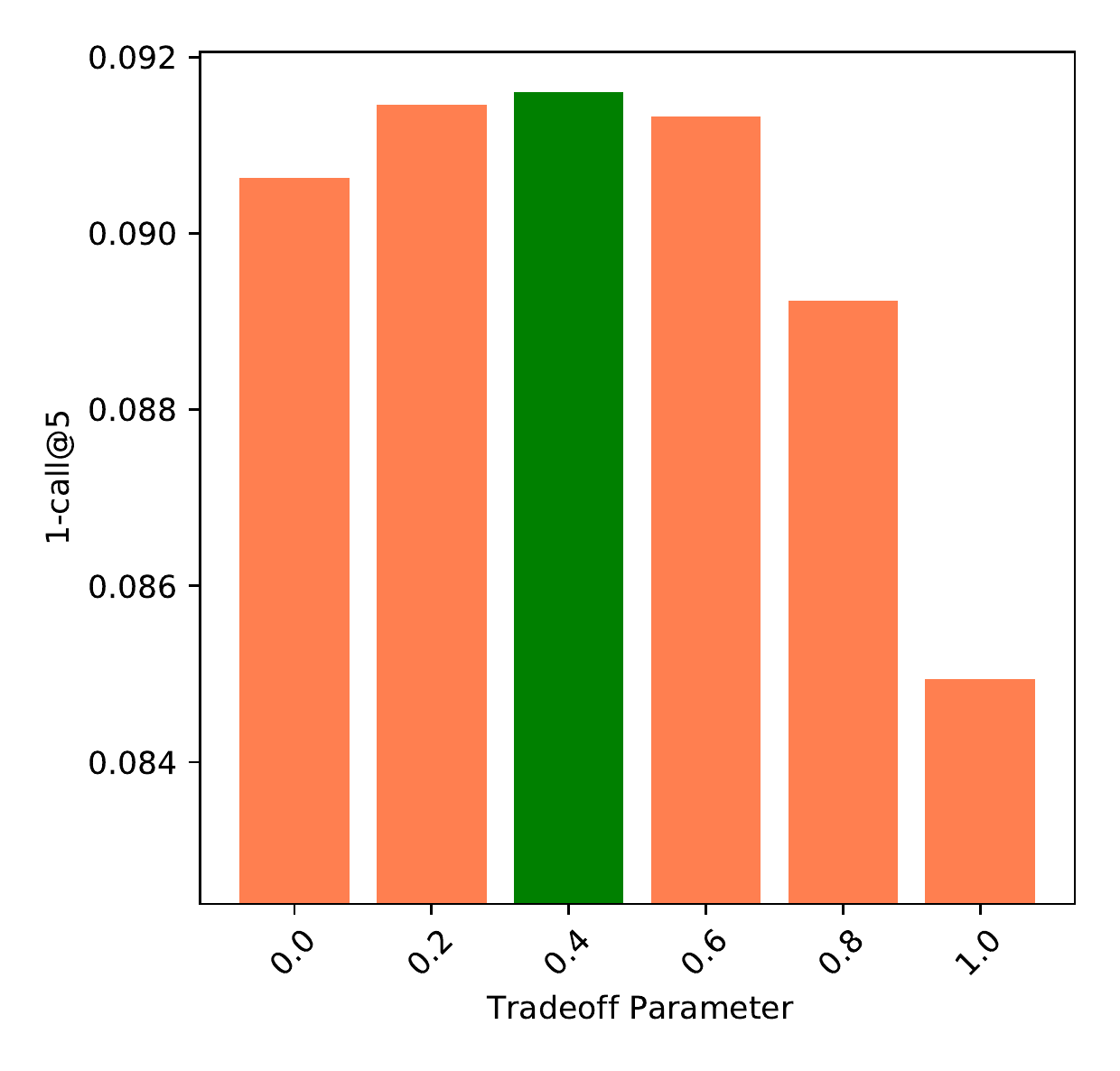,height=0.60in,width=0.41in }&
			\psfig{figure=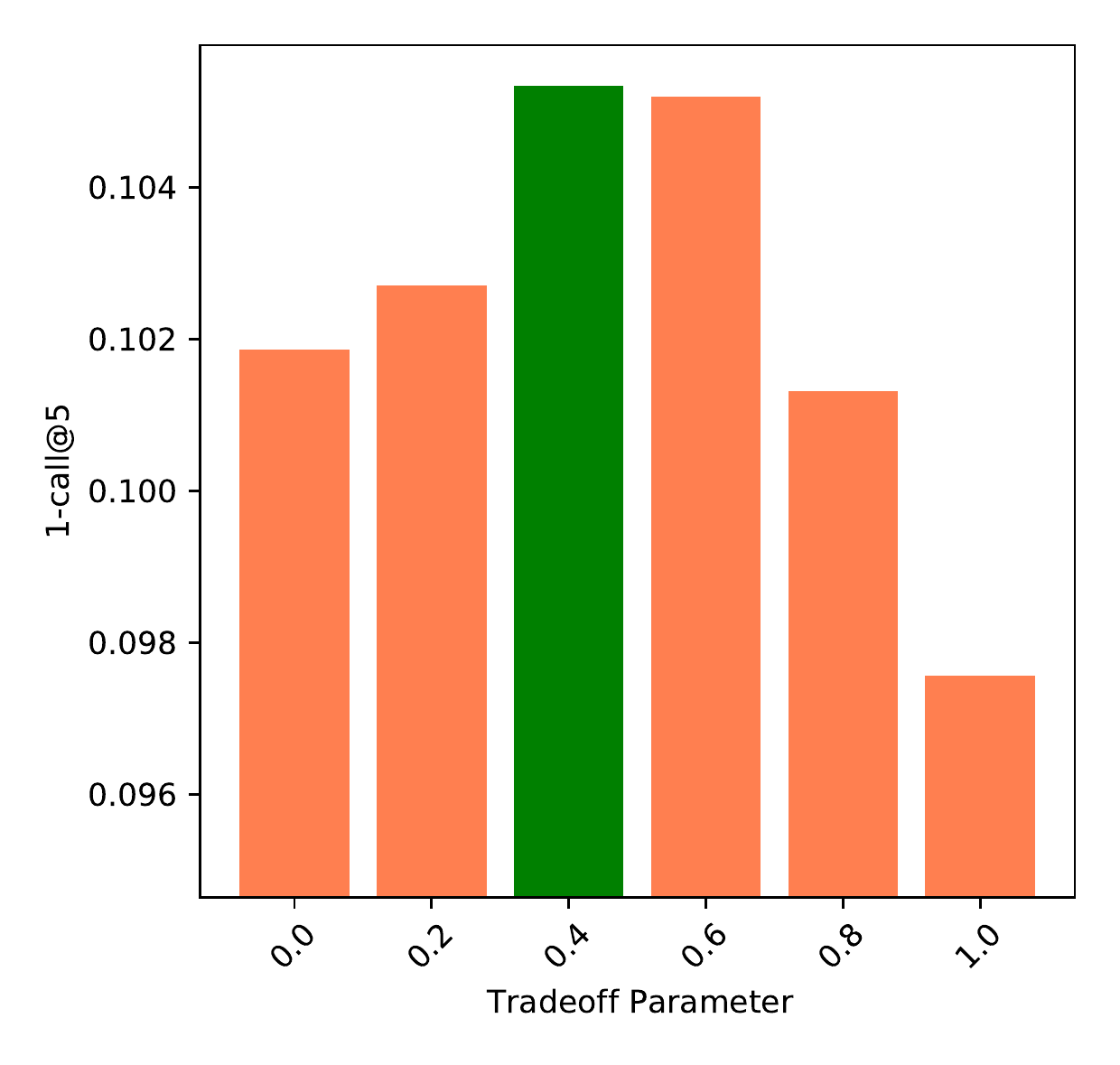,height=0.60in,width=0.41in }&
			\psfig{figure=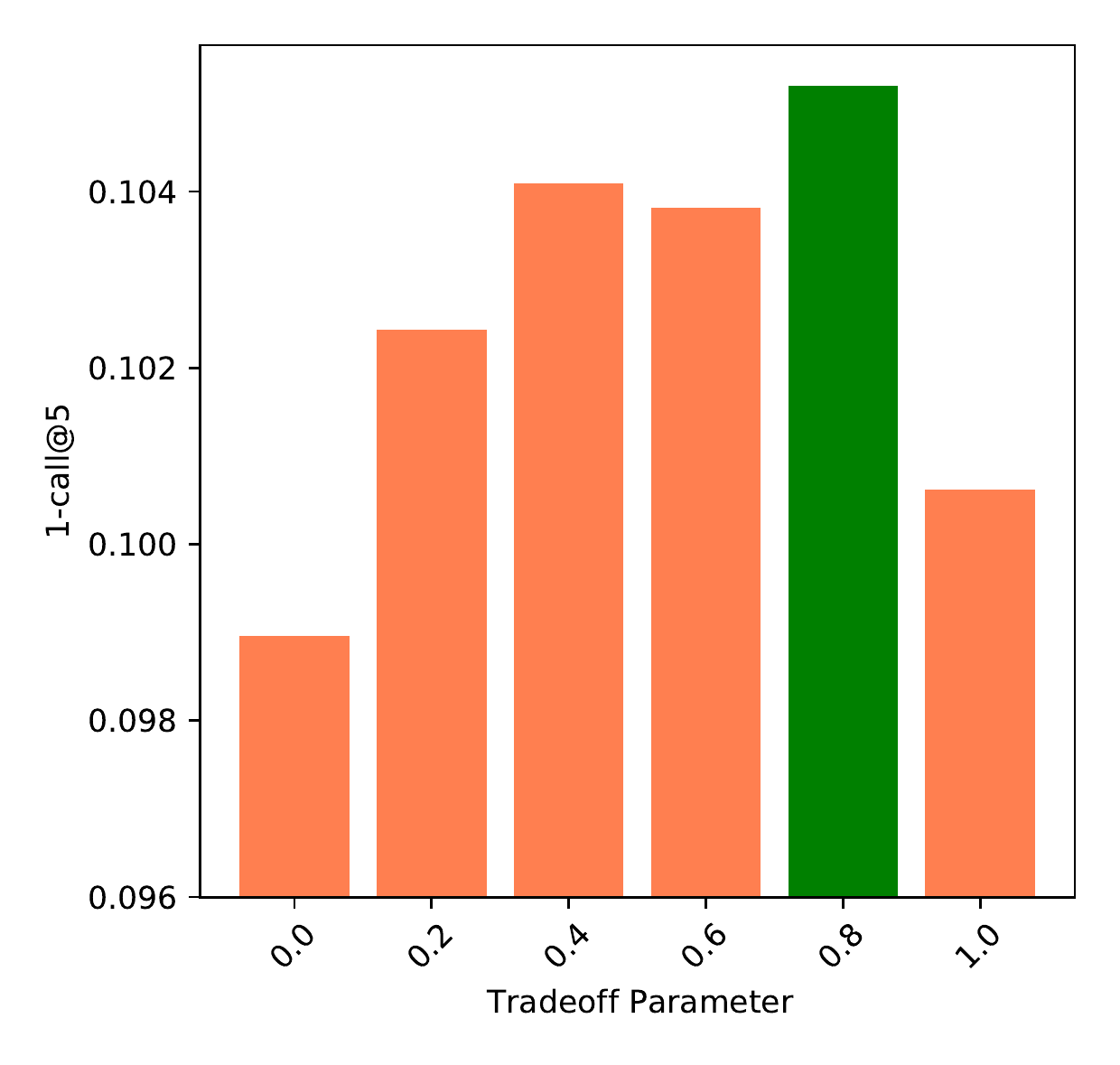,height=0.60in,width=0.41in }&
			\psfig{figure=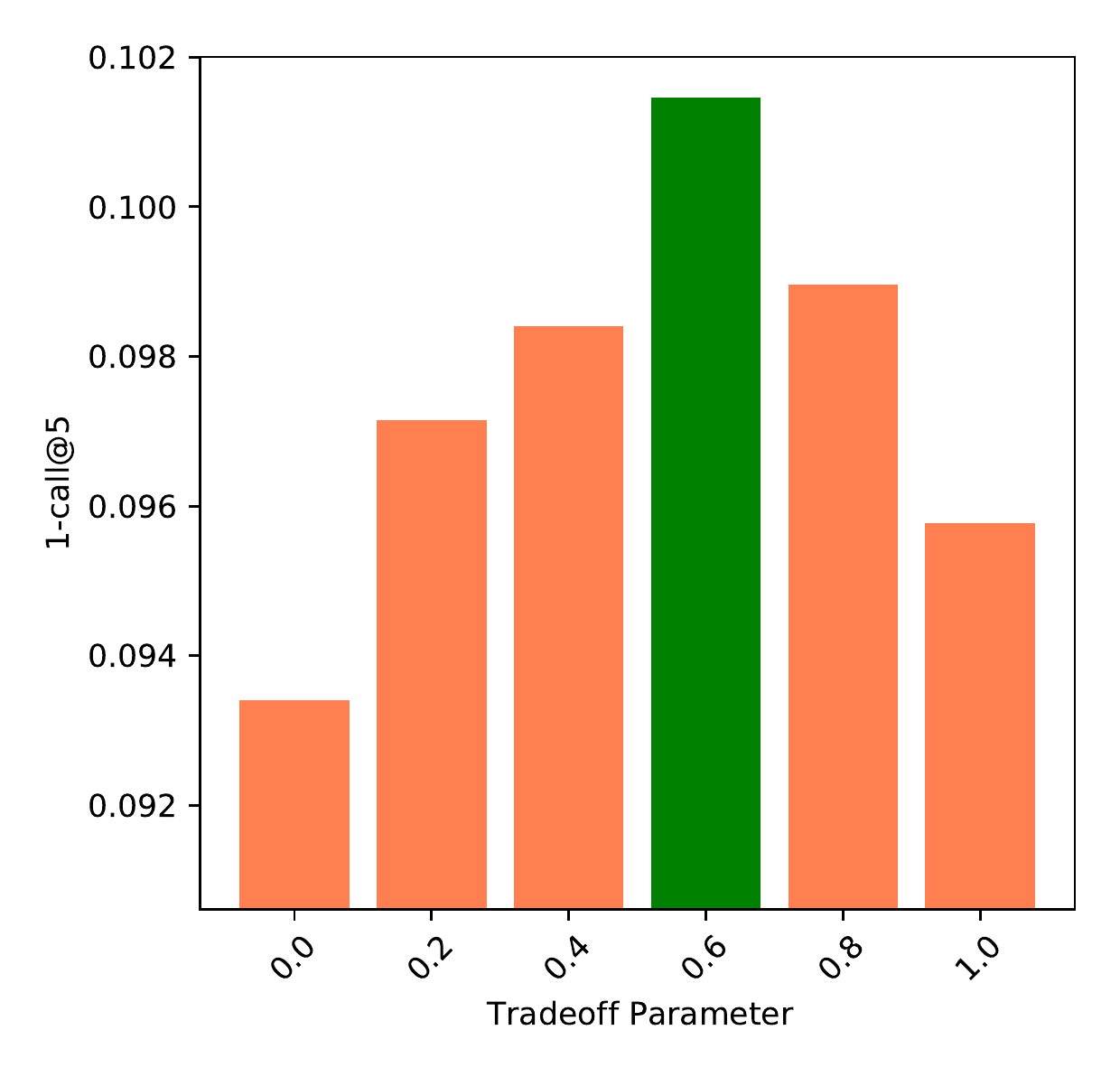,height=0.60in,width=0.41in }&
			
			&
			
			\psfig{figure=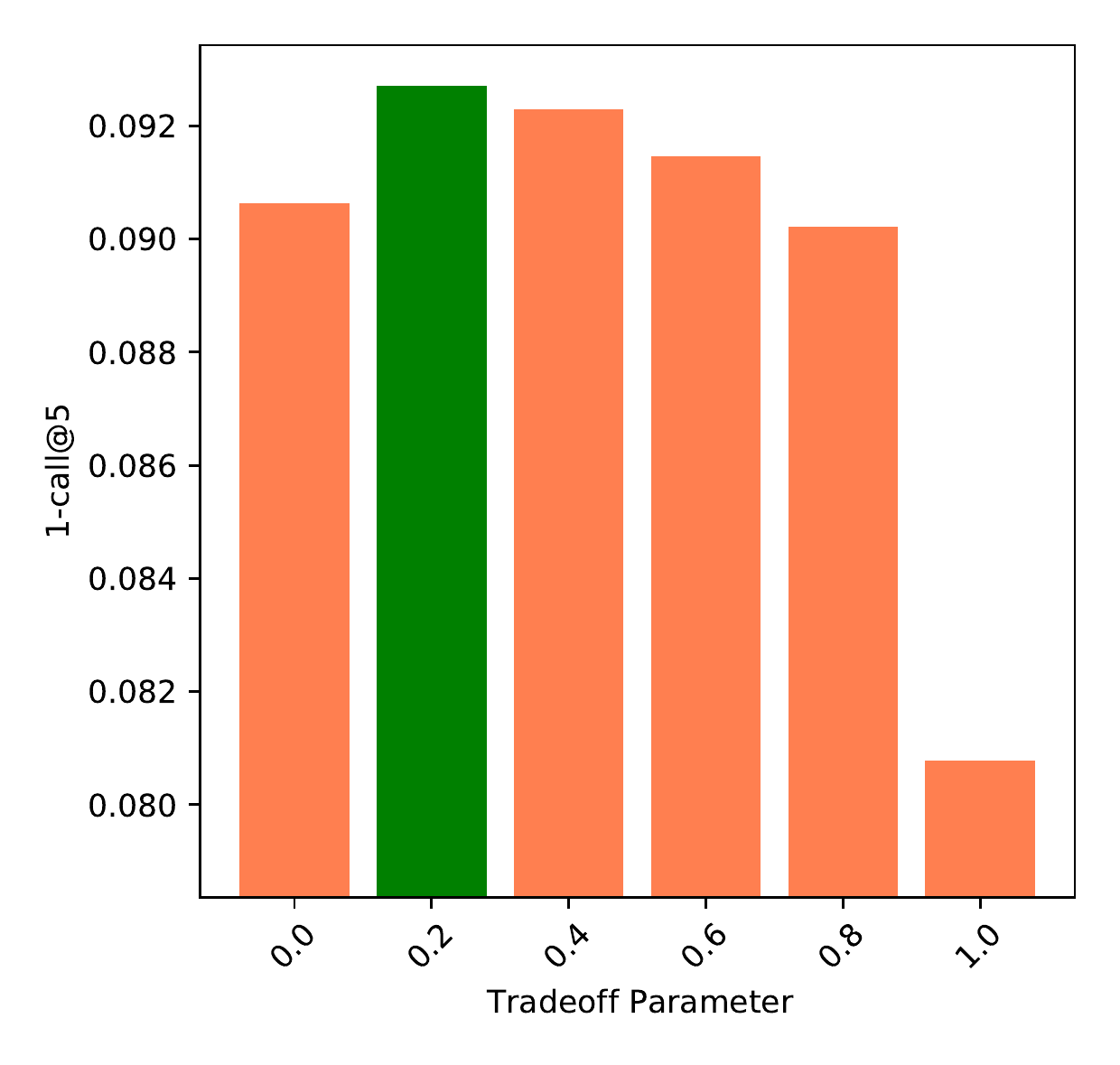,height=0.60in,width=0.41in }&
			\psfig{figure=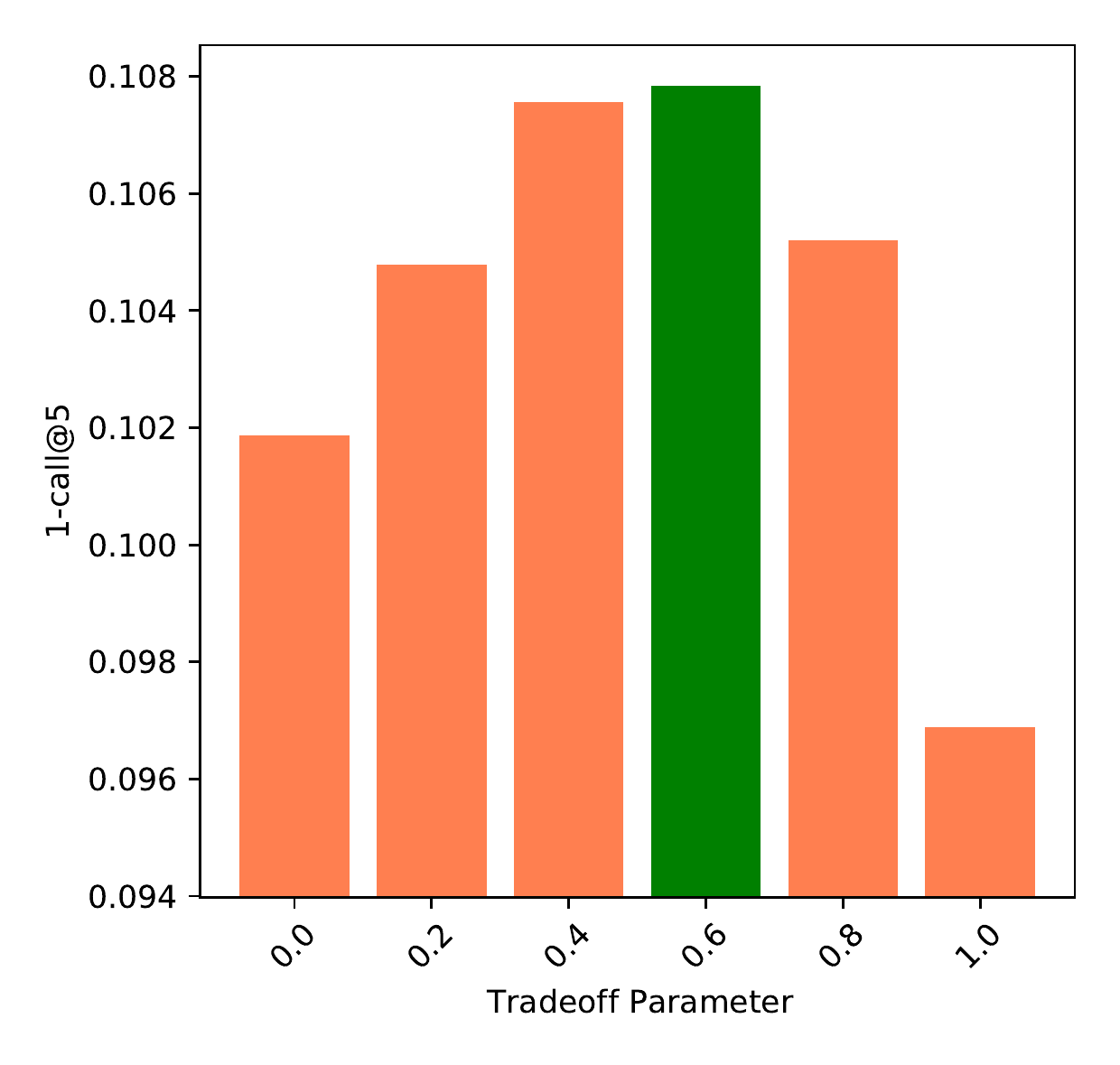,height=0.60in,width=0.41in }&
			\psfig{figure=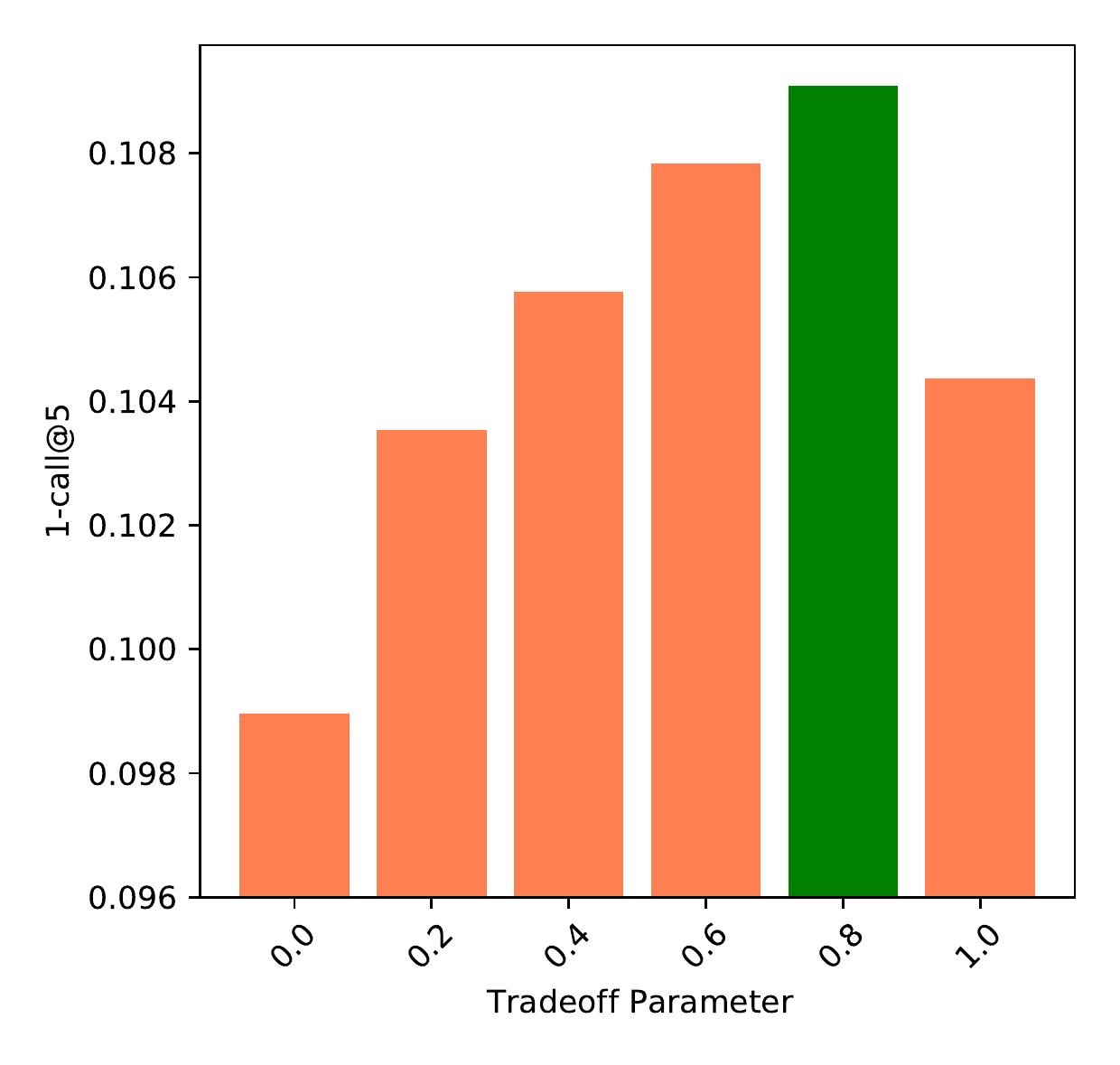,height=0.60in,width=0.41in }&
			\psfig{figure=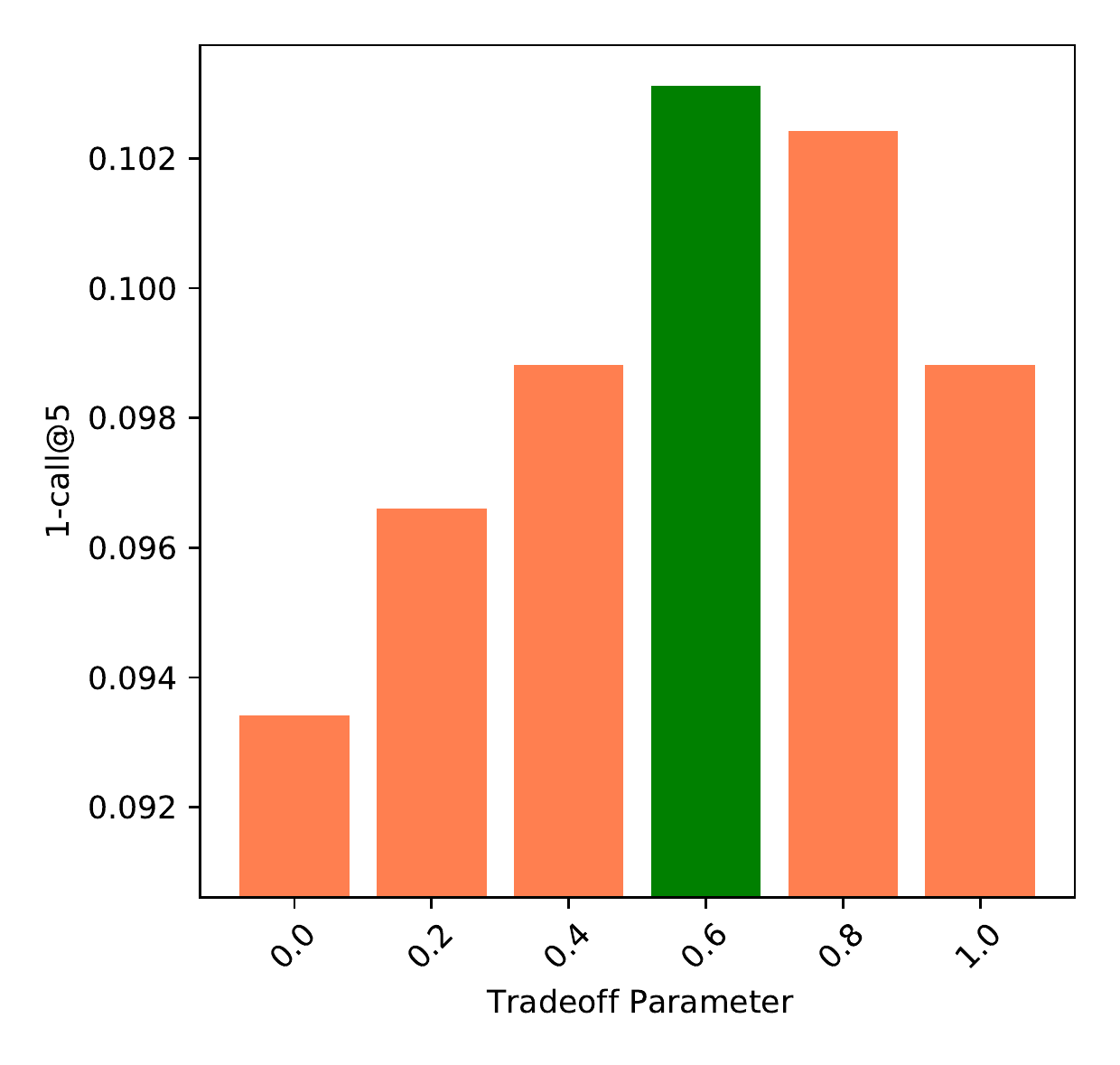,height=0.60in,width=0.41in }
			
			\\
			
			\psfig{figure=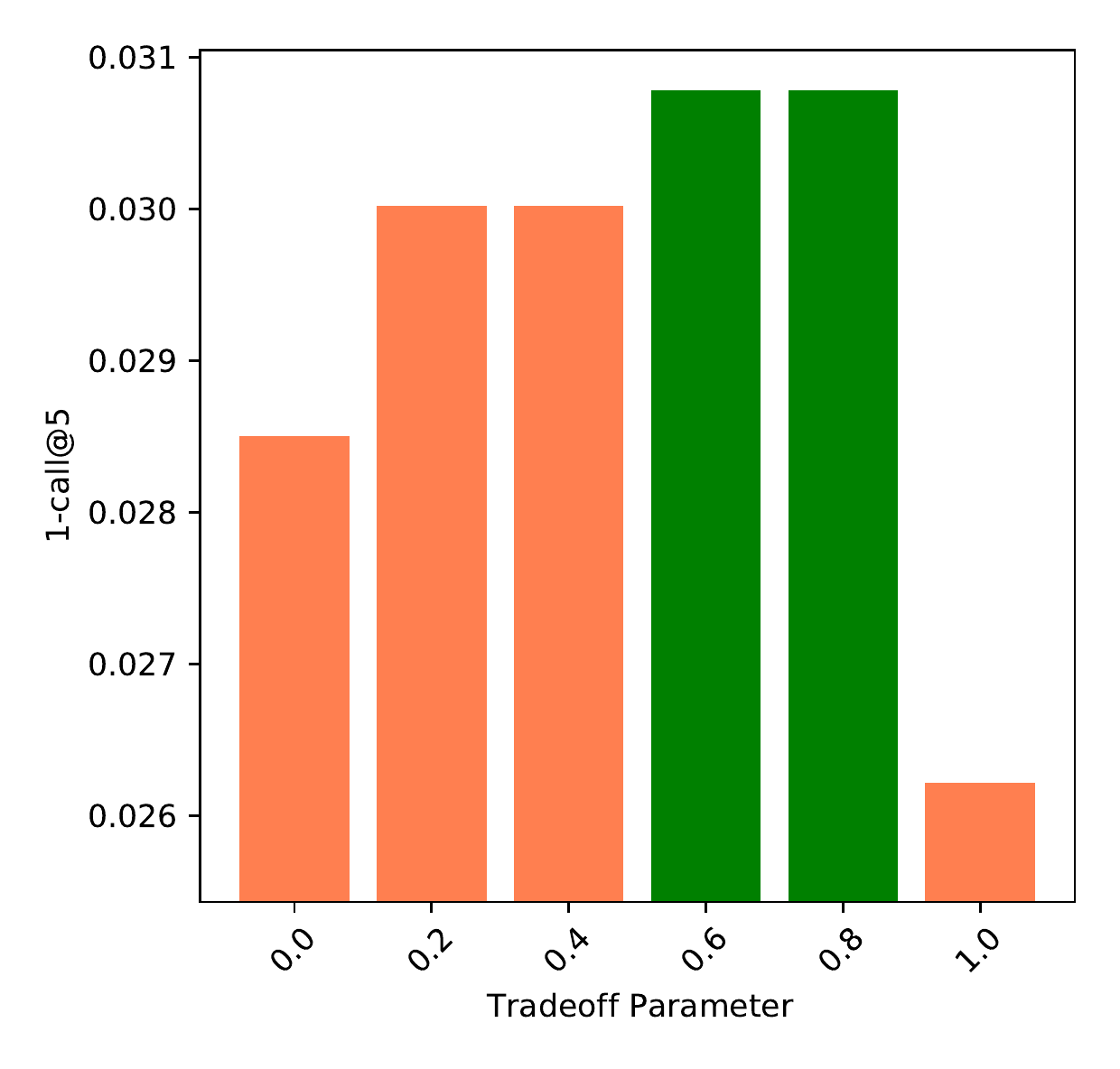,height=0.60in,width=0.41in }&
			\psfig{figure=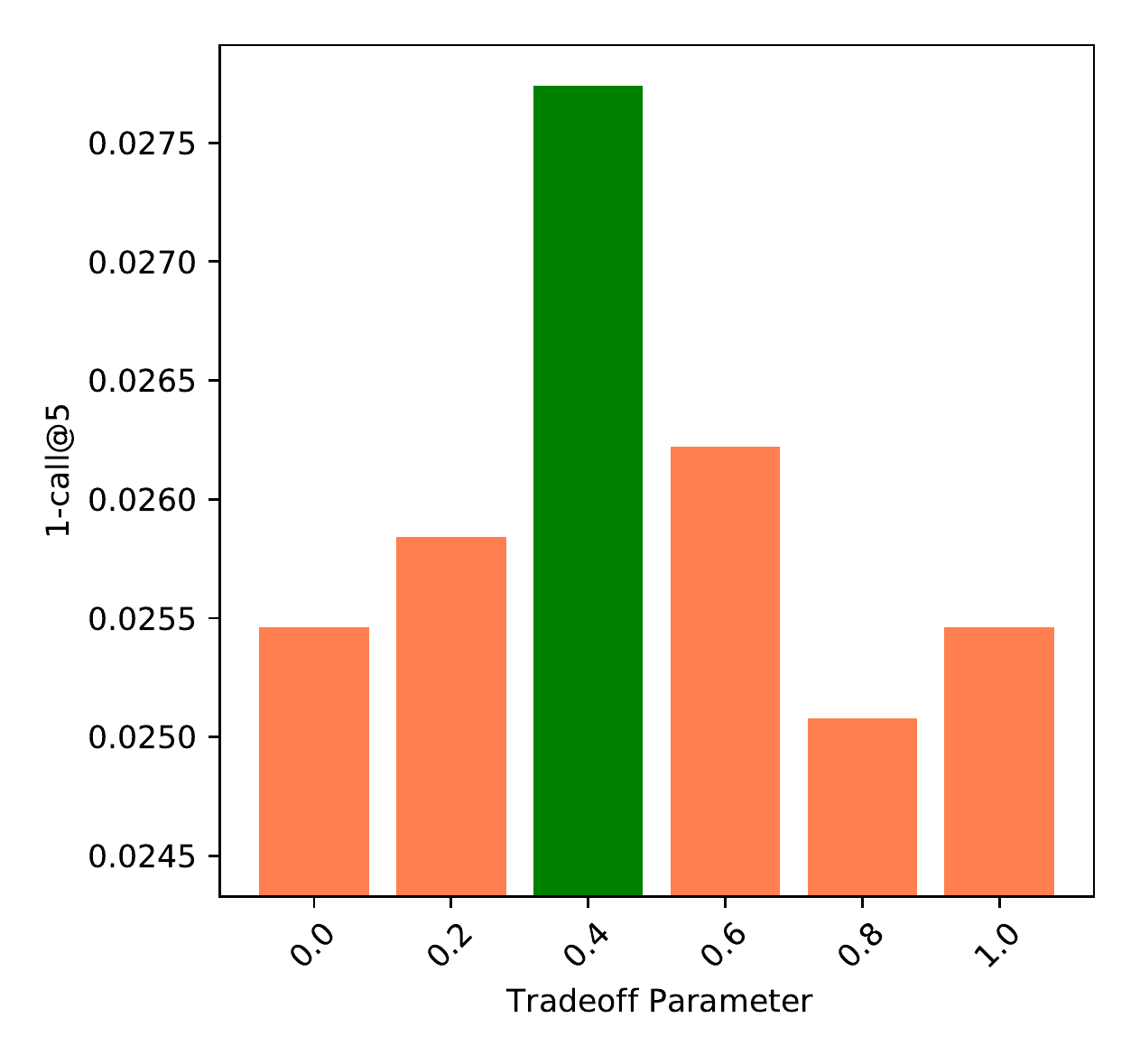,height=0.60in,width=0.41in }&
			\psfig{figure=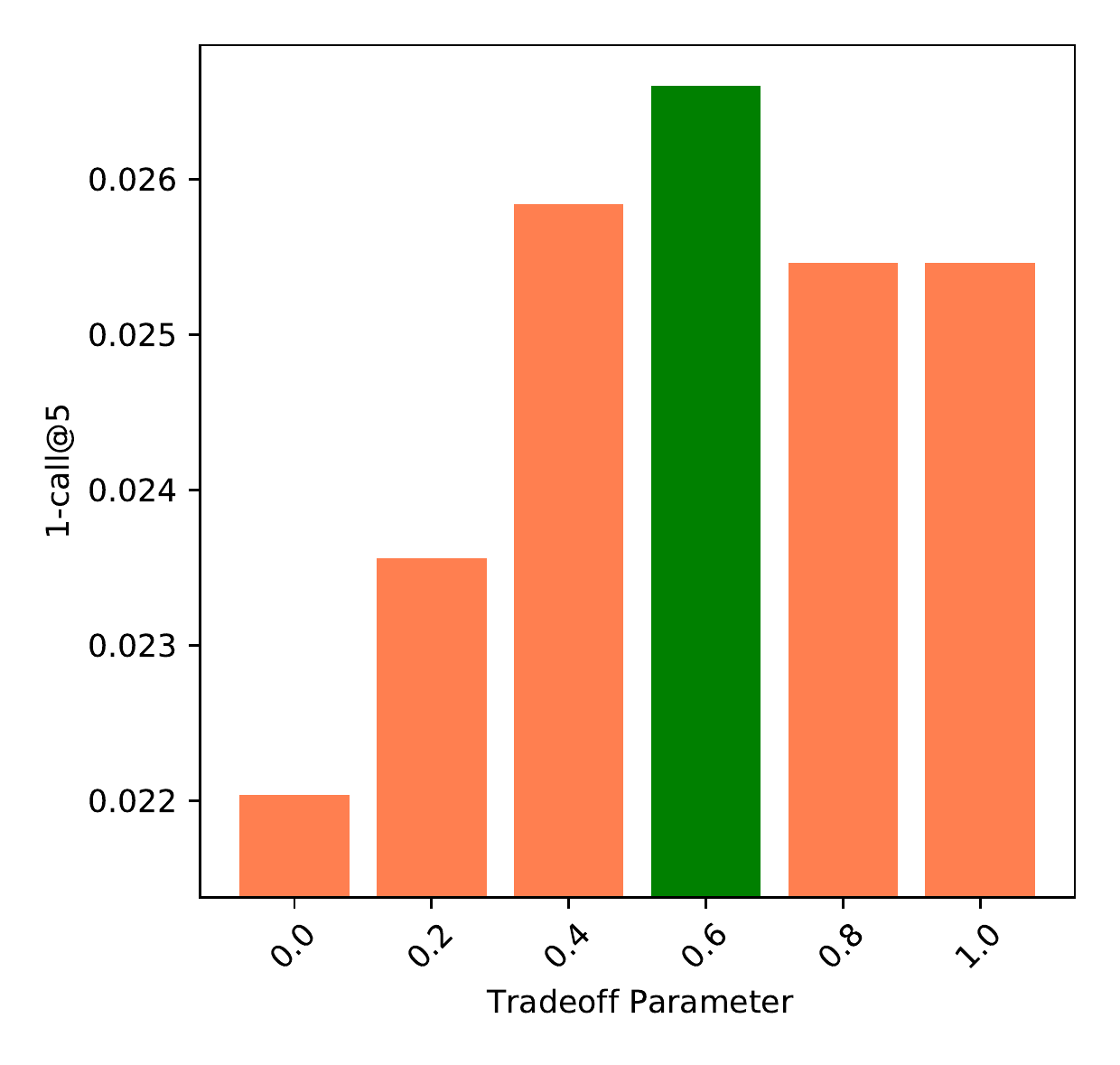,height=0.60in,width=0.41in }&
			\psfig{figure=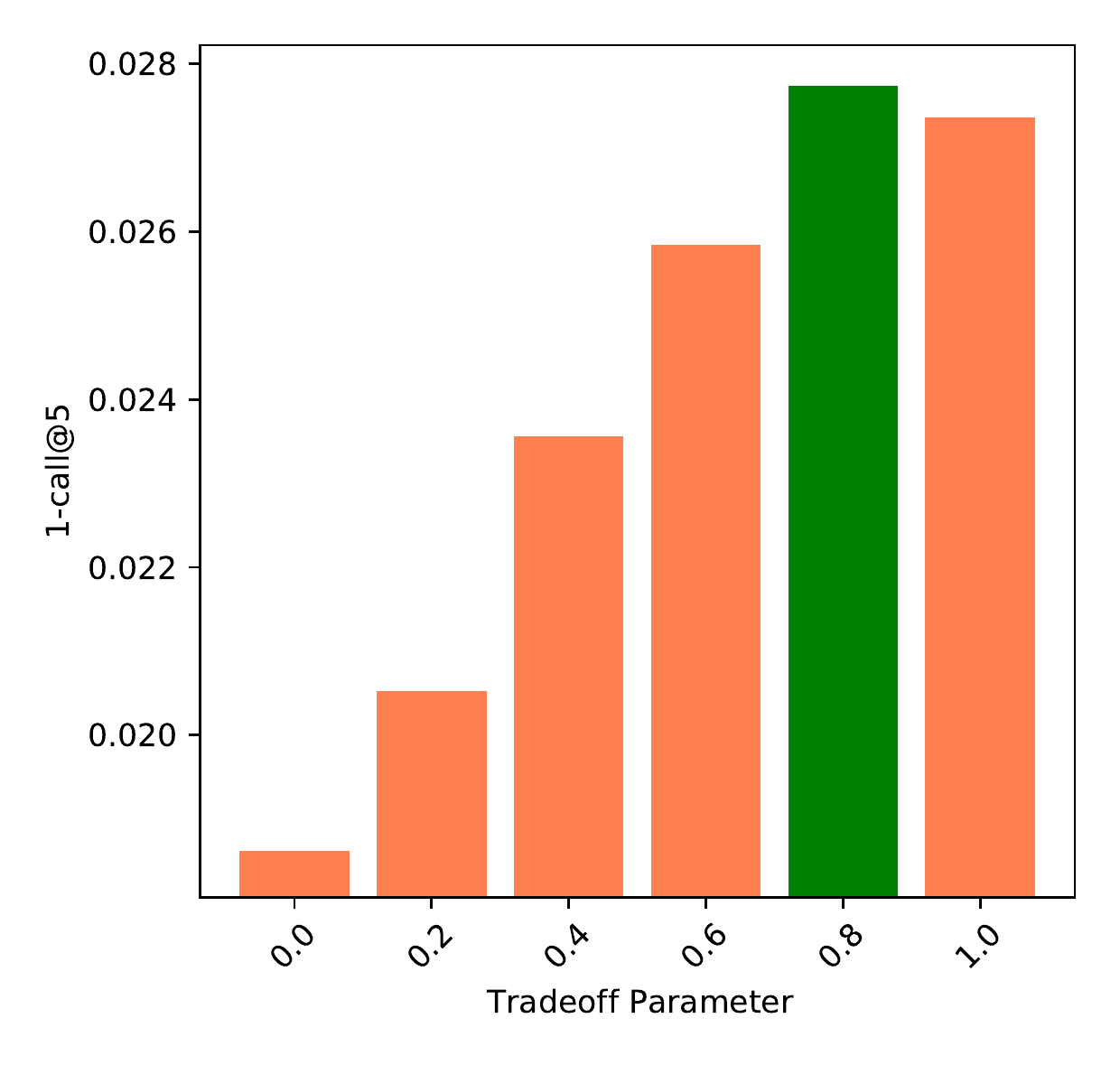,height=0.60in,width=0.41in }&
			&
			
			\psfig{figure=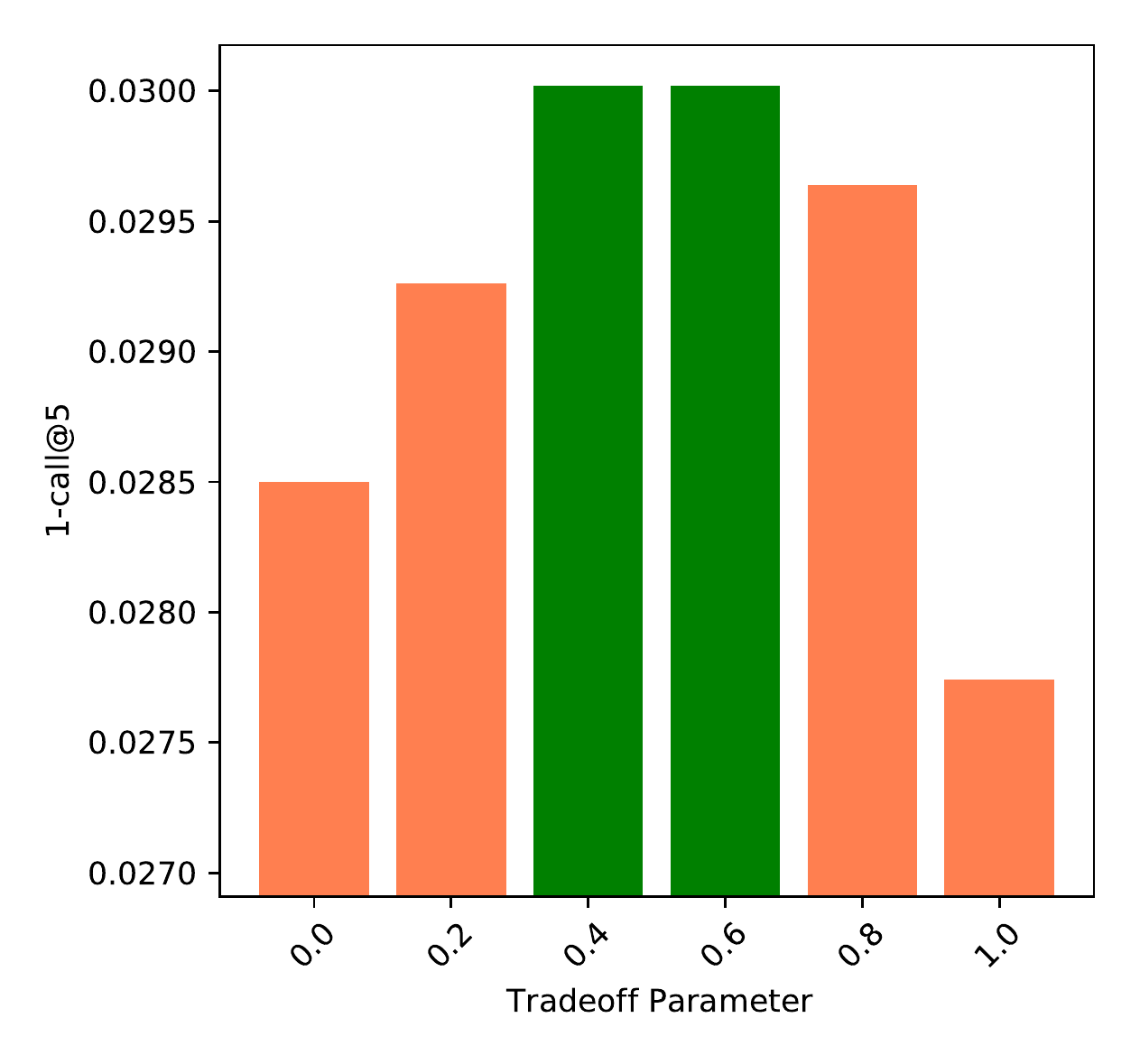,height=0.60in,width=0.41in }&
			\psfig{figure=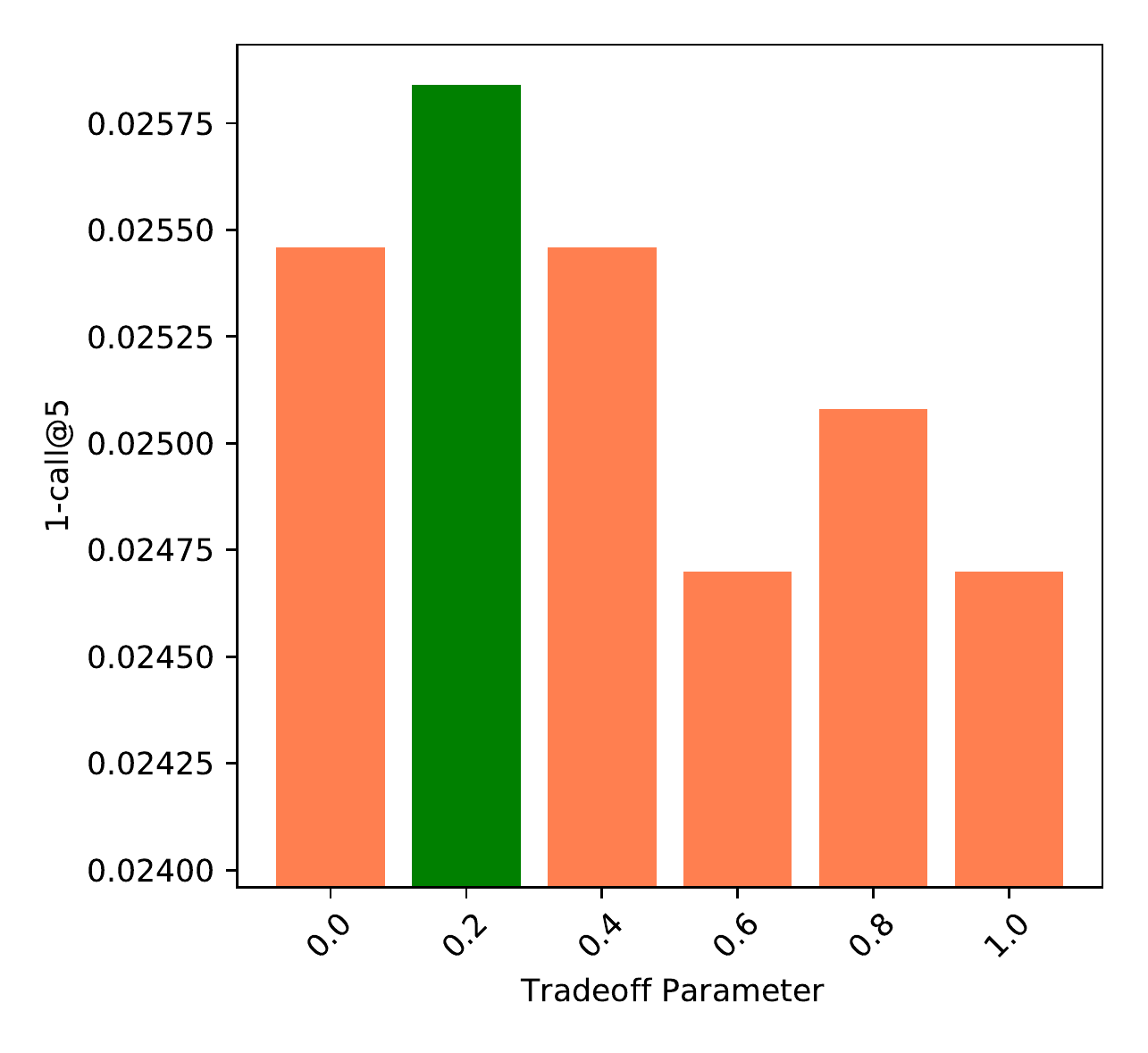,height=0.60in,width=0.41in }&
			\psfig{figure=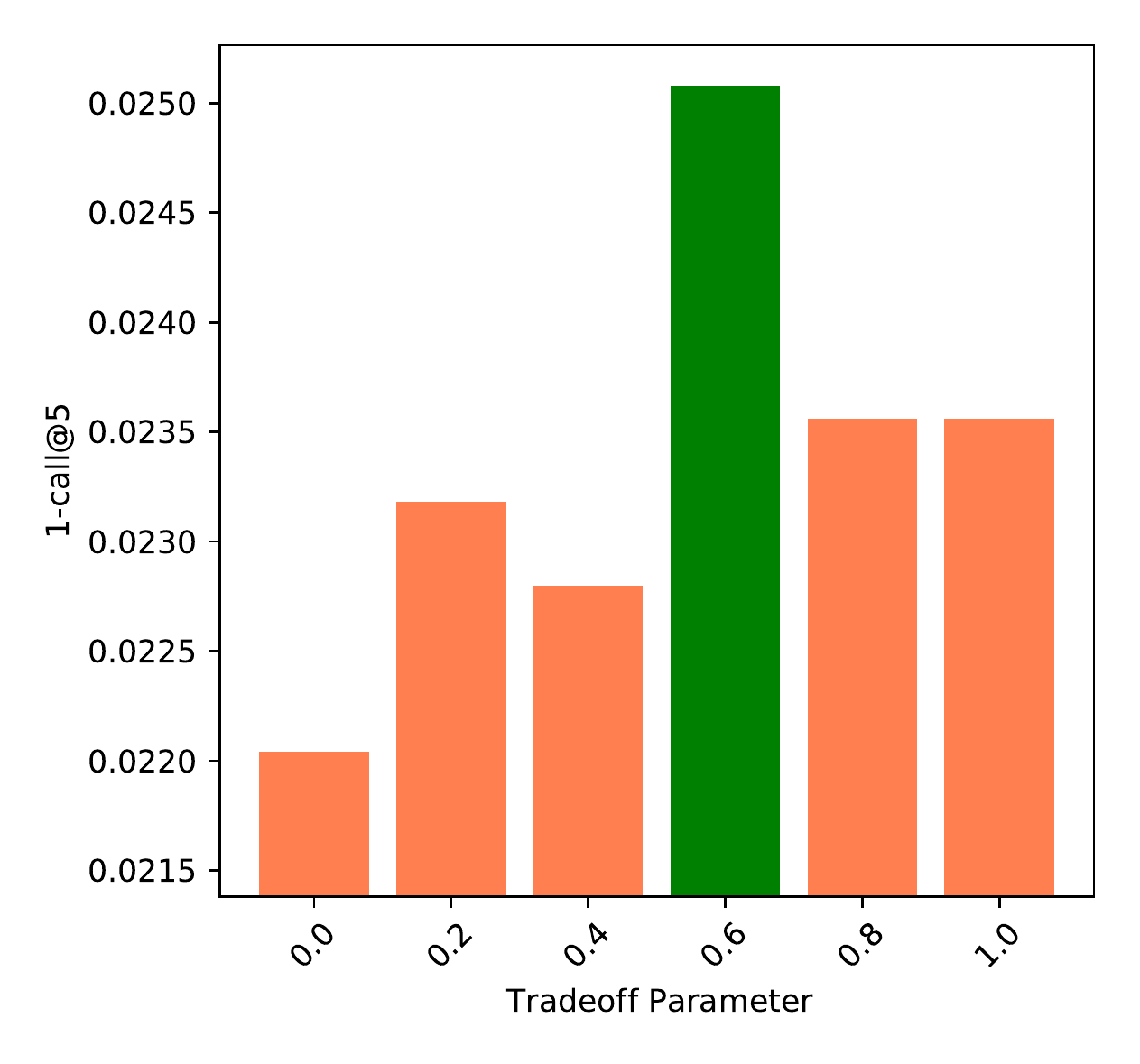,height=0.60in,width=0.41in }&
			\psfig{figure=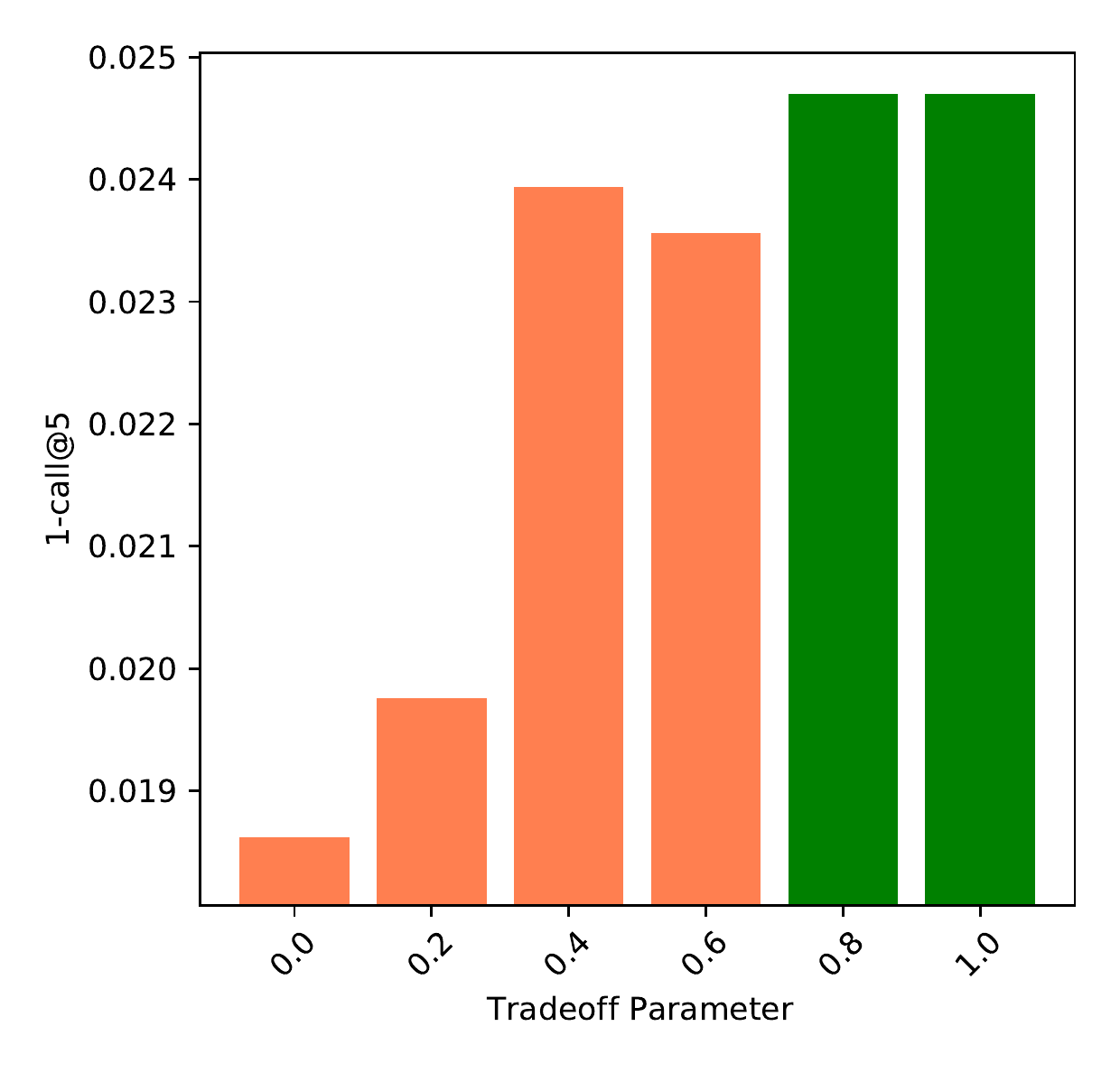,height=0.60in,width=0.41in }&
			&
			
			\psfig{figure=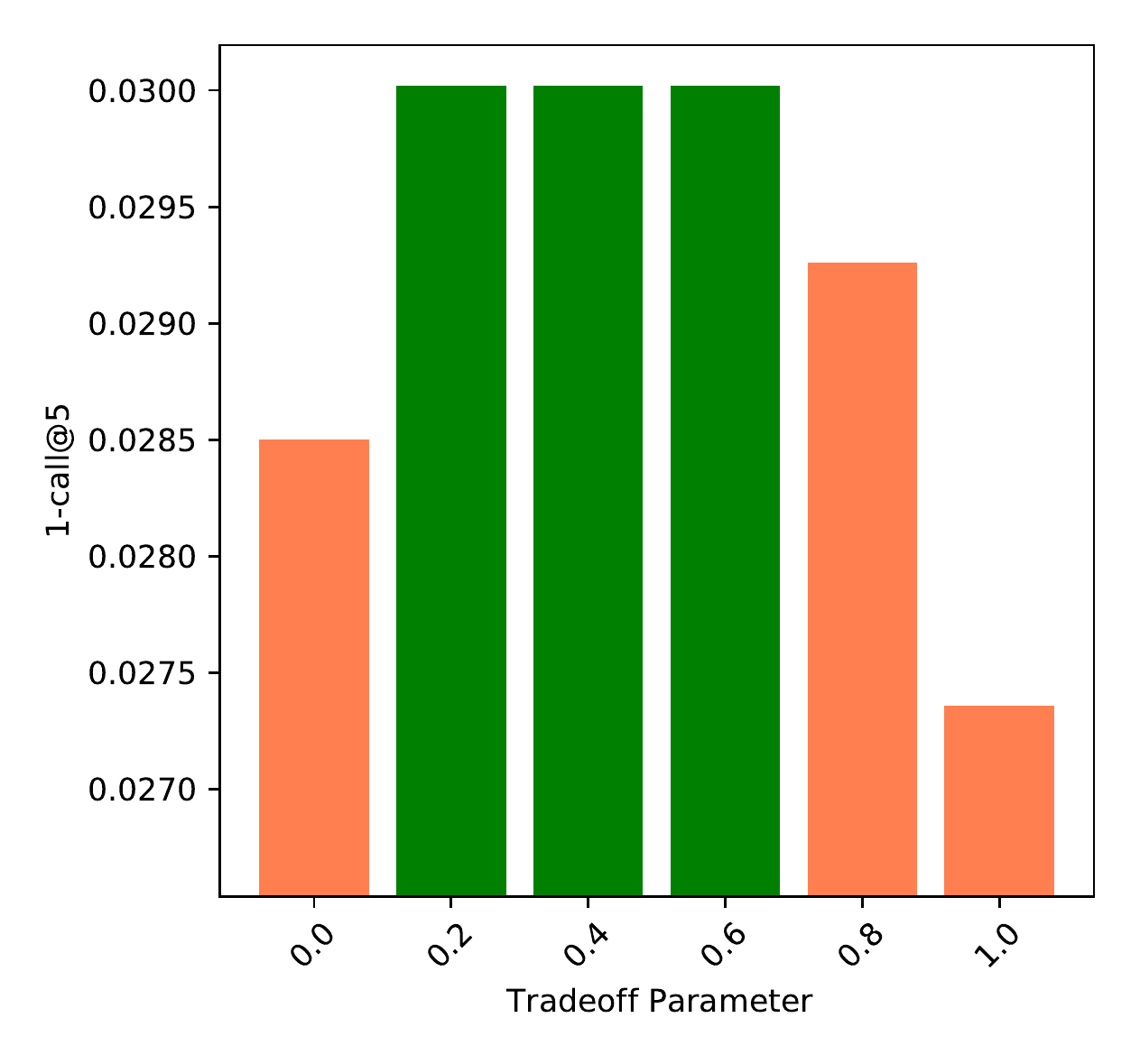,height=0.60in,width=0.41in }&
			\psfig{figure=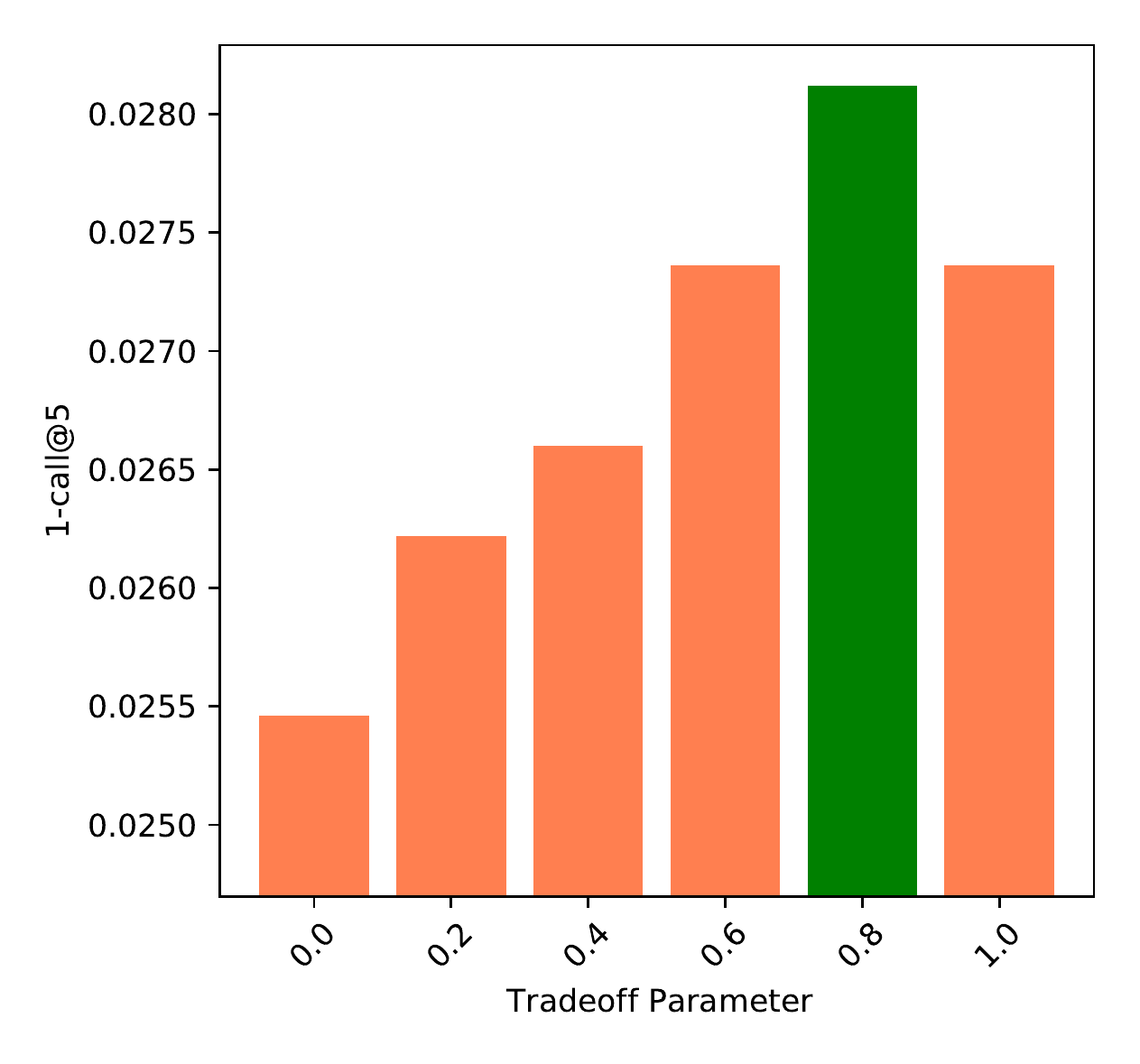,height=0.60in,width=0.41in }&
			\psfig{figure=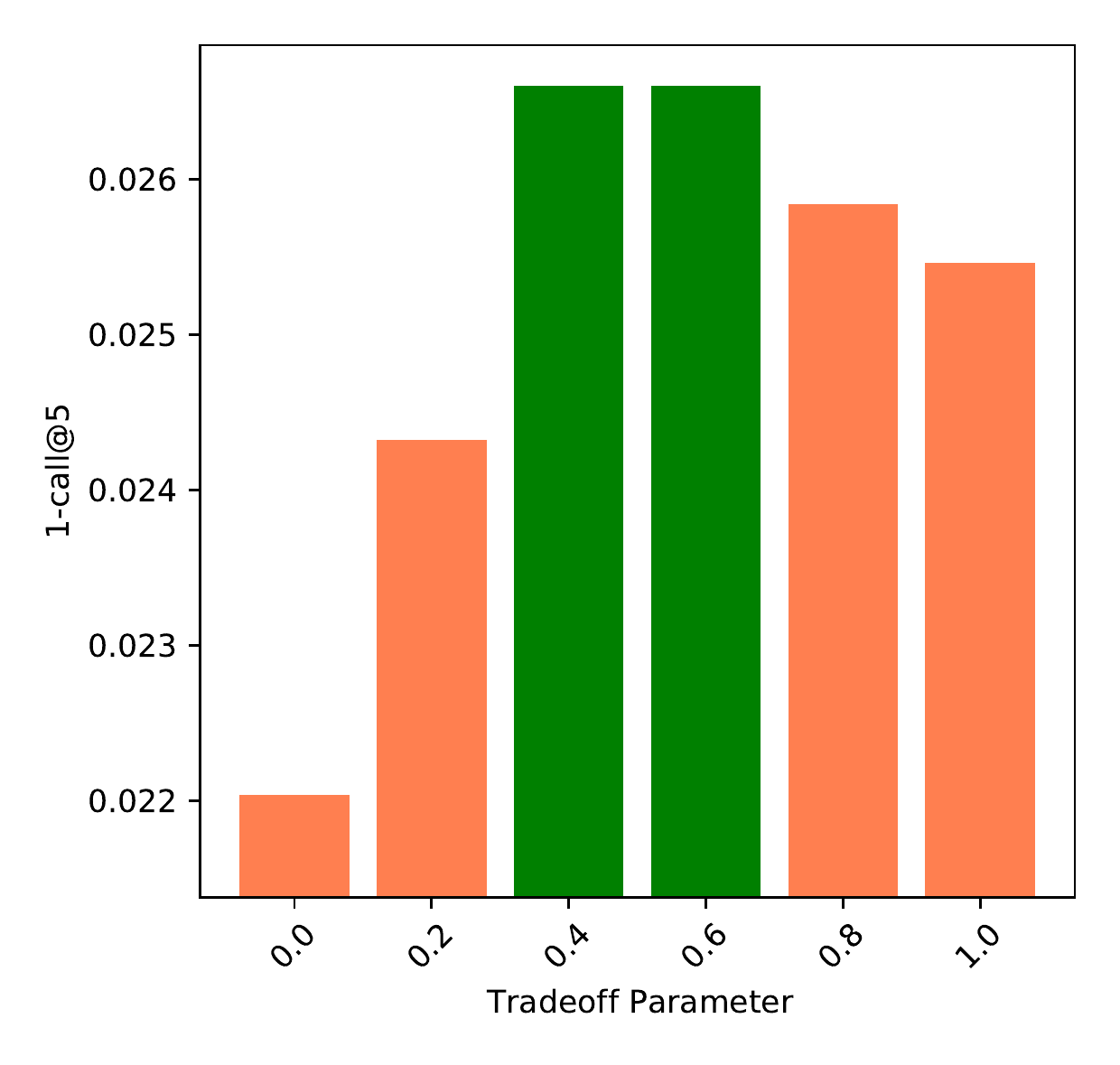,height=0.60in,width=0.41in }&
			\psfig{figure=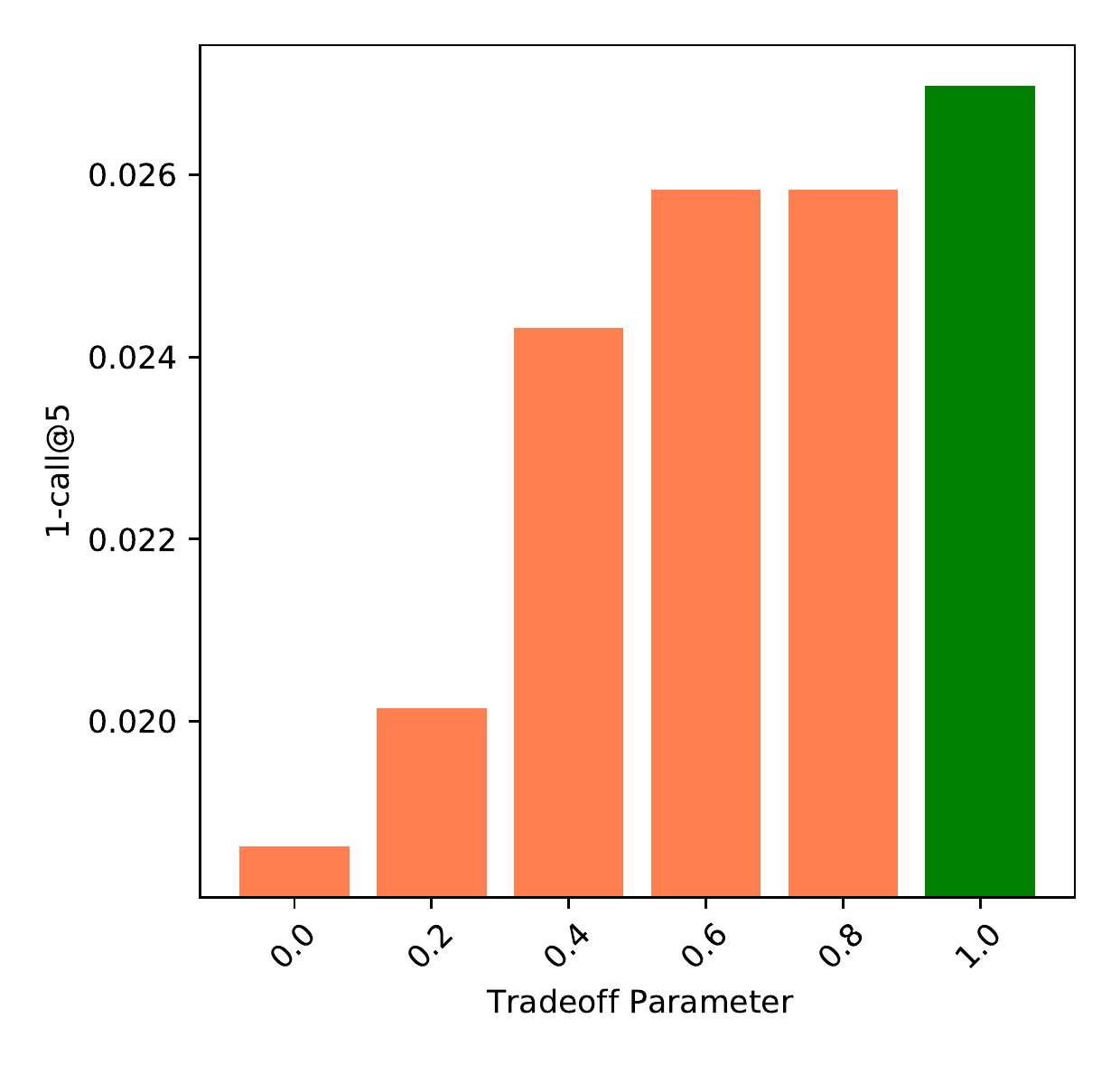,height=0.60in,width=0.41in }
			
			\\
			
			\psfig{figure=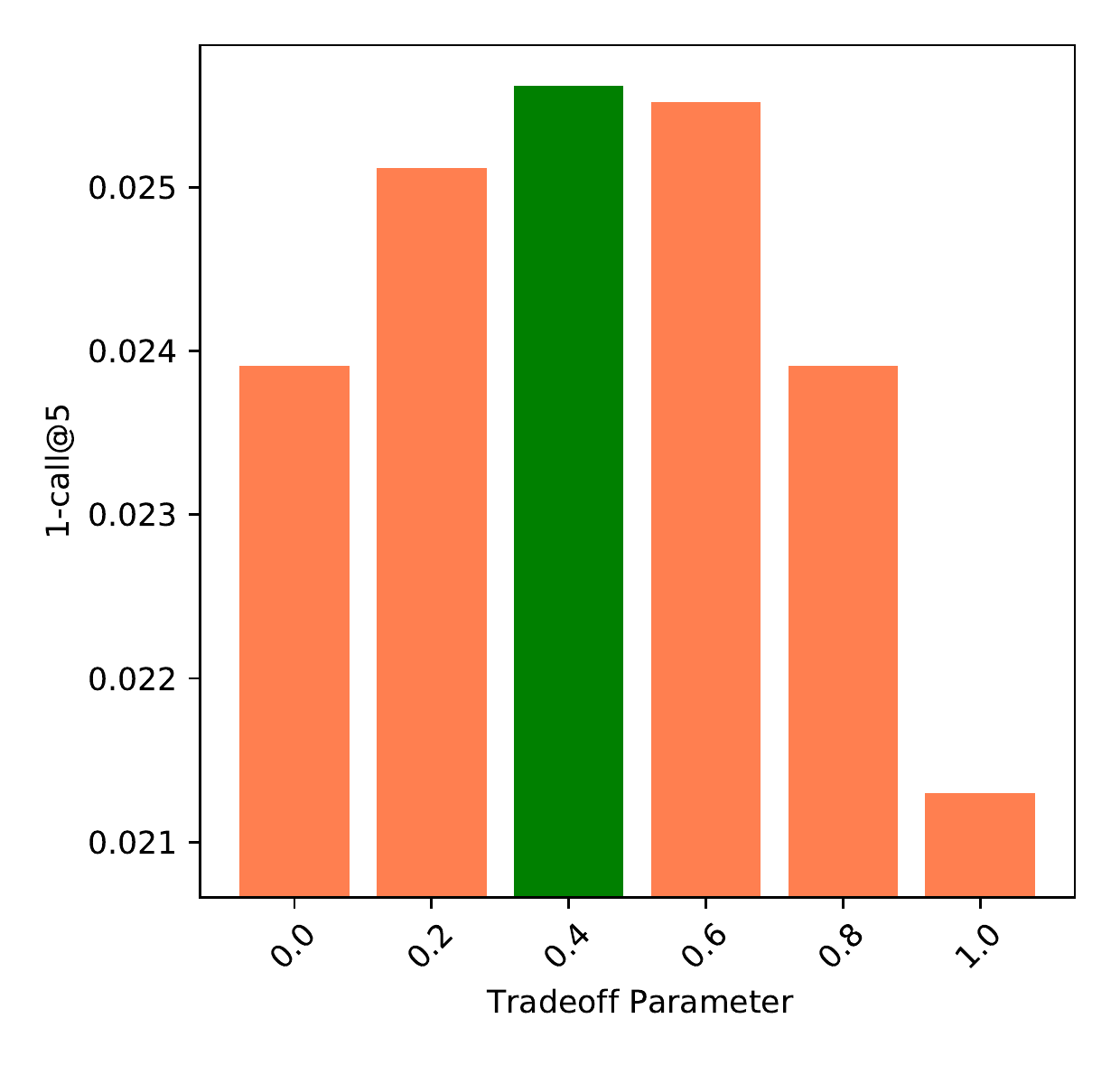,height=0.60in,width=0.41in }&
			\psfig{figure=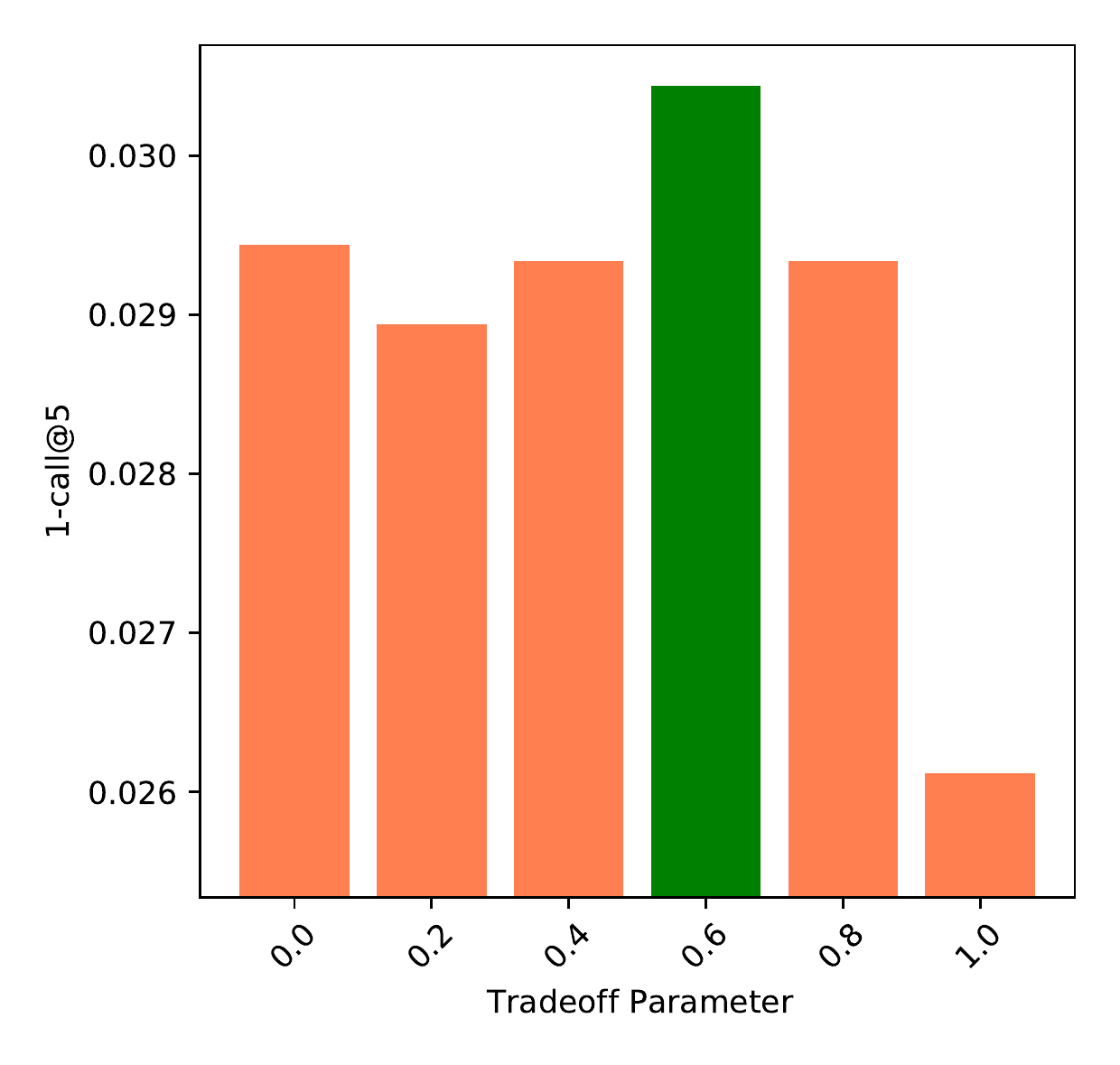,height=0.60in,width=0.41in }&
			\psfig{figure=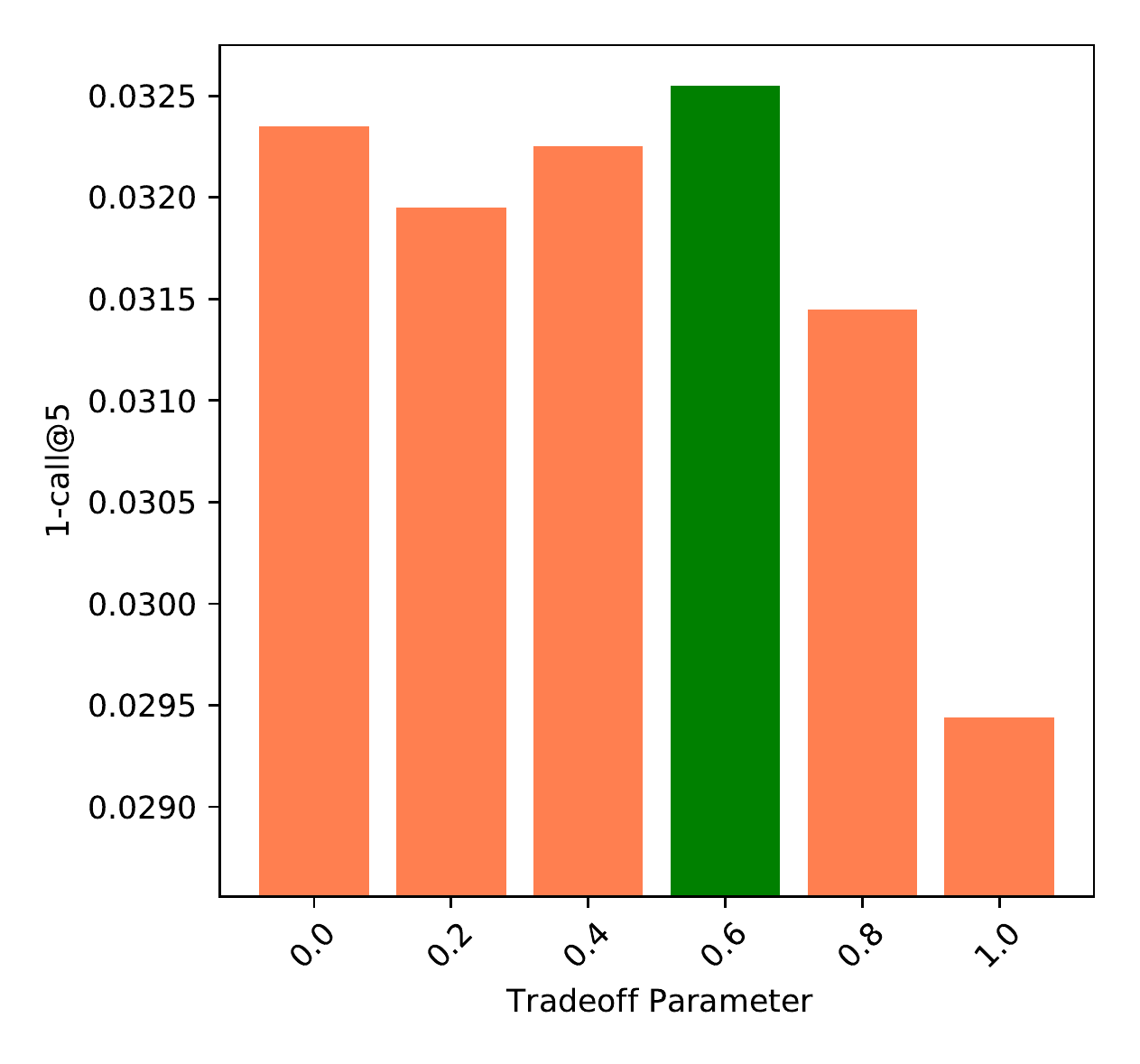,height=0.60in,width=0.41in }&
			\psfig{figure=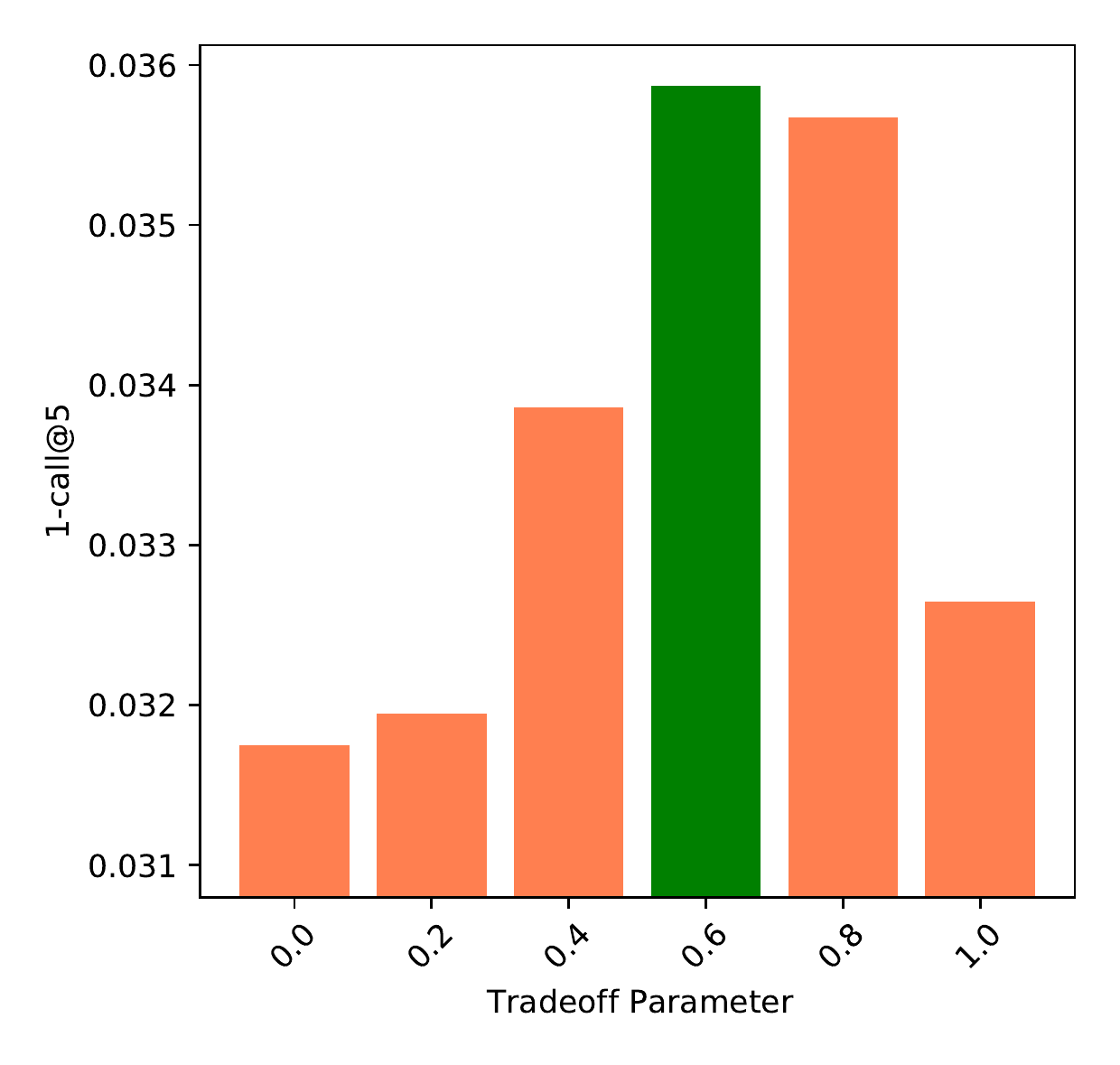,height=0.60in,width=0.41in }&
			
			&
			
			\psfig{figure=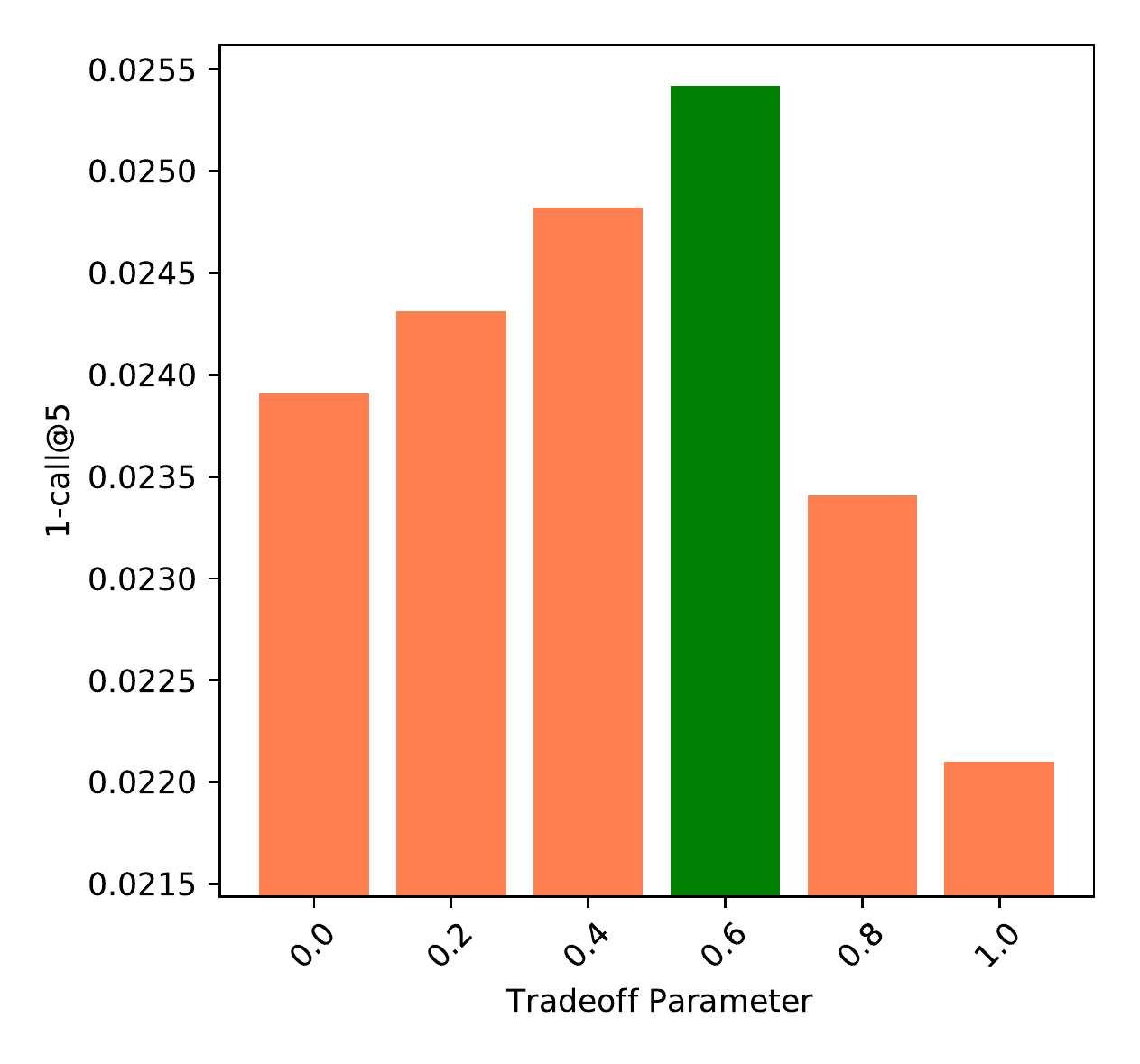,height=0.60in,width=0.41in }&
			\psfig{figure=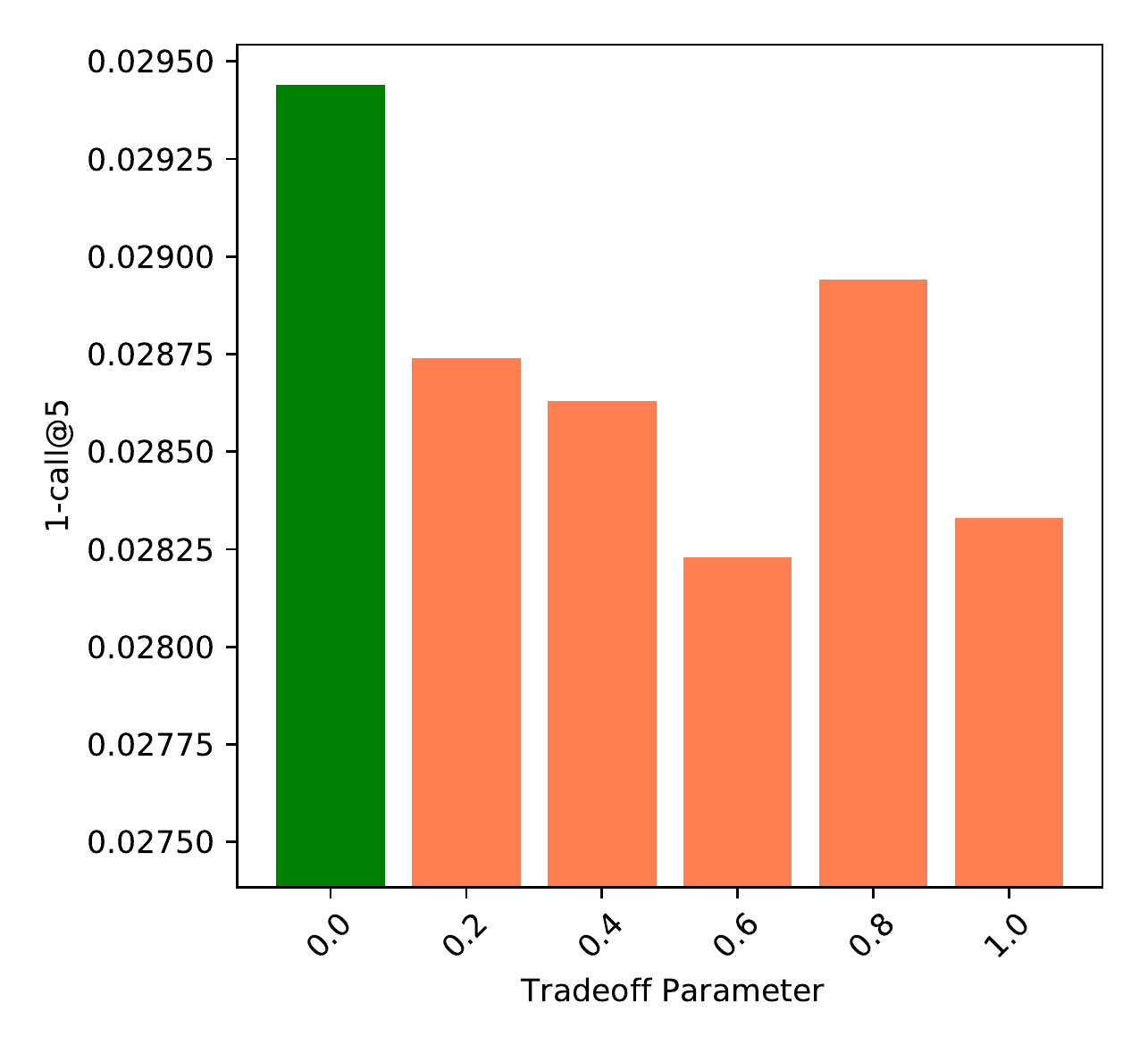,height=0.60in,width=0.41in }&
			\psfig{figure=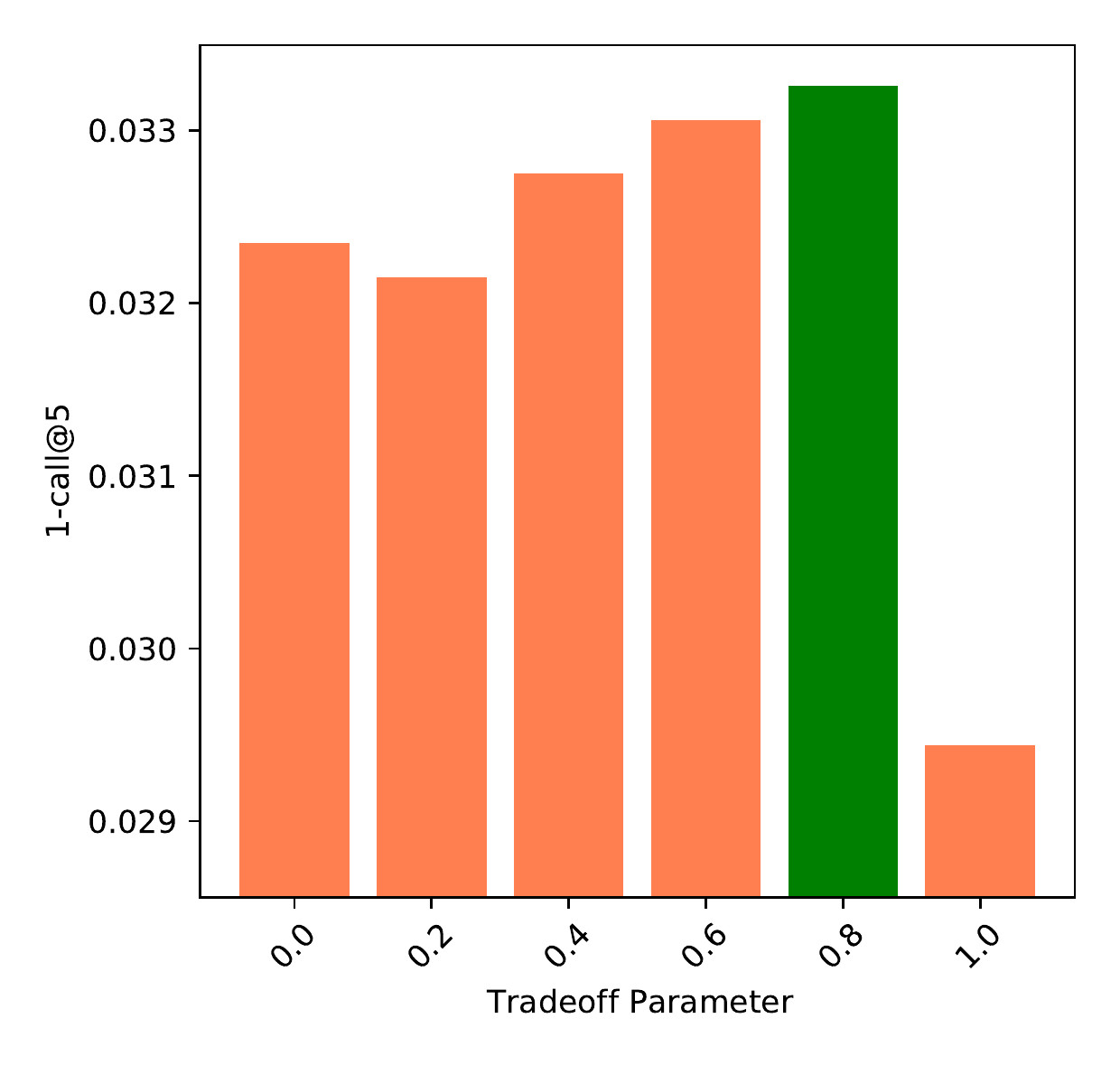,height=0.60in,width=0.41in }&
			\psfig{figure=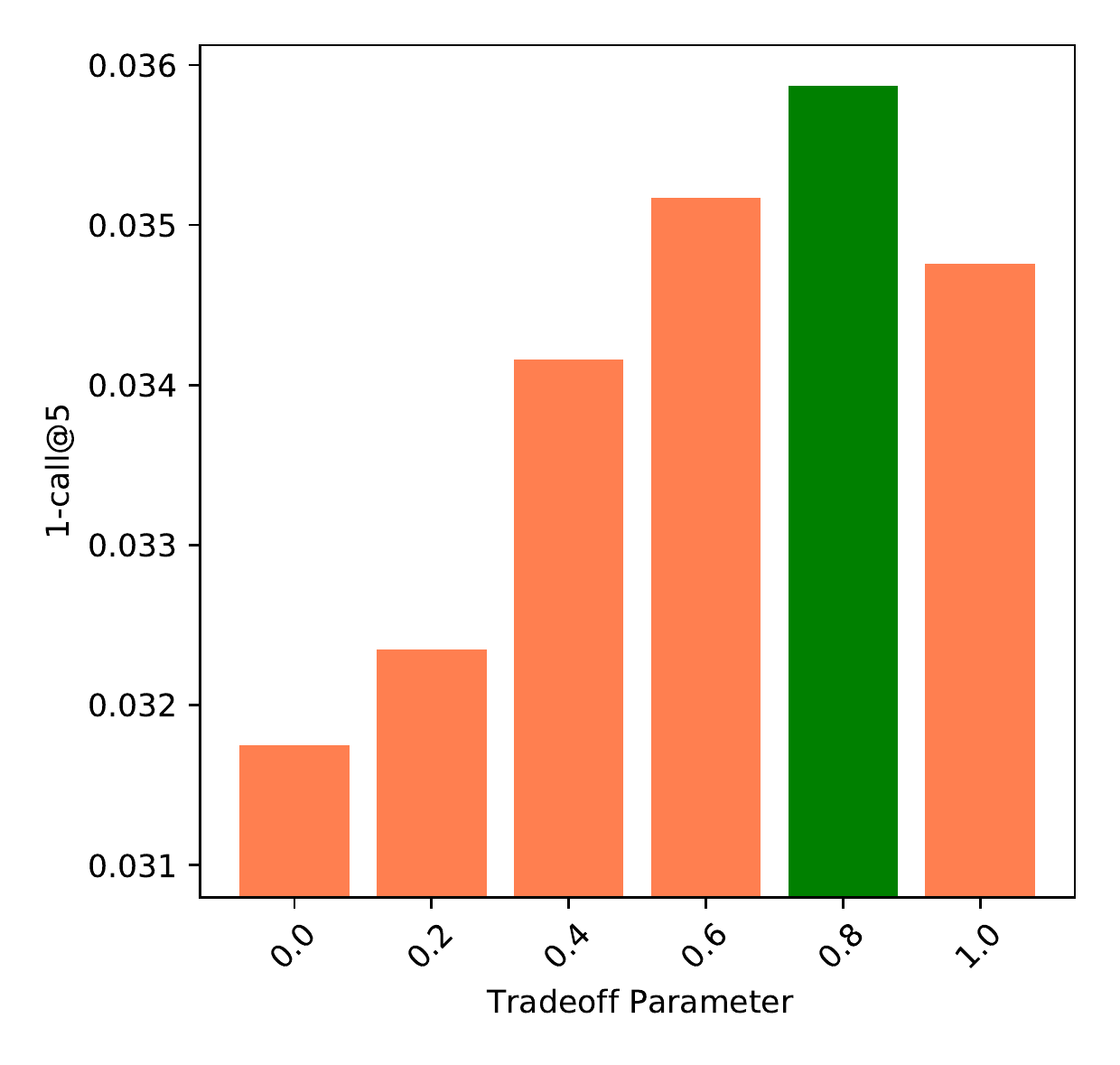,height=0.60in,width=0.41in }&
			
			&
			
			\psfig{figure=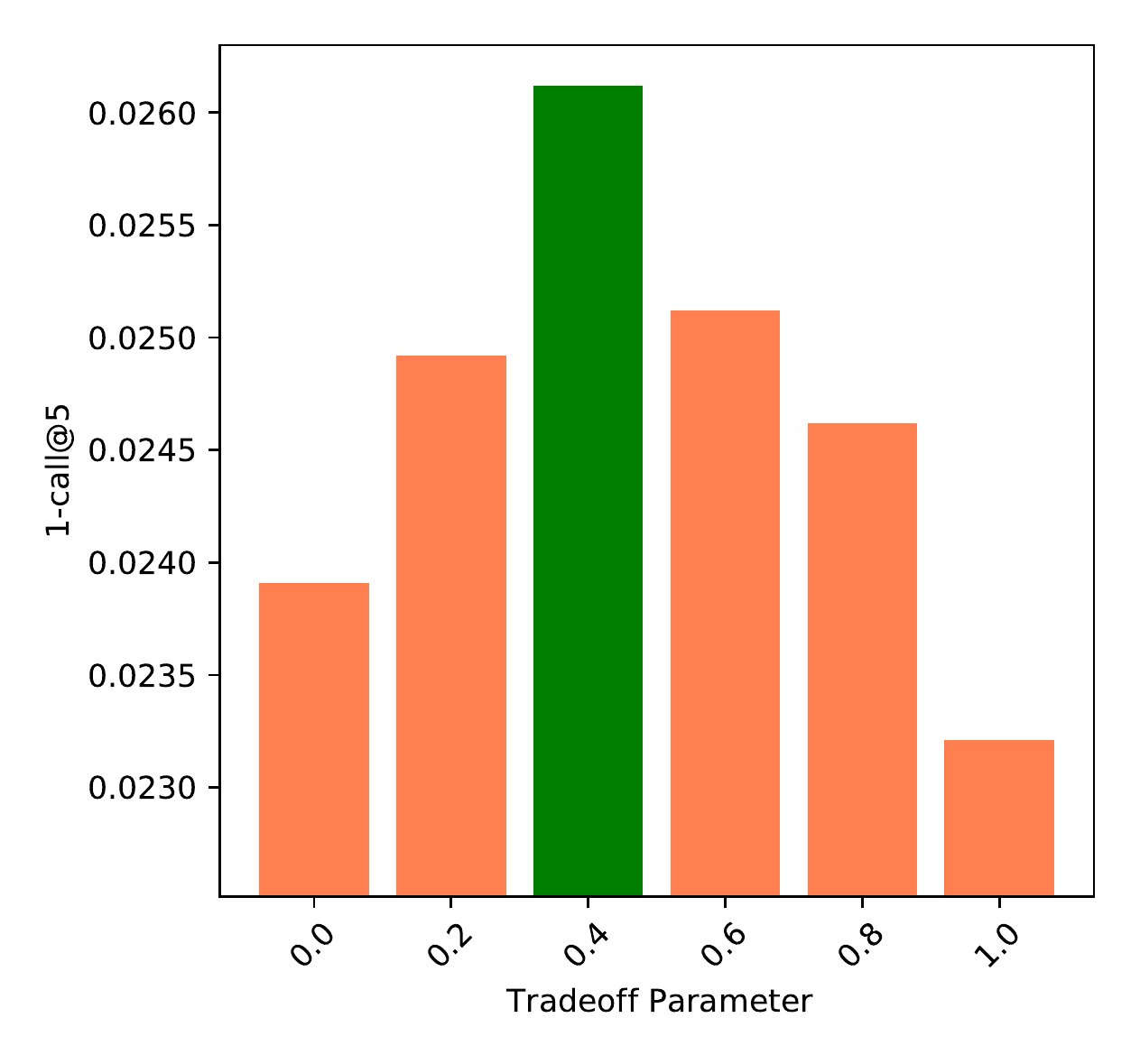,height=0.60in,width=0.41in }&
			\psfig{figure=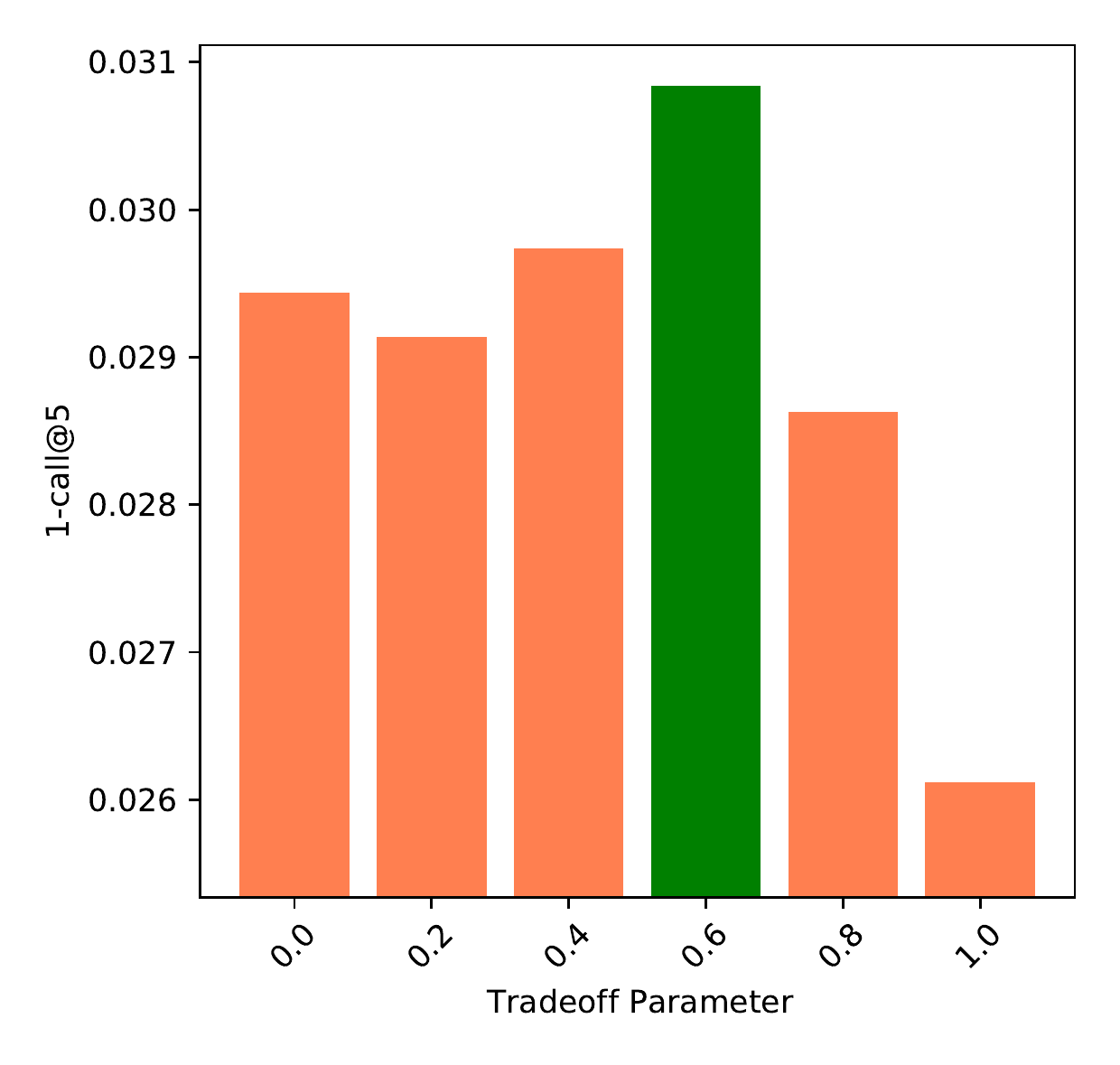,height=0.60in,width=0.41in }&
			\psfig{figure=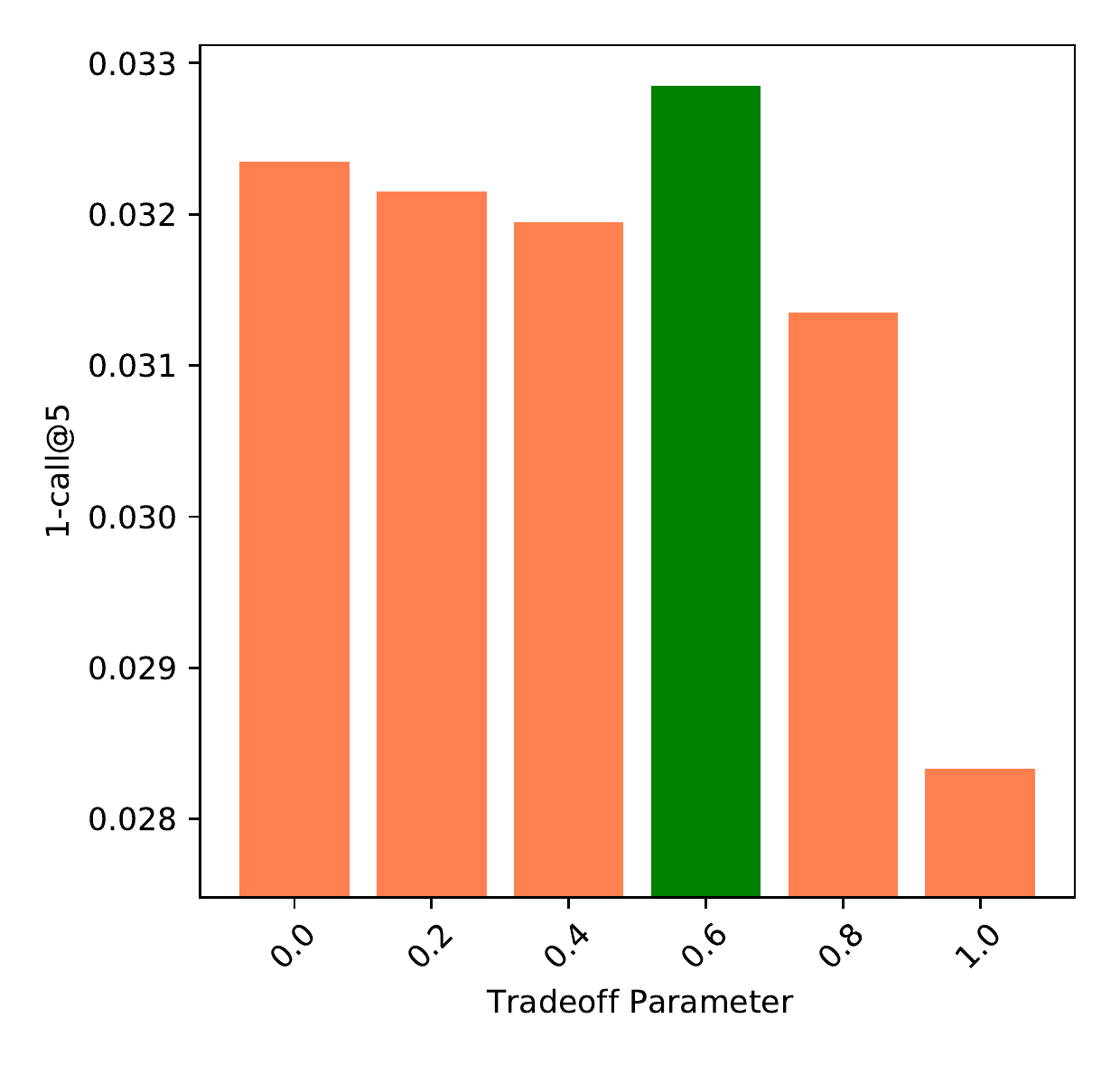,height=0.60in,width=0.41in }&
			\psfig{figure=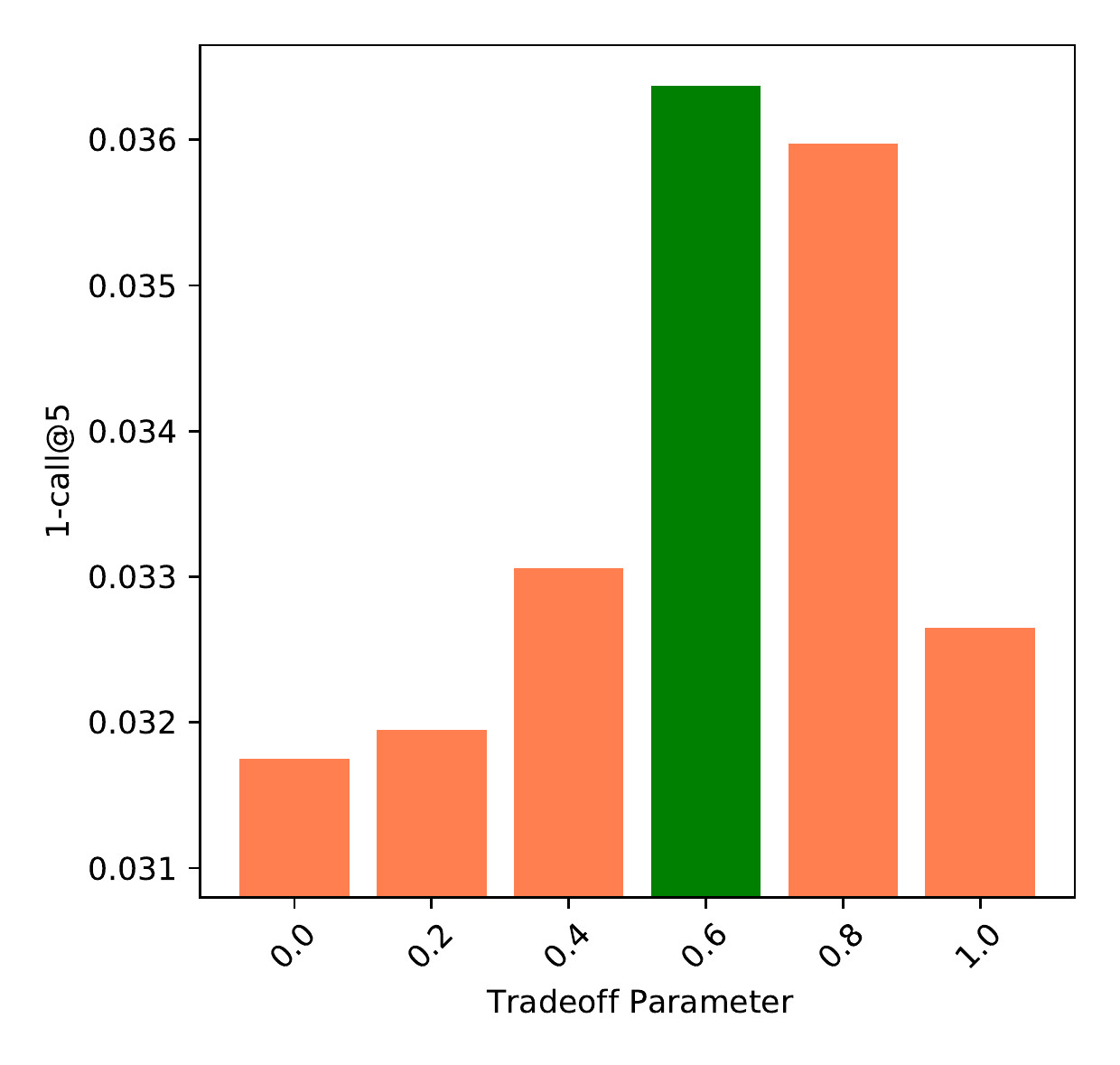,height=0.60in,width=0.41in }
			
			\\
			$k=5$ & $k=10$ & $k=20$ & $k=40$ && $k=5$ & $k=10$ & $k=20$ & $k=40$ && $k=5$ & $k=10$ & $k=20$ & $k=40$\\

		\end{tabular}
	}
	\end{center}
	\caption{
		Exploration of the influence of the tradeoff parameter $\lambda$ on the proposed PAS-based collaborative filtering method (from top to bottom: ML10M, Netflix, Beauty and Steam). columns 1-4,  columns 5-8 and columns 9-12 correspond to the empirical results of PAS$_a(\lambda)$, PAS$_b(\lambda)$ and PAS$_c(\lambda)$, respectively. The bar in green in each subfigure denotes the best performance on the metric $\text{1-call}@5$. 
	}\label{fig:lambda}
\end{figure*}

Table \ref{tbl:summary2} is a summary of some statistics observed from the experimental results shown in Figure \ref{fig:lambda}. Specifically, for the empirical results of PAS$_a(\lambda)$, we observe that among the total $ 16 $ cases, $18.75\%$ ($ 3/16 $) of the best $\lambda$ w.r.t. recommendation performance fall on the range $[0,0.4]$ while $81.25\%$ ($ 13/16 $) of the best $\lambda$ fall on the range $[0.6,1.0]$. We can refer to Table \ref{tbl:summary2} to see the similar observations of PAS$_b(\lambda)$ and  PAS$_c(\lambda)$. In conclusion, we observe that among the total $ 48 $ cases, $25\%$ ($ 12/48 $) of the best $\lambda$ w.r.t. recommendation performance fall on the range $[0,0.4]$ while $81.25\%$ ($ 39/48 $) of the best $\lambda$ fall on the range $[0.6,1.0]$. Kindly note that for some cases, there can be more than one best $\lambda$, which means the best $\lambda$ can fall on $[0,0.4]$ and $[0.6,1.0]$ at the same time. We believe that such observations indicate the significant role that the position-aware binary indicator (i.e., the term $\widetilde{\delta}_{ p_{vii'} }^{ (\ell, L_u(i')) } $ in Equation (\ref{eq:PAS})) plays in the proposed PAS framework.

\begin{table}[!htb]
	

	\caption{
		\MakeUppercase{Statistics observed from Figure \ref{fig:lambda}. Among the total $ 16 $ cases of PAS$_\MakeLowercase{a}(\lambda)$, $18.75\%$ ($ 3/16 $) of the best $\lambda$ w.r.t. recommendation performance fall on the range $[0,0.4]$ while $81.25\%$ ($ 13/16 $) of the best $\lambda$ fall on the range $[0.6,1.0]$. Kindly note that for some cases, there can be more than one best $\lambda$, which means the best $\lambda$ can fall on $[0,0.4]$ and $[0.6,1.0]$ at the same time. The observations of PAS$_\MakeLowercase{b}(\lambda)$ and PAS$_\MakeLowercase{c}(\lambda)$ can be interpreted similarly.}	
	}
	
	\label{tbl:summary2}
	\vspace{-2mm}
	\begin{center}
		
		
		\begin{tabular}{ c|ccc  }
			\toprule
			& PAS$_a(\lambda)$ &PAS$_b(\lambda)$&PAS$_c(\lambda)$\\\midrule
			
			$\lambda\in[0,0.4]$&$ 18.75\%(3/16) $&$ 31.25\%(5/16) $&$ 25\%(4/16) $\\
			
			$\lambda\in[0.6,1.0]$&$ 81.25\%(13/16) $&$ 75\%(12/16) $&$ 87.5\%(14/16) $\\
			
			\bottomrule
			
		\end{tabular}
		
	\end{center}
	\vspace{-3mm}
\end{table}

\section{Conclusions and Future Work}\label{sec:conclusion}

In this paper, we devise a novel similarity measurement called position-aware similarity (PAS) and propose a novel collaborative filtering method for sequential recommendation based on the proposed PAS. We first notice that a very recent similarity measurement called BIS, though being able to capture the sequential patterns, is calculated independent of the position information within the current input sequence. Besides, we find that by introducing the position-aware binary indicator alone, the proposed PAS(uni) may suffer from a sparsity problem and then become unreliable. We therefore (\romannumeral1) devise a novel position-aware similarity measurement, which considers the position information within the current input sequence when leveraging the advantages of BIS to avoid the potential sparsity problem and (\romannumeral2) introduce a scaling function $ h(\cdot) $ to alleviate the sparsity problem. We also conduct comprehensive empirical studies to verify the advantages of our PAS over BIS and show the competitiveness of the proposed method compared with the popular factorization-based and the state-of-the-art GNN-based sequential recommendation methods. 

For future works, firstly we are interested in devising the position-aware binary indicator in a new way that is not susceptible to the sparsity problem.
Moreover, we intend to explore how different levels of granularities with respect to the item positions may affect the recommendation performance of our PAS-based method.

\section*{Acknowledgment}
We thank the support of National Natural Science Foundation of China No. 62172283.


\bibliographystyle{IEEEtran}
\bibliography{sample-base}

\end{document}